\documentclass[aps,superscriptaddress,rmp,reprint,amsmath,amssymb,graphicx,longbibliography]{revtex4-1}
\usepackage[T1]{fontenc}
\usepackage[utf8]{inputenc}
\usepackage{graphicx}
\usepackage{amsmath}
\usepackage{braket}
\usepackage{xcolor}
\usepackage{amsfonts}

\makeatletter

\usepackage{hyperref}
 \hypersetup{
pdftitle={},
pdfauthor={},
colorlinks=true, 
linkcolor=black, 
citecolor=blue, 
filecolor=magenta, 
urlcolor=black 
}

\makeatother

\usepackage[english]{babel}
\begin{document}

\title{Topologically Protected Transport in Engineered Mechanical Systems}

\author{Tirth Shah}

\email{tirth.shah@mpl.mpg.de}
\affiliation{Max Planck Institute for the Science of Light, Staudtstrasse 2, 91058 Erlangen, Germany}
\affiliation{Department of Physics, Friedrich-Alexander Universit\"at Erlangen-N\"urnberg, Staudtstrasse 7, 91058 Erlangen, Germany}

\author{Christian Brendel}

\affiliation{Max Planck Institute for the Science of Light, Staudtstrasse 2, 91058 Erlangen, Germany}

\author{Vittorio Peano}

\affiliation{Max Planck Institute for the Science of Light, Staudtstrasse 2, 91058 Erlangen, Germany}

\author{Florian Marquardt}

\affiliation{Max Planck Institute for the Science of Light, Staudtstrasse 2, 91058 Erlangen, Germany}
\affiliation{Department of Physics, Friedrich-Alexander Universit\"at Erlangen-N\"urnberg, Staudtstrasse 7, 91058 Erlangen, Germany}

\begin{abstract}
Mechanical vibrations are being harnessed for a variety of purposes and at many length scales, from the macroscopic world down to the nanoscale. The considerable design freedom in mechanical structures allows to engineer new functionalities. In recent years, this has been exploited to generate setups that offer topologically protected transport of vibrational waves, both in the solid state and in fluids. Borrowing concepts from electronic physics and being cross-fertilized by concurrent studies for cold atoms and electromagnetic waves, this field of topological transport in engineered mechanical systems offers a rich variety of phenomena and platforms. In this review, we provide a unifying overview of the various ideas employed in this area, summarize the different approaches and experimental implementations, and comment on the challenges as well as the prospects.

\end{abstract}
\maketitle

\tableofcontents

\section{Introduction}

% ...

Topology is a branch of mathematics that deals with properties of objects that are invariant under smooth deformation. Even in the earliest days of research in topology, potential connections to physics were already pointed out. In the 1860s, for example, Lord Kelvin speculated that atoms might be knots in the aether, and this inspired the first attempts at a complete classification of knots. While this idea was not borne out, topological defects are now known to occur in a large variety of important physical settings, ranging from vortices in fluids and superfluids and dislocations in crystalline solids to liquid crystals as well as magnetic skyrmions. Even on the largest scales, topological defects may play a role in the form of cosmological strings. 

These are all examples of topology directly present in real-space structures. A more subtle but far-reaching example of topological physics was uncovered upon closer inspection of the Quantum Hall Effect. After its discovery \cite{klitzing_new_1980} in 1980, the astonishing precision of the quantization of Hall resistance in a 2D electron sample in a magnetic field -- even in the presence of disorder -- called for a fundamental explanation. Already the early explanation given by Laughlin was at its heart a topological argument \cite{laughlin_quantized_1981}. However, the entirety of the deep connection to mathematical topology was uncovered a bit later, when Thouless et al. showed that the quantized Hall conductance is directly related to a topological invariant, the so-called Chern number \cite{thouless_quantized_1982, kohmoto_topological_1985}. In general, topological invariants, i.e. quantities that do not change under smooth deformations, play a prominent role in any analysis of topology. Simple examples include winding numbers for vortices and the number of holes in an object. Remarkably, while the Chern number can be calculated for the infinitely extended bulk system, it is directly related to the number of edge channels that carry electrical current along the boundary of the sample. This is the celebrated bulk-boundary correspondence. These edge channels are topologically protected against backscattering, which explains the precision of conductance quantization even in the presence of disorder.

In 1988, Haldane \cite{haldane_model_1988} showed that it is the breaking of time-reversal symmetry, and not a finite magnetic field, which is responsible for non-zero Chern numbers and the unidirectional edge channels, leading to the general discussion of so-called Chern insulators. Later, in 2005, Kane and Mele pointed out that it is even possible to obtain topological transport for time-reversal-invariant systems \cite{kane_quantum_2005}, introducing the study of topological insulators. In such materials, the intrinsic spin-orbit coupling leads to spin-polarised undirectional edge channels (so-called "helical" channels). This phenomenon is now called Quantum Spin Hall effect. A review of these developments can be found in \cite{hasan_colloquium_2010}.

Given the importance of topologically protected transport in the domain of electronic systems, it was natural to ask whether other systems could show the same kind of physics. However, this raised a number of challenges, since many interesting particles and excitations are neither affected by magnetic fields nor do they naturally exhibit some kind of spin-orbit coupling. This pertains to cold atoms as well as photons and phonons. In all of these cases, some engineering is required to make progress towards the realization of topological transport.

First ideas for the design of artificial magnetic fields for neutral particles emerged in the field of cold atoms in optical lattices \cite{jaksch_creation_2003}. A few years later, the breaking of time-reversal symmetry for microwave photons using the magneto-optical effect as a means to produce chiral edge channels  was suggested and subsequently realized \cite{haldane_possible_2008,wang_observation_2009}. Since those days, considerable progress has been made in the design and exploitation of artificial chiral transport in cold atomic systems \cite{aidelsburger_artificial_2018,cooper_topological_2019} and photonic systems \cite{lu_topological_2014,ozawa_topological_2019}.

In this review, we will devote our attention to another important excitation: phonons, i.e. vibrations in solids or sound waves in fluids. Just like for photons and neutral cold atoms, the basic underlying mathematics is the same as that for electrons. This shared language across platforms has proven to be very beneficial, as it has enabled researchers to learn from ideas first advocated in other settings and adapt them in a suitable manner. On the other hand, despite this joint basis, there are important differences which make every platform unique with its own challenges and opportunities. For example, the design capabilities, the readout modalities, the possibilities of creating excitations and of coupling them to other systems are all vastly different, in addition to the large range in physical parameters. More specifically, for phononic systems, their  technological application potential is greatly aided by the compactness of the resulting devices when fabricated in the form of nanomechanical systems \cite{bachtold_mesoscopic_2022}. This is due to the relatively slow wave speed, allowing for devices that are a million times smaller than photonic systems at the same frequencies. As a solid-state platform for quantum technologies, they also offer unique efficient coupling to localized spins and other solid-state qubits (like superconducting qubits or quantum dots) \cite{clerk_hybrid_2020,barzanjeh_optomechanics_2022,safavi-naeini_controlling_2019}.
This could turn topologically protected phononic edge channels into a particularly promising way of interconnecting such qubits or quantum sensors. We may add that optomechanical interactions \cite{aspelmeyer_cavity_2014,barzanjeh_optomechanics_2022} represent one of the most promising ways to turn on-chip quantum information into photons for long-distance communication, adding to the power of on-chip phononic networks. All of these are reasons for studying topologically protected transport of phonons. Ideas like uni-directional amplification inside chiral edge channels add to these prospects.

As we will describe in detail in this review, the design of vibrational topological transport started in the macroscopic domain, including pendula and sound waves in flowing fluids, but is now moving into the micro- and nanoscopic domains which will unlock the promising applications mentioned above.

In the present review, we will concentrate on topologically protected transport along phononic edge channels, since that is the area most promising for applications. This means we will be dealing specifically with 1D edge channels at the boundary of 2D systems (which in the nanoscopic domain are engineered chip platforms). There are other interesting aspects of phonon topology that we will not cover, like zero-frequency modes in isostatic lattices \cite{kane_topological_2014},  0D localized states (e.g. at the boundary of the 1D Su-Schrieffer-Heeger model \cite{xiao_geometric_2015,yang_acoustic_2016,chaunsali_demonstrating_2017},  in the form of corner states \cite{serra-garcia_observation_2018}, or Dirac vortices \cite{ma_nanomechanical_2021}), and Weyl cones in 3D systems \cite{xiao_synthetic_2015,yang_acoustic_2016,li_weyl_2018,ge_experimental_2018,he_topological_2018,wang_multiple_2018,fruchart_soft_2018}. A few reviews on phononic topology already exist \cite{huber_topological_2016,zhang_topological_2018,liu_topological_2020,nassar_nonreciprocity_2020,miniaci_design_2021} but have a different scope compared to our work: \cite{huber_topological_2016} offered a short early review of the field, \cite{zhang_topological_2018}  covered acoustic systems, \cite{miniaci_design_2021} elastic systems, and \cite{nassar_nonreciprocity_2020} has its focus mainly on non-reciprocal systems. In the present review, we aim to present an up-to-date snapshot of the field, covering a wide variety of systems including both acoustic- and elastic-wave systems, as well as discrete mechanical systems like arrays of gyroscopes and pendula. We also give a comprehensive and unifying discussion of the physics behind the different design possibilities, emphasizing the conceptual connections and differences between  the various design schemes. 




Our review is organized as follows: We will first briefly describe the general (platform-independent) mathematics behind topologically protected transport: topology in band structures, the bulk-boundary correspondence, Chern numbers in systems with broken time-reversal symmetry, and topological insulators with intact time-reversal symmetry. We will then provide a first overview of the entire field of topological phonon transport, highlighting the general trends and some early works in this domain. The bulk of this review can be found in the subsequent Section \ref{sec:approaches-for-engineering}. There, we first discuss in some more detail the similarities and differences between electrons, electromagnetic waves and vibrations in the context of topological transport. We then describe each of the different approaches to designing topological transport: from engineering the breaking of time-reversal symmetry, thus leading to phononic Chern insulators, to the various design schemes that rely purely on suitable geometry and connectivity and do not require broken time-reversal symmetry. 
We finally conclude with a discussion of challenges, like mechanical dissipation, and future applications.

\section{General Background: Topological Transport of Waves}
\label{sec:background}

\begin{figure}
\includegraphics[width=1\columnwidth]{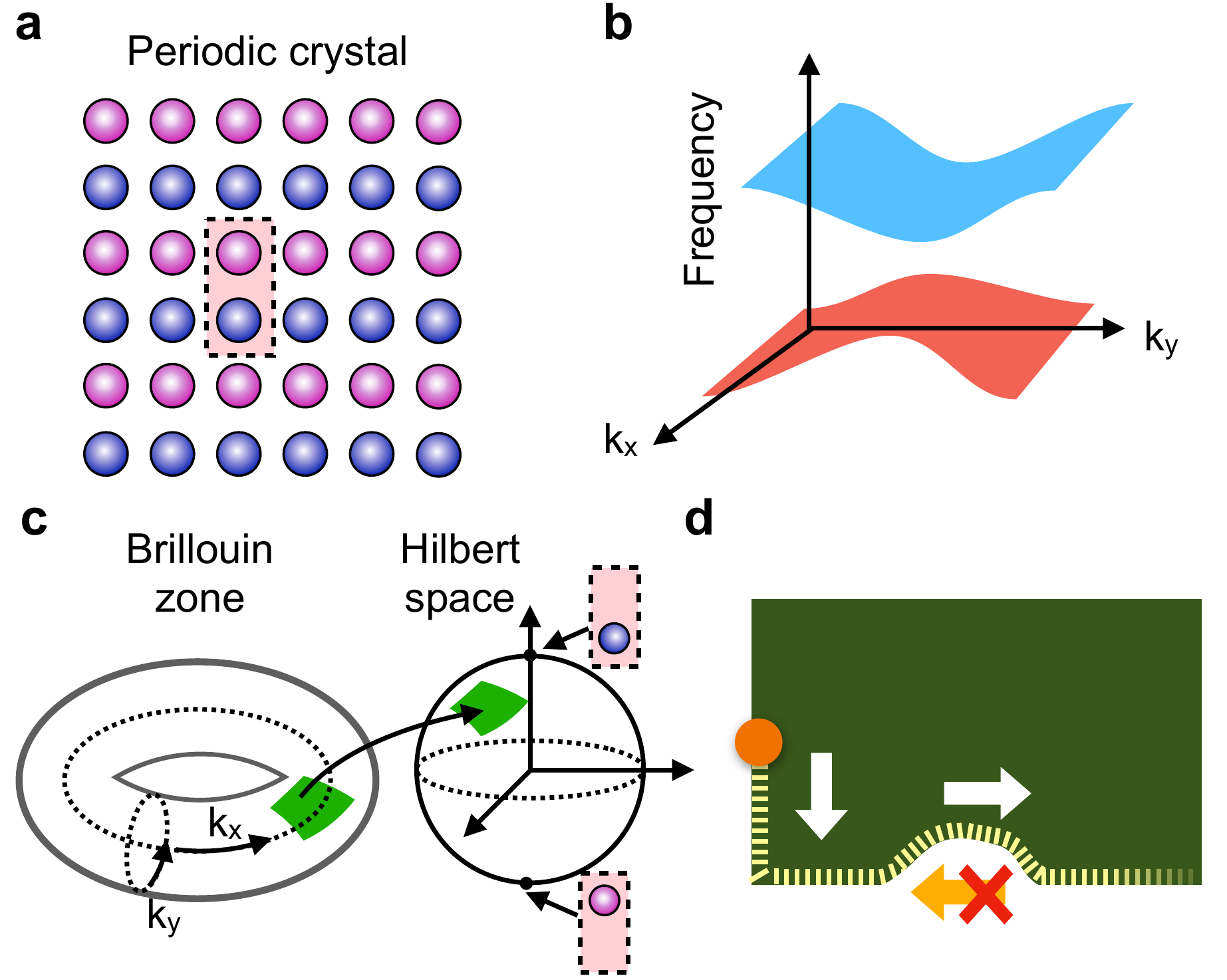}
\caption{
Illustration of topological transport. 
\textbf{a}, Schematic of a periodic crystal with two different sublattices. 
\textbf{b}, Fictitious band structure of (\textbf{a}), featuring two bands with a band gap in the middle. 
\textbf{c}, Topological features can be inspected by investigating the map from the Brillouin zone (torus) to the Hilbert space (here visualized as a Bloch sphere, for a two-band model), for each band. 
\textbf{d}, Topologically protected edge state featuring transport in only one direction (for a Chern insulator). The transport is immune to backscattering even at irregularities.
}
\label{fig:basic_concept}
\end{figure}


\subsection{Overview: Topology in Band Structures and Bulk-Boundary Correspondence}

In mathematics, topological properties of geometrical objects are those that are invariant under smooth deformations, such as the number of holes in the object. Even before the discussion of topological transport, the existence of topological features in physics had long been recognized. Early examples include topological defects in fields, such as vortices or skyrmions. Mathematically, a continuous deformation of one function (e.g., a field) into another function (a smoothly distorted version of the field) is denoted a "homotopy" and establishes that the two functions are topologically equivalent.

When we turn to the propagation of waves in an infinitely extended periodic medium (Fig.~\ref{fig:basic_concept}a), we arrive at the concept of Bloch waves as stationary solutions of the underlying wave equation. Attention often tends to focus on the band structure, i.e. the eigenfrequencies of these waves as a function of their quasimomentum inside the Brillouin zone, cf Fig.~\ref{fig:basic_concept}b. However, to discover topological features, one needs to inspect the behaviour of the Bloch waves themselves. The Brillouin zone is formally equivalent to a torus, i.e. a compact manifold. The Bloch waves live in a Hilbert space of suitably normalized wave functions, again a compact manifold. One may thus ask what happens to this map (see Fig.~\ref{fig:basic_concept}c) from Brillouin zone to Hilbert space when the underlying periodic medium is smoothly deformed (e.g. changing the geometry or potential), and thereby arrive at topological properties. These properties, typically expressed via so-called topological indices, remain invariant as long as the smooth deformations leave the map uniquely defined. This breaks down only whenever a band gap closes, leaving to a degeneracy at some point in the Brillouin zone. This is when topological properties can change abruptly.

Studying these topological properties by inspecting Bloch waves in the bulk can of course be done mathematically, but is rather hard in experiments, since it requires measurement access to the wave functions \cite{li_bloch_2016}. However, one of the most remarkable aspects of Bloch wave topology is that the mathematical properties of the bulk have immediate consequences for what happens at the boundary of a finite sample, or at an interface between two domains of the medium that have different topological features. In particular, lower-dimensional stationary states develop at the boundary/interface, the so-called edge states, cf Fig.~\ref{fig:basic_concept}d. This is the celebrated bulk-boundary correspondence. It is these edge states that permit topologically protected robust transport and which are of greatest importance for potential applications.

We now review the more detailed description of edge states and bulk topological indices. Two qualitatively different situations have to be distinguished, according to whether time-reversal symmetry is broken or not. For the time-symmetric case, we further distinguish between the schemes that allow to implement a Spin-Hall Hamiltonian supporting spin Chern numbers for each of two local phononic pseudo-spin directions, and those in which each pseudo-spin direction and the corresponding topological excitations are described by a Dirac Hamiltonian derived within a smooth-envelope approximation.
While our description should be sufficient to provide the reader with enough background to be able to understand the remainder of this review, with the applications to topological transport in mechanical systems, there are much more extensive mathematically oriented reviews available that may be consulted for additional details 
\cite{hasan_colloquium_2010,ozawa_topological_2019}.

\subsection{Broken time-reversal symmetry}
\label{subsec:mechanism_chern_insulator}
\begin{figure}
\includegraphics[width=\columnwidth]{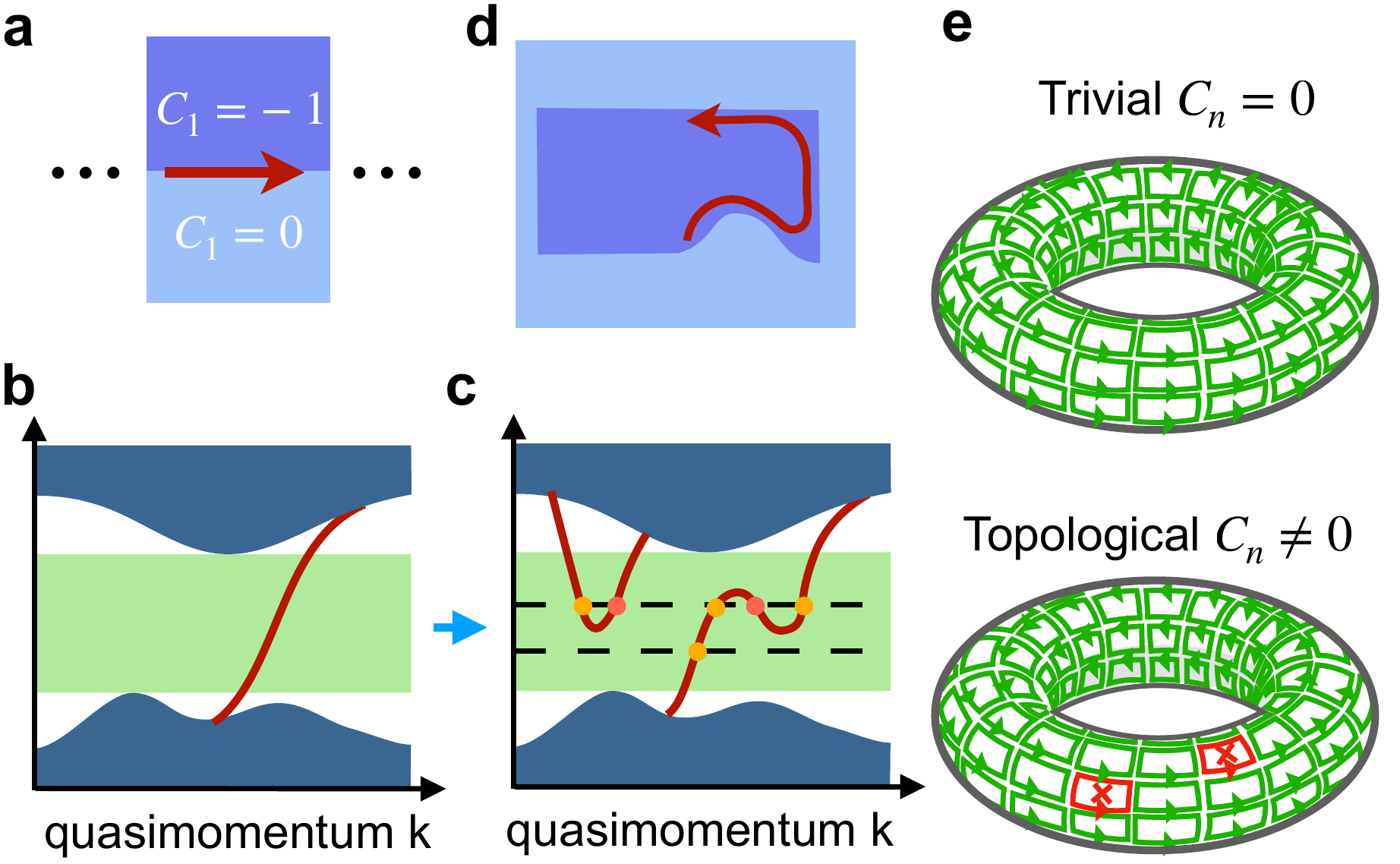}
\caption{Topological transport in a Chern insulator. \textbf{a}, Domain wall geometry. Also indicated  are the Chern number of the lower band for the two domains and the edge state propagation direction (red arrow).  \textbf{b}, Example of topological  band structure for the domain-wall geometry shown in \textbf{a}.   \textbf{c}, Band structure of a smoothly deformed system, with the same resulting net number of right-movers at any frequency. \textbf{d}, Robustness against deformations of the domain wall. \textbf{e}, The Chern number becomes nonzero if it is impossible to fix the same gauge (indicated by the contour's colour) for all plaquettes in the Brillouin zone (here depicted as a torus) without creating any obstruction (depicted as a cross). 
}
\label{fig:chern}
\end{figure}

The topological properties of interest for the general field of topological band structures are properties of a band gap that are invariant under continuous modifications of the underlying Hamiltonian. As already discussed above, these properties are encoded in so-called topological invariants, integer quantities that are  a function  of the bulk normal modes, but that can also be inferred directly from the spectrum in a system with a physical boundary. Here, we want to highlight this aspect for the special case of  systems with broken time-reversal symmetry.
For this purpose, we initially consider the simplest geometry comprising a physical boundary, a semi-infinite plane Fig.~\ref{fig:chern}(a).   This approach allows  also to provide some simple physical intuition for the resulting topologically robust transport and to understand why a $\mathbb{Z}$-topological invariant (assuming arbitrary integer values) is required to classify topological band gaps in systems with broken time-reversal symmetry. 

For a semi-infinite plane, the frequency can be plotted as a function of the quasimomentum in the direction longitudinal to the edge, which is a conserved quantity. A typical example of such a band structure is shown in Fig.~\ref{fig:chern}(b). In this example, two bulks bands (blue)  separated by a bulk band gap (green) are connected via a gapless edge state (red).
The edge state is a right mover because the slope, which sets the group velocity, is always positive, Fig.~\ref{fig:chern}(a-b).  
The chiral nature of the wave transport  is clearly robust even once weak disorder (compared to the width of the bulk band gap) is introduced into the system: no backscattering can occur simply because the system does not support any left-moving states in the bandwidth of interest. This argument applies even when the  edge changes direction at a corner in a finite system  with a closed boundary Fig.~\ref{fig:chern}(d). We note that the breaking of the time-reversal symmetry is a precondition to realize this kind of physics because it opens up the possibility of engineering systems with chiral edge states without time-reversed counter-propagating solutions.

Next, we  consider any arbitrary translationally-invariant continuous modification of the underlying Hamiltonian. The  band structure will then also change continuously, with the only constraint that the edge bands should  remain  (single-valued) periodic functions of the quasimomentum or start from a bulk band and end in  the neighboring bulk band (such topological edge bands would also be periodic functions of the quasimomentum in a strip configuration with a finite number of bands).  Possible modifications include introducing one or more local maxima in the edge band, or pulling  one or more edge bands into the bulk bands, cf Fig.~\ref{fig:chern}(c). An allowed  modification might lead to an frequency-dependent number of edge states giving rise also to  left-moving states. However, it never changes the difference  of the numbers of right movers and left movers which remains independent of the frequency inside the band gap, cf Fig.~\ref{fig:chern}(c). 
Thus, the net number of edge states $N_R-N_L$ (with $N_R$ and $N_L$  the number of right and left movers, respectively) is a topological invariant. Since it can assume any arbitrary integer value, it is a so-called $\mathbb{Z}$-topological invariant.

\begin{figure*}
\includegraphics[width=2\columnwidth]{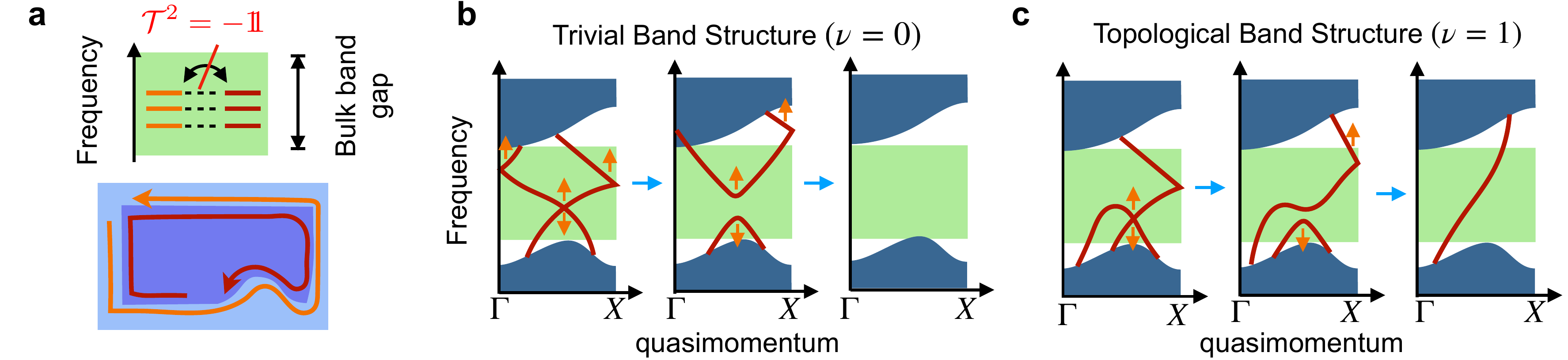}
\caption{\textbf{a}, (Top) Sketch of the spectrum of a (finite-size) time-symmetric topological insulator. Each pair of degenerate levels corresponds to a pair of counter-propagating edge waves (bottom). Any coupling is forbidden  because it would violate Kramers degeneracy.   \textbf{b}, and \textbf{c} Examples of, respectively, a trivial and a topological band structure in a semi-infinite plane geometry.  The degeneracies at the time-symmetric high-symmetry points $\boldsymbol{\Gamma}$ and $\mathbf{X}$ correspond to Kramers doublets. The trivial  and topological band structures can be smoothly modified, respectively, into a gapped band structure and a band structure with a single edge state connecting the two bulk bands, without lifting any Kramer doublet.}
\label{fig:kramers}
\end{figure*}

The net number of edge states is related to the band's Chern numbers $C_n$ via the bulk-boundary correspondence, $N_R-N_L=-\sum_{n} C_n$ where the sum is taken over the bands below the band gap of interest. In a two-material scenario, in which two semi-infinite domains with different bulk Hamiltonians are divided by a straight domain wall, the same formula holds with the difference of Chern numbers across the surface replacing the Chern number. The  band Chern number is a function of the bulk normal modes: 
\begin{equation}
\label{Eq:chern_number}
C_n = \frac{1}{2\pi} \int _{\rm BZ} d^2k \left(\nabla _{\mathbf{k}} \times {\cal A}_n(\mathbf{k})\right) \cdot \mathbf{e}_z.
\end{equation}
Here the integral is over the Brillouin zone (BZ), ${\cal A}_n(\mathbf{k})=i\braket{\mathbf{k}_n|\nabla_\mathbf{k}|\mathbf{k}_n}$ is the Berry connection,  and $|\mathbf{k}_n\rangle$ are the normal modes for the $n$-th band. By dividing the BZ as a sum of plaquettes and applying Stoke's theorem separately to the surface integrals over each plaquette, we can rewrite the Chern number as a sum of Berry phases 
\begin{equation}
\label{Eq:chern_number_berry}
C_n=\frac{1}{2\pi}\sum_{i \in {\rm BZ}}\gamma_{i,n},\quad \gamma_{i,n}= \oint_i {\cal A}_n(\mathbf{k}) \cdot d\mathbf{k}.
\end{equation}
We note that special care has to be taken in evaluating  Eq.~(\ref{Eq:chern_number_berry}). Stoke's theorem applied to a plaquette is only valid for a  choice of  gauge  that is well defined across the whole plaquette. On the other hand, the  contour integral $\gamma_{i,n}$ differs by a multiple of $2\pi$ from the corresponding surface integral in the presence of so-called obstructions,  quasi-momenta for which  a gauge is ill-defined \cite{kohmoto_topological_1985}. In this case, $\gamma_{i,n}$ can still be viewed as representing  the same Berry phase, however,  one would now arrive at a result for Eq.~(\ref{Eq:chern_number_berry}) that differs from the definition Eq.~(\ref{Eq:chern_number}) by an integer. 
Whenever one can fix the same gauge  across the whole BZ, Stoke's theorem can be directly applied  to the integral over the BZ. In this special case, the Chern number is zero because the BZ is a closed (boundaryless) surface, cf Fig.~\ref{fig:chern}(e). For topological band structures, any attempt to fix a global gauge leads to one or more obstructions, cf Fig.~\ref{fig:chern}(e). In this scenario,  one can not evaluate Eq.~(\ref{Eq:chern_number_berry})  using the same gauge for all plaquettes (which would give zero) and using an appropriate gauge for each plaquette one arrives at a non-zero integer Chern number \cite{kohmoto_topological_1985}.

\subsection{Time-reversal-symmetric topological insulators}
\label{sec:Time-reversal-mechanism}

\subsubsection{Edge states protected by Kramers degeneracy}
\label{sec:Time-reversal-mechanism_kramers}

Topological systems with time-reversal symmetry support so-called helical edge states and have been initially discovered in electronic systems. In electronic and other fermionic systems, the time-reversal symmetry operator $\cal T$ squares to minus the identity, ${\cal T}^2=-1\!\!1$. This leads to Kramers degeneracy: For every normal mode $|\psi\rangle$ there is a time-reversed partner normal mode  ${\cal T}|\psi\rangle$ with equal frequency. For excitations in the bulk band gap of a topological insulator, these Kramers partner solutions are pairs of counter-propagating  waves  travelling along the physical boundary (or the interface to a trivial material). In this setting, any coupling between the running waves is forbidden because it would break Kramers degeneracy, cf Fig.~\ref{fig:kramers}a. This results in transport that is robust against backscattering in the presence of weak time-reversal-symmetric disorder. 

Next, we discuss how a topological invariant can be inferred from the spectrum of a system with  a semi-infinite geometry.
Without loss of information, we display only positive quasi-momenta ($E_n(k)=E_n(-k)$ because of the time-reversal symmetry), cf Fig.~\ref{fig:kramers}(b-c). We remark that for the band structure of a time-reversal symmetric Hamiltonian with ${\cal T}^2=-1\!\!1$, only the band crossings that occur at a time-reversal-invariant high-symmetry point are essential (symmetry-protected) degeneracies.  In a semi-infinite configuration these are the $\boldsymbol{\Gamma}$ and  $\mathbf{X}$ points with quasimomenta $k=0$ and  $k=\pi/a$ (with $a$  the lattice constant), respectively. Accidental band crossings away from these special quasi-momenta are not a consequence of Kramers degeneracy (the underlying normal modes are obviously not Kramers pairs)  and can be lifted without breaking the time-reversal symmetry.  We now consider a general arbitrary complicated band structure, see two examples in Fig.~\ref{fig:kramers}b,c. By lifting all accidental degeneracies and pulling edge bands (pairs of edge bands) into the bulk bands, we arrive at just two topologically distinct configurations: (i) A trivial configuration in which no edge band is present, see Fig.~\ref{fig:kramers}b. (ii) A topological configuration in which a single edge band connects  the bulk bands below and above the band gap, cf Fig.~\ref{fig:kramers}c. Thus,  the topology of  $2$D time-symmetric topological insulators  are encoded in a so-called $\mathbb{Z}_2$  topological invariant, assuming only two integer values. By convention, the value $\nu=1$ ($\nu=0$) is assigned to the topological (trivial) configuration.



However, for bosonic excitations and, in particular, the phononic excitations  of interest in this review, the physical time-reversal symmetry  squares to the identity, ${\cal T}^2=1\!\!1$. Thus, Kramers degeneracy does not naturally emerge in the presence of time-reversal symmetry in bosonic systems. Nevertheless, no fundamental principle prevents us from engineering the effective Hamiltonian of  a bosonic system  to be identical to the single-particle Hamiltonian of a  fermionic system of interest (here, a topological insulator). In this scenario, Kramers degeneracy will be produced via an engineered local anti-unitary symmetry ${\cal T}_{\rm en}$ with ${\cal T}_{\rm en}^2=-1\!\!1$. We note that ${\cal U}={\cal T}_{\rm en}{\cal T}$ is then a local unitary symmetry which  can be interpreted as a conserved pseudo-spin.  For the important special case of a binary pseudo-spin (equivalent to a spin $1/2$ particle)  we have 
\begin{equation}\label{eq:Ham_Top_Ins}
    \hat{H}=
    \begin{pmatrix}
    \hat{H}_\uparrow & 0 \\
    0 & \hat{H}_\downarrow
    \end{pmatrix},
\end{equation}
with $\hat{H}_\downarrow = \hat{H}_\uparrow ^*$.
For a topological bosonic insulator, each  pseudo-spin polarized block  supports arbitrary integer non-zero Chern numbers. An explicit example is given in Section \ref{sec:kramers_degeneracy_implementation}. We note that for electronic systems, this form of the Quantum Spin-Hall Hamiltonian corresponds to the special case in which the out-of-plane mirror transormation is a symmetry. Without this symmetry, a Rashba spin-orbit interaction leads to a coupling of the two blocks \cite{kane_quantum_2005}. This more general form can be implemented in a bosonic system only by breaking the physical time-reversal symmetry ${\cal T}$. Vice versa, any perturbation that couples the different pseudo-spin directions without breaking ${\cal T}$ will break ${\cal T}_{\rm en}$ and, thus, evades the topological protection.

\begin{figure}
\includegraphics[width=\columnwidth]{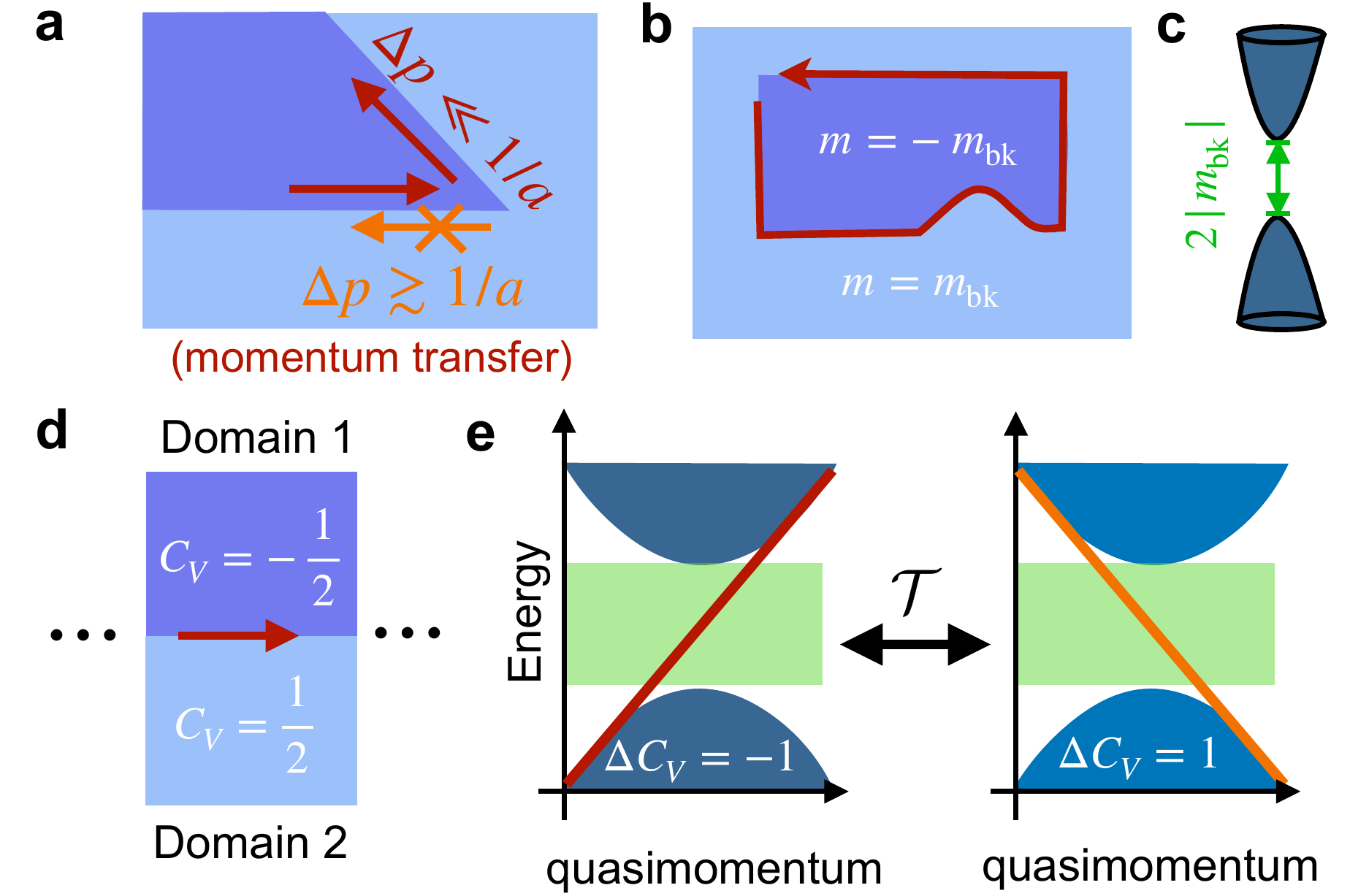}
\caption{Generating edge states via an engineered Dirac Hamiltonian. \textbf{a}, Edge excitations can travel around arbitrarily shaped corners. The change of direction involves only a small momentum transfer $\Delta p$. Backscattering requires large $\Delta p$ and is, thus, suppressed. \textbf{b}, An edge channel is localized along a domain wall separating two domains governed by the Dirac Hamiltonian Eq.~(\ref{eq:Dirac_Hamiltonian}) with opposite values $\pm m_{\rm bk}$ of the mass parameter $m$. \textbf{c}, The two domains share the same gapped Dirac cone band structure. \textbf{d}, Straight domain wall geometry. Indicated are also the valley Chern numbers $C_V$. \textbf{e}, Band structure for the geometry and valley Chern numbers displayed in (\textbf{d}) (left) and for the time-reversed Dirac Hamiltonian (right). The change of valley Chern number $\Delta C_v$ across the domain wall is indicated. The edge states (red and orange) obey the bulk-boundary correspondence.}
\label{fig:smooth_envelope}
\end{figure}

\subsubsection{Engineered Dirac Hamiltonians}

\label{sec:Time-reversal-mechanism_dirac_engineering}

In many experimental platforms, it is not possible to engineer a desired effective Hamiltonian across the whole BZ. To address this challenge, a variety of design strategies,  reviewed in Section \ref{sec:preserved-time-reversal-systems}, have been invented that allow to engineer robust edge states that remain well localized in  quasimomentum space but can nonetheless propagate along arbitrarily-shaped domain walls with negligible backscattering, even turning  sharp corners. In this setting, backscattering is  suppressed because it requires large momentum transfer,
cf Fig.~\ref{fig:smooth_envelope}(a). In a smooth-envelope approximation, these helical edge states are described by the large-wavelength limit of a Spin-Hall Hamiltonian (see below).

The description of these systems starts from an effective Dirac  equation
$\omega \boldsymbol{\Psi}(\mathbf{x})=\hat{H}\boldsymbol{\Psi}(\mathbf{x})$,
\begin{equation}\label{eq:Dirac_Hamiltonian}
\hat{H}=m(\mathbf{x})\hat{\sigma}_z+v\mathbf{q}\cdot\hat{\boldsymbol{\sigma}},
\end{equation}
where $v$ is the Dirac velocity, $\hat{\sigma}_{x,y,z}$ are the Pauli matrices, $\hat{\boldsymbol{\sigma}}=(\hat{\sigma}_x,\hat{\sigma}_y)$, $\mathbf{x}=(x,y)$ is the position, and $\mathbf{q}=(-id/dx,-id/dy)$. The mass function $m(\mathbf{x})$ is chosen to assume opposite values $m(\mathbf{x})=\pm m_{\rm bk}$ in  two  distinct bulk regions and, thus, vanishes along a one-dimensional domain wall dividing the two regions, cf Fig.~\ref{fig:smooth_envelope}(b). In this way, the two bulk regions share the same gapped Dirac cone band structure, cf Fig.~\ref{fig:smooth_envelope}(c),  but are topologically distinct, as discussed below.
The Dirac equation  is derived within a smooth-envelope approximation (or, equivalently, the so-called $\mathbf{k}\cdot\mathbf{p}$ perturbation theory). Each of the two components of the vector $\boldsymbol{\Psi}(\mathbf{x})$ is a smooth envelope which multiplies a different carrier Bloch wave  to define a (position-dependent) superposition. The two carrier Bloch waves are quasi-degenerate and share the same quasimomentum $\mathbf{q}_0$. Thus,  $\mathbf{q}$ is the quasimomentum distance from ${\mathbf q}_0$, and the smooth-envelope approximation is valid in a small quasimomentum region about ${\mathbf q}_0$, $|\mathbf{q}|\ll 1/a$ ($a$ is the lattice constant), commonly referred to as a valley. If one  formally considers $\mathbf{q}\in \mathbb{R}^2$, one can define  so-called Valley Chern numbers applying the standard definition  Eq.~(\ref{Eq:chern_number}) but now integrating  over the  2D plane instead of the BZ. An explicit calculation shows that the valley Chern number is a half integer, $$C_V={\rm sign}(m)/2$$ for the lower band.
We note that for $|m|a\ll |v|$ the integrand is strongly peaked in a small  quasimomentum region, $|\mathbf{q}|\ll 1/a$, within the limit of validity of the Dirac equation. In this case, the Valley Chern number gives an accurate estimate of the contribution from the valley to the overall Chern number (beyond the smooth-envelope approximation leading to the Dirac equation). 

The bulk-boundary correspondence applied to the Dirac equation with a straight domain wall,  cf Fig.~\ref{fig:smooth_envelope}(d), predicts that the net number of edge states is
\begin{equation}\label{bulk-boundary_valley}
N_R-N_L=-\Delta C_V={\rm sign} (m_{\rm bk})   
\end{equation}
where $m_{\rm bk}$ is the bulk mass in the lower half-plane. This is compatible with a single edge band connecting the lower and upper bulk bands, whose propagation direction is set by the sign of $m_{\rm bk}$.  Indeed, Eq.~(\ref{bulk-boundary_valley}) is fulfilled by an edge-band solution originally derived by Jackiw and Rebbi \cite{jackiw_solitons_1976} (see  Fig.~\ref{fig:smooth_envelope}(d-e)) with dispersion  $\omega(q_x)={\rm sign} (m_{\rm bk})vq_x$   and wavefunction
\begin{equation}
\boldsymbol{\Psi}(\mathbf{x})=\begin{pmatrix}1\\
{\rm sign} (m_{\rm bk})
\end{pmatrix}e^{iq_x  x}e^{{\rm sign} (m_{\rm bk} )\int_0^ym(y')dy'/v}.
\end{equation}
Additional edge states starting  and ending in the {\em same} bulk band (see e.g. \cite{wang_valley-locked_2020}) do not change the net number of edge states and, thus, are also compatible with Eq.~(\ref{bulk-boundary_valley}).  

The Dirac Hamiltonian Eq.~(\ref{eq:Dirac_Hamiltonian}) is not invariant under any anti-unitary symmetry ${\cal T}$ (with ${\cal T}^2=1$). For this reason,  Dirac cones must  always appear in pairs in time-symmetric bosonic  systems.  It is possible to describe each cone with a copy of Eq.~(\ref{eq:Dirac_Hamiltonian}). Because of the time-reversal symmetry the two copies must have  opposite mass parameter. This ensures that the two Dirac Hamiltonians have also opposite valley Chern numbers  and, thus, the domain wall supports  counter-propagating edge states, cf Fig.~\ref{fig:smooth_envelope}(e). Introducing a binary valley pseudo-spin degree of freedom $\tau_z=\pm 1$, we arrive at the effective Hamiltonian
\begin{equation}\label{eq:Dirac_TRS}
\hat{H}=m(\mathbf{x})\hat{\tau}_z\hat{\sigma}_z+v\mathbf{q}\cdot\hat{\boldsymbol{\sigma}}. \end{equation}
We note that this is equivalent to the large-wave-length limit of the Bernevig-Hughes-Zhang model \cite{bernevig_quantum_2006}, a well known model for topological insulators. In addition to the time-reversal operator   ${\cal T}=\hat{\tau}_y\hat{\sigma}_y{\cal K}$, it supports also an anti-unitary symmetry  ${\cal T}_{\rm en}={\cal T}\hat{\tau}_z=i\hat{\tau}_x\hat{\sigma}_y{\cal K}$ with  ${\cal T}_{\rm en}^2=-1$  (${\cal K}$ is the complex conjugation and $\hat{\tau}_{x,y,z}$ a set of Pauli matrices). We emphasize, however, that in this setting ${\cal T}_{\rm en}$ is defined only for smooth-envelope solutions, i.e. assuming no large momentum transfer, $\Delta p\ll 1/a$. Beyond the smooth-envelope approximation, large momentum transfer can create a gap in the edge bands (in a straight-domain-wall geometry) and backscattering transitions (in any closed domain geometry). These effects are suppressed if the edge states' transverse localization length $\xi$  exceeds a few lattice constants. For sharp domain walls we have $\xi=v/m_{\rm bk}$,  and therefore $\xi\gg a$ imposes a limit on the mass parameter, $m_{\rm bk}\ll v/a$, and, thus, on the bandwidth available for  topological transport. This trade-off between backscattering suppression and bandwidth is eliminated for smooth domain walls \cite{shah_tunneling_2021}.

Until now we have referred to the pseudospin degree of freedom
$\hat{\tau}_z$ as denoting the valley in the BZ. Nevertheless, our discussion clearly applies also to the case where both pseudo-spin sectors have quasimomentum localized about the $\boldsymbol{\Gamma}$-point (for $\mathbf{q}_0=0$). This then refers to so-called zone-folding schemes, see \ref{sec:zone-folding-scheme}. It also applies to scenarios in which   more than one copy of Eq.~(\ref{eq:Dirac_TRS}) is required to describe all normal modes in a bandwidth about the same bulk band gap, see section \ref{Section:accidental}.

\section{Overview: Topological Phonon Transport}

\begin{figure*}
\centering
\begin{tabular}{ccccc}
\includegraphics[width=28mm]{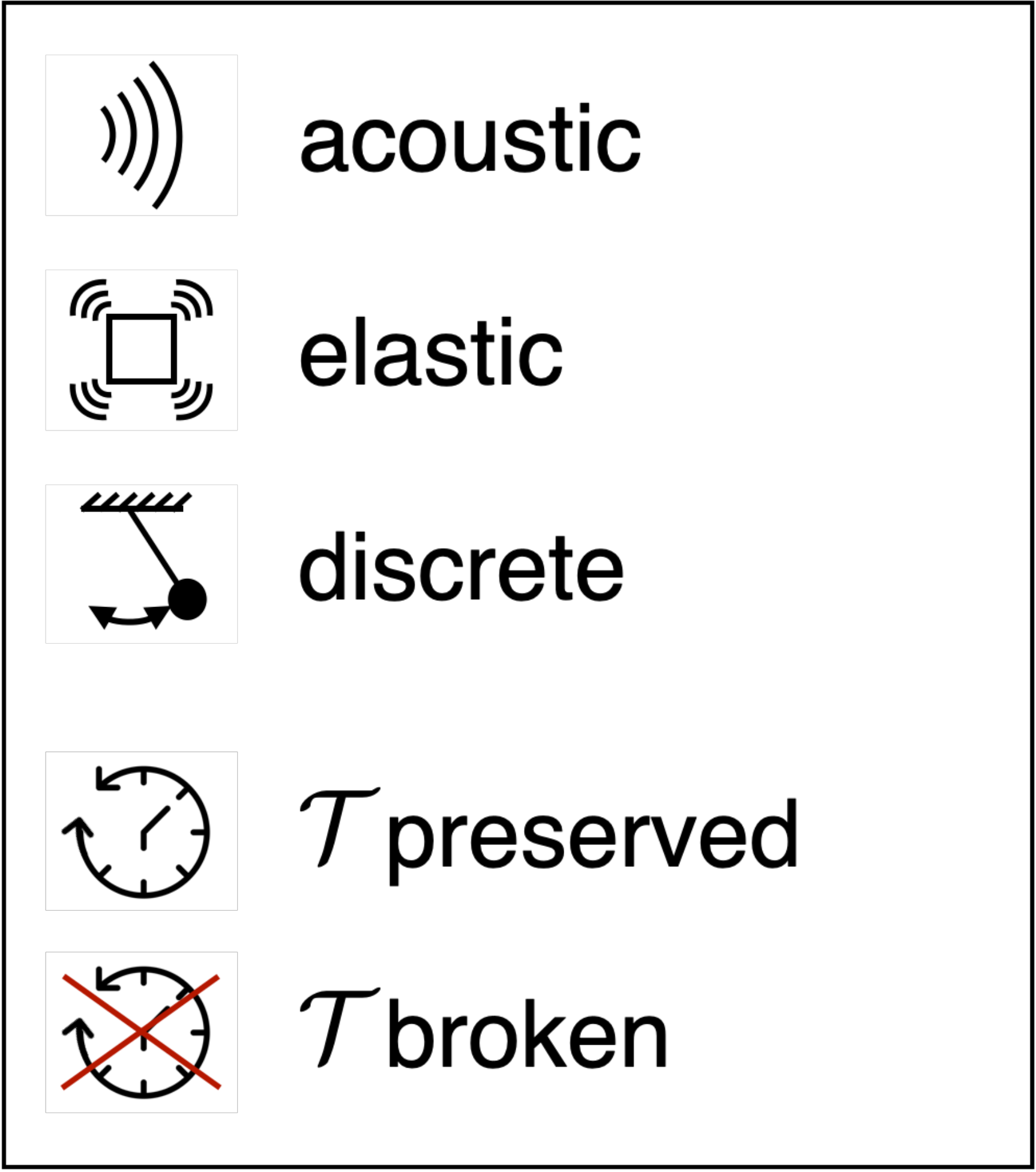}
\includegraphics[width=28mm]{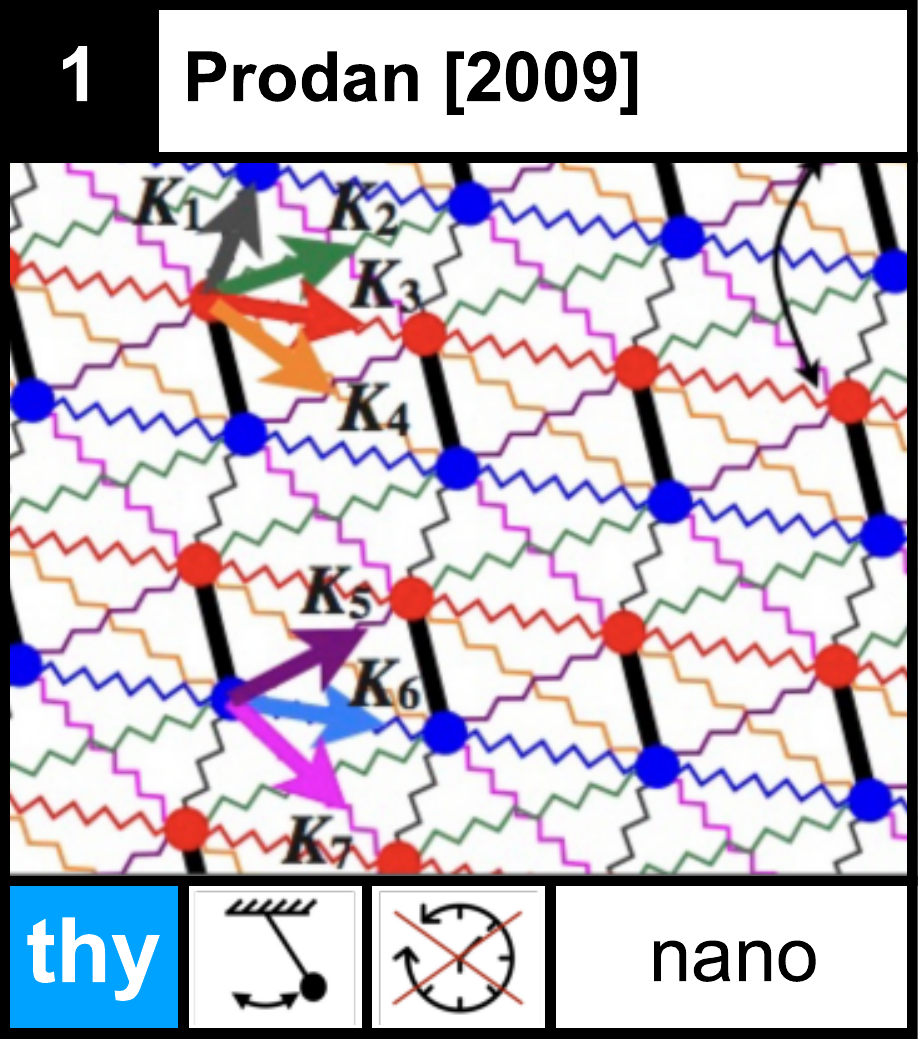}
\includegraphics[width=28mm]{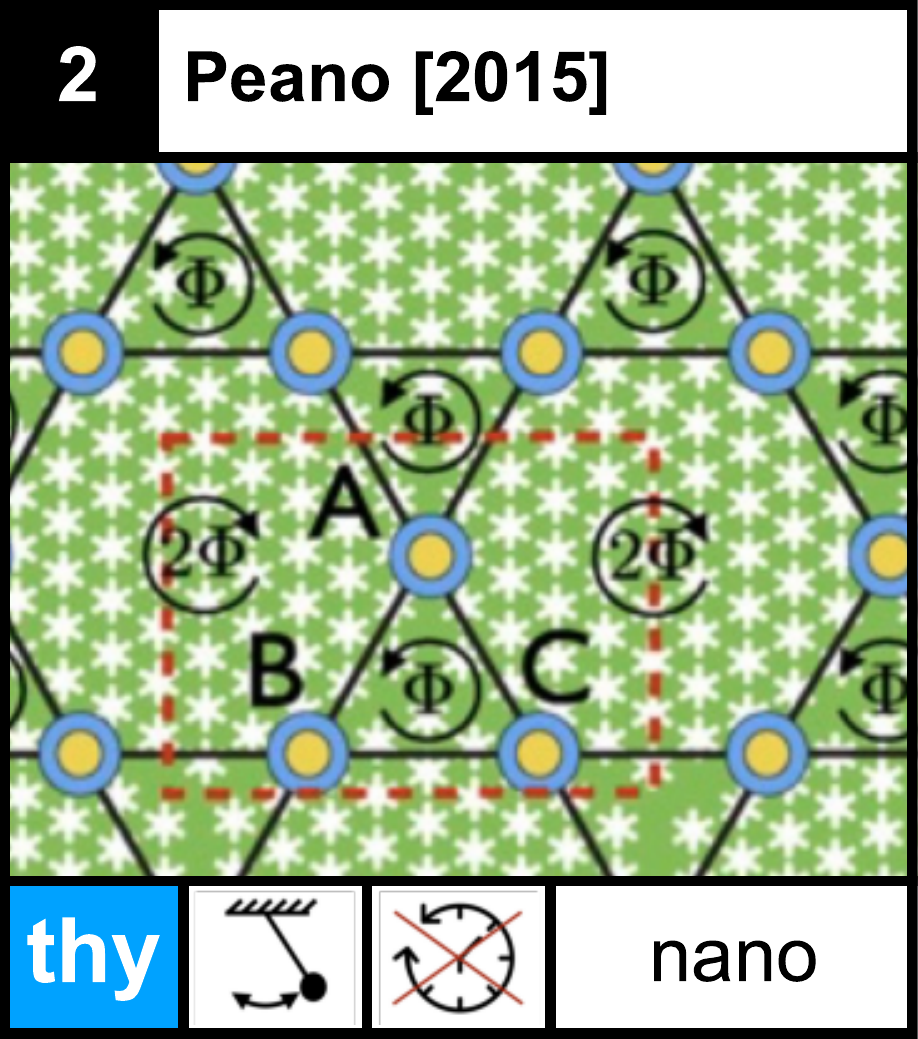}
\includegraphics[width=28mm]{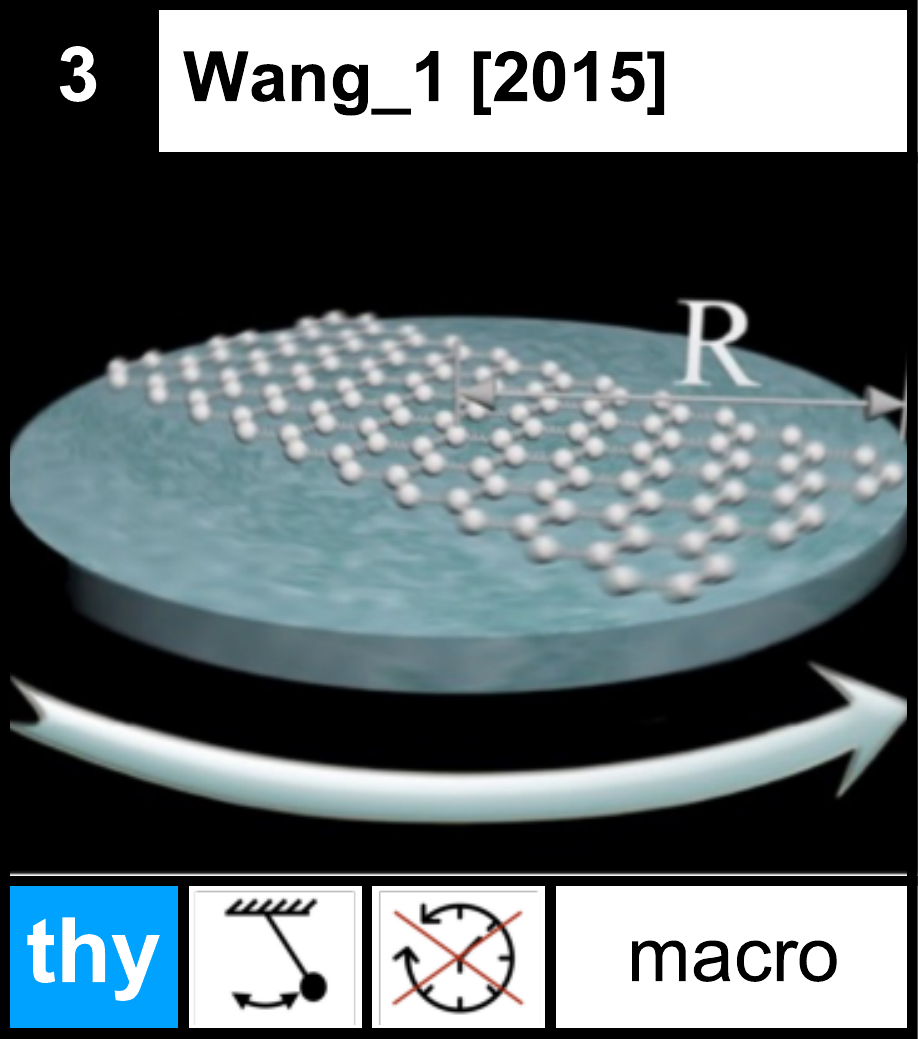}
\includegraphics[width=28mm]{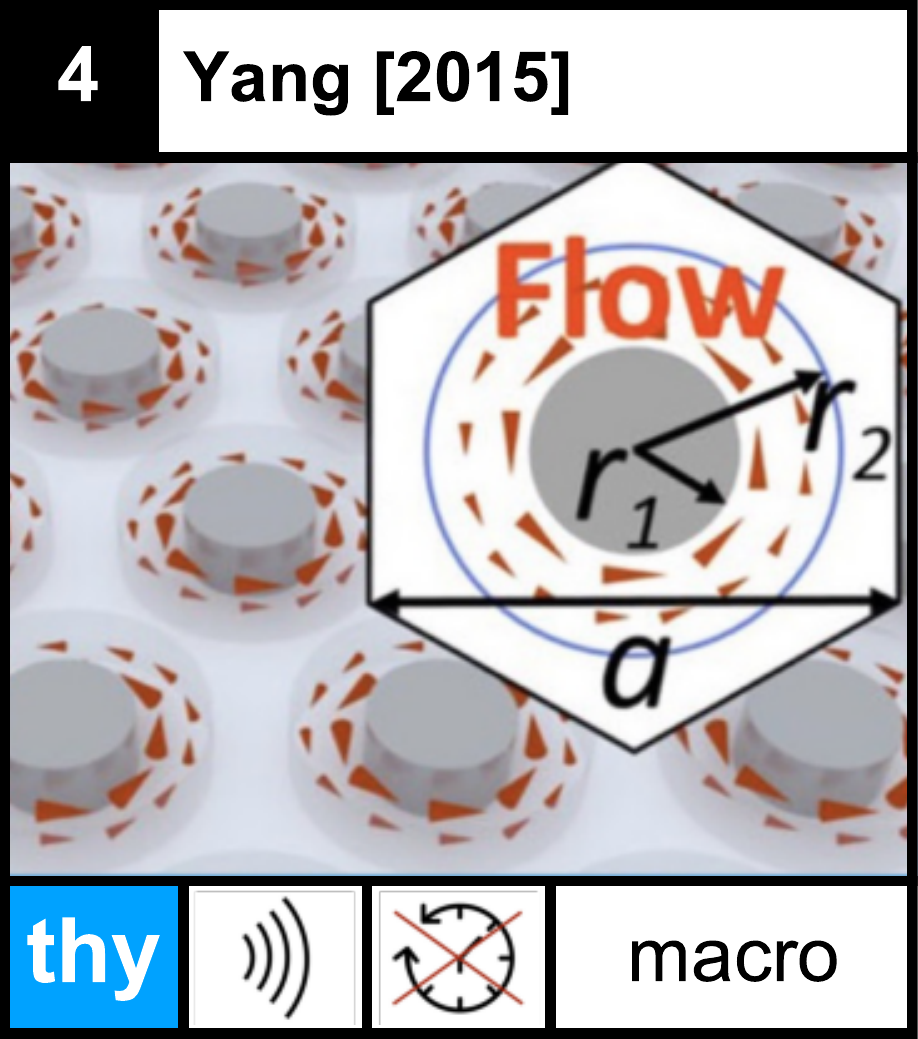}
\includegraphics[width=28mm]{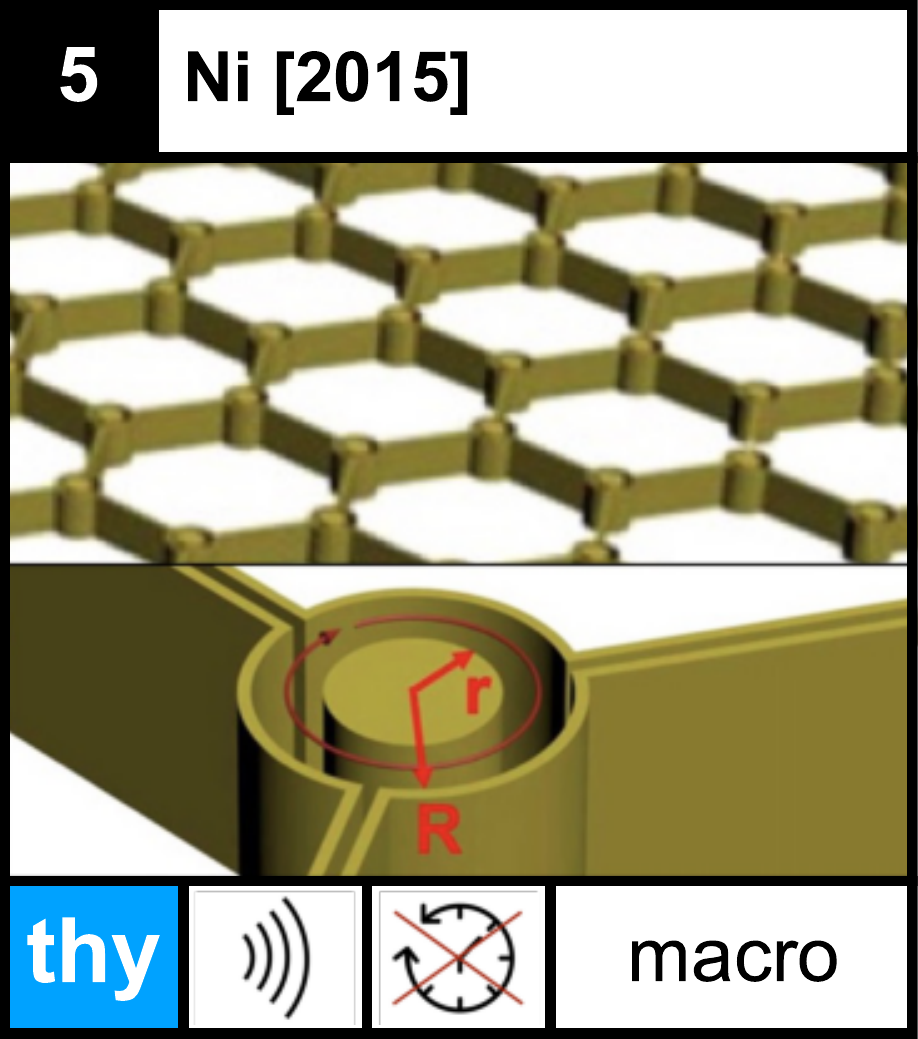}
\\
\includegraphics[width=28mm]{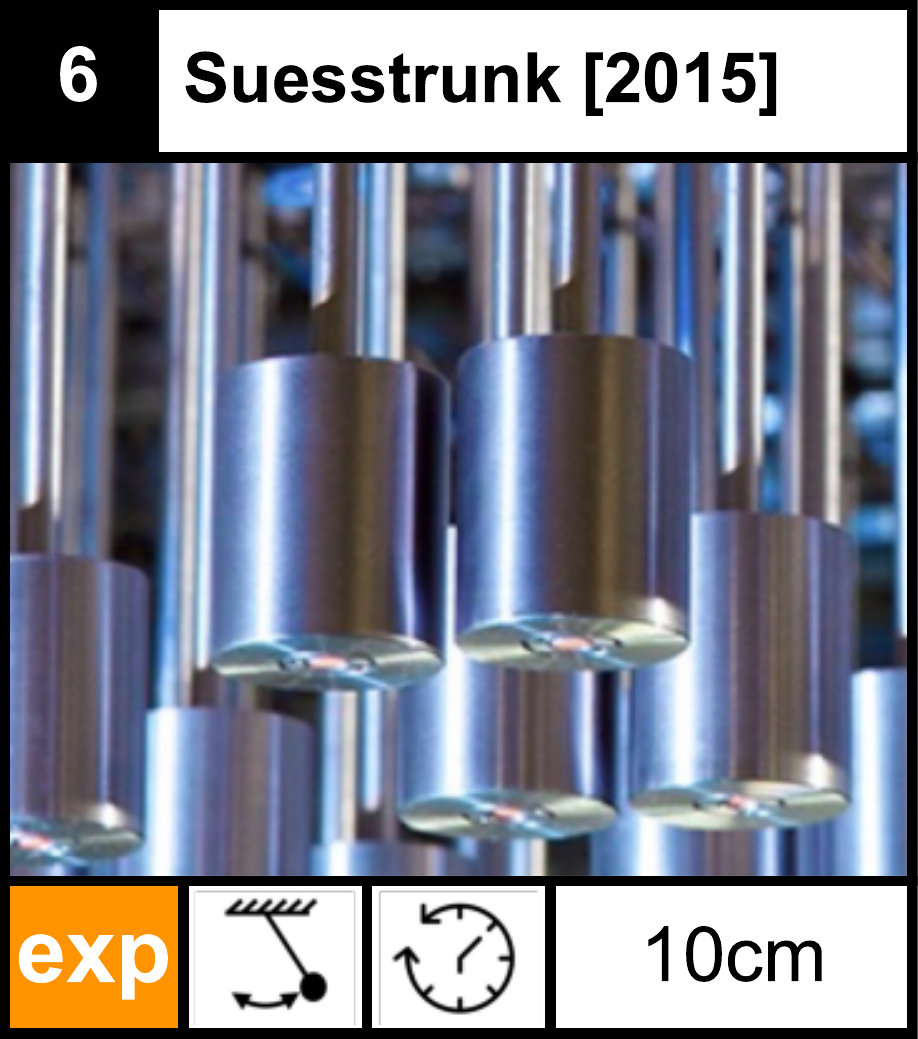}
\includegraphics[width=28mm]{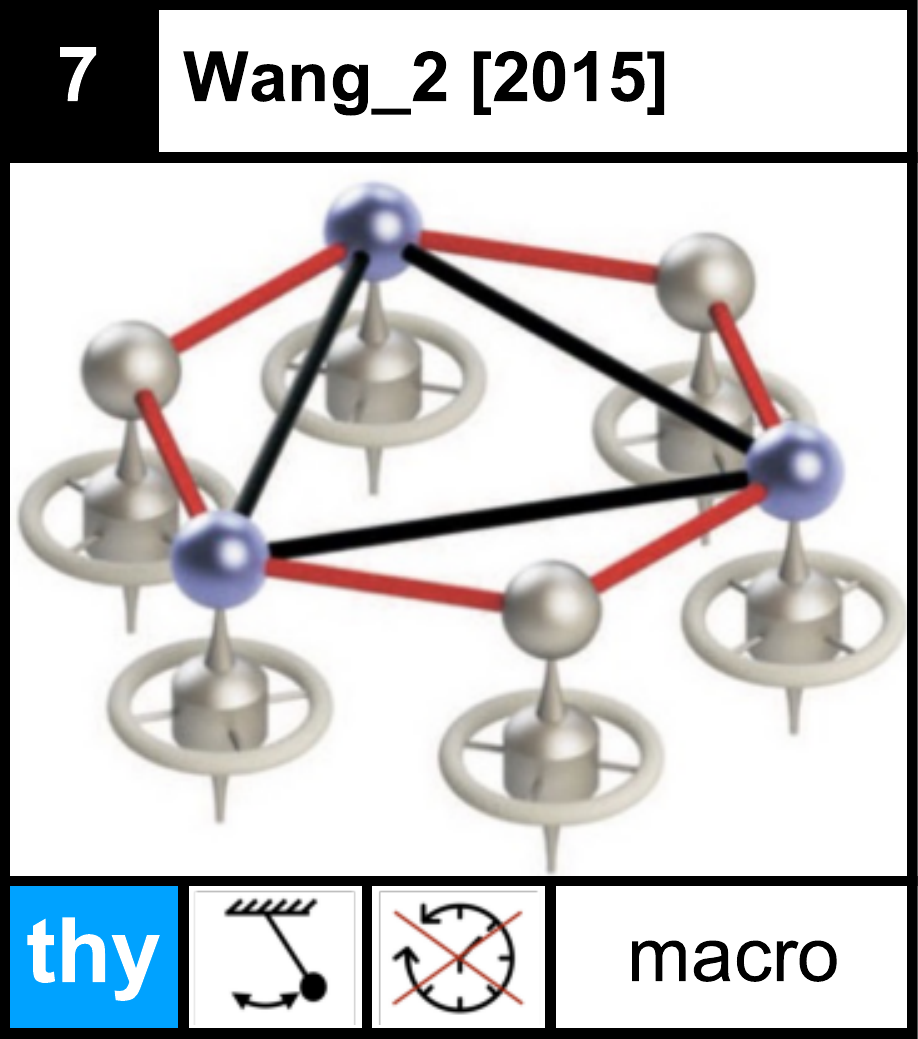}
\includegraphics[width=28mm]{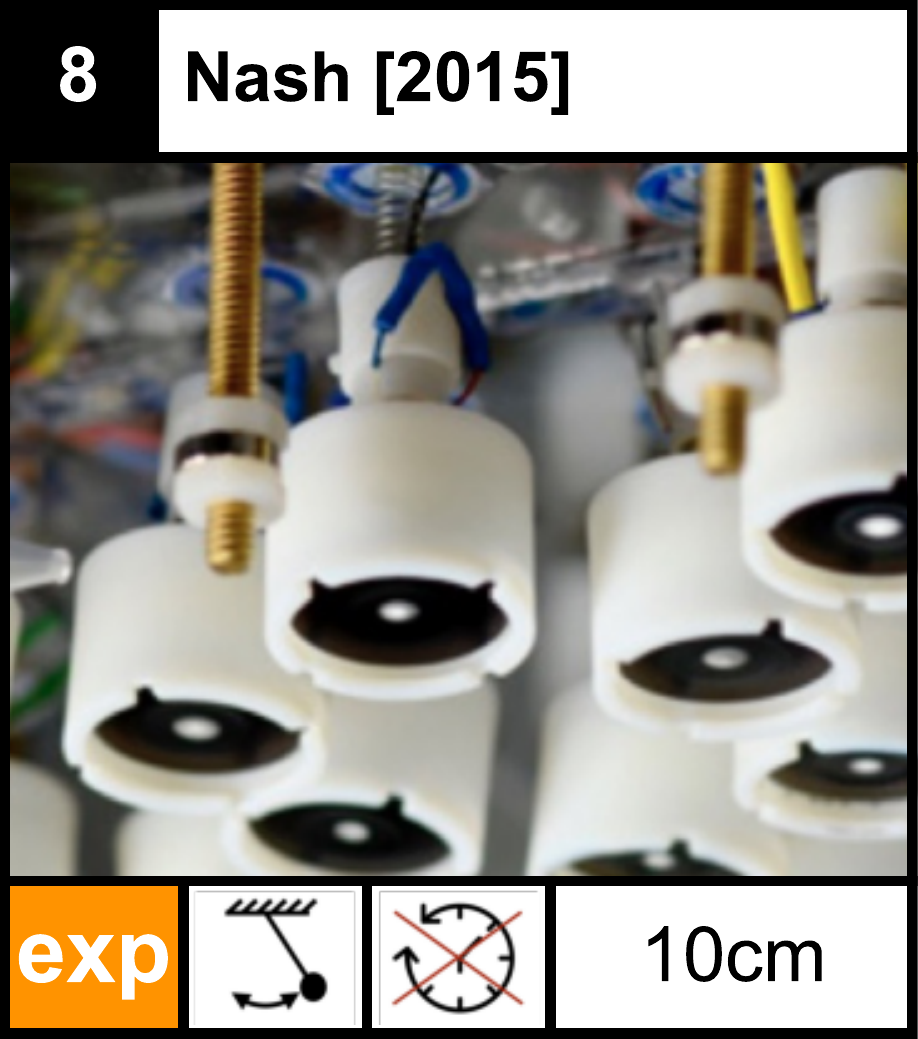}
\includegraphics[width=28mm]{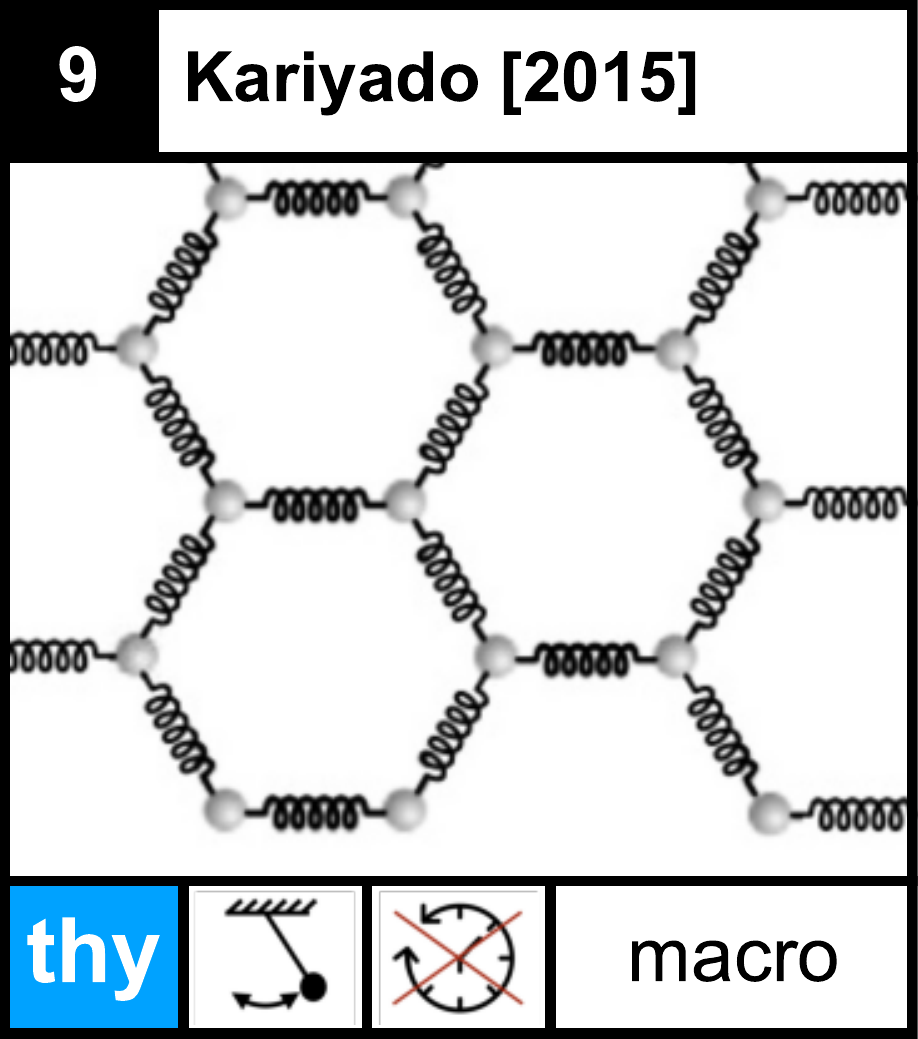}
\includegraphics[width=28mm]{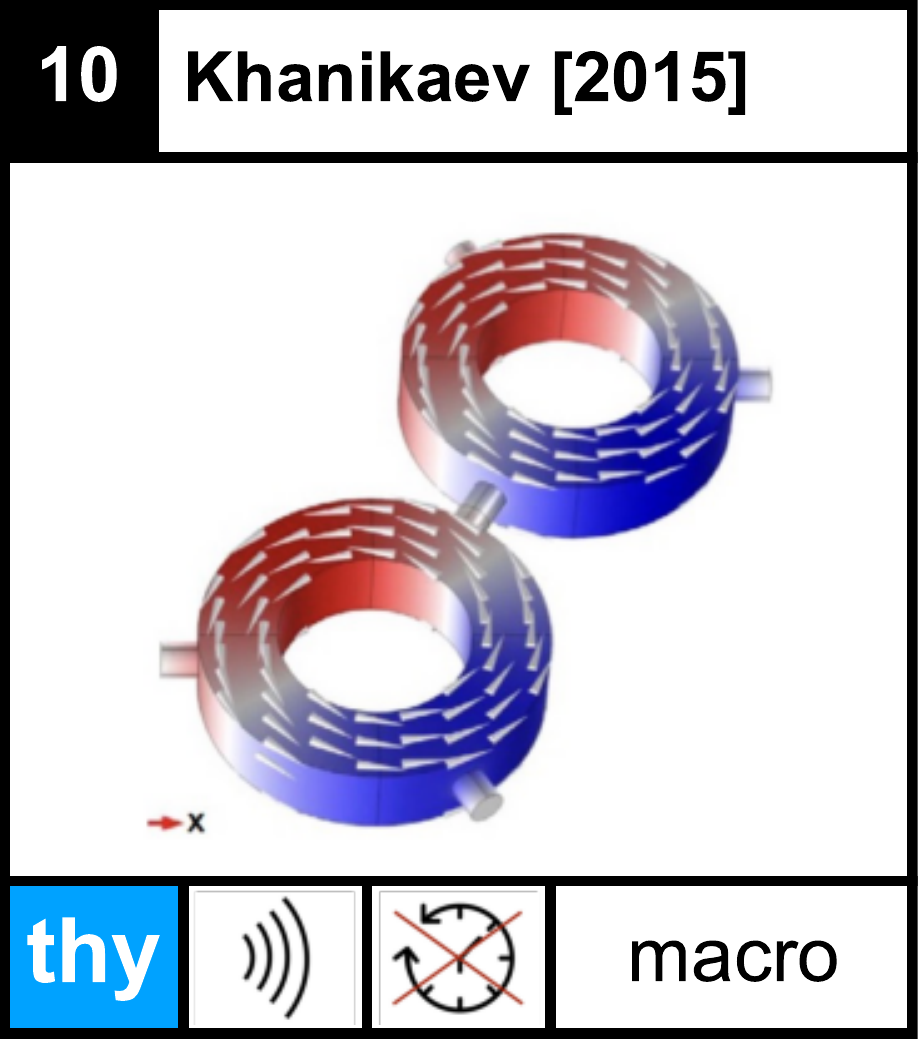}
\includegraphics[width=28mm]{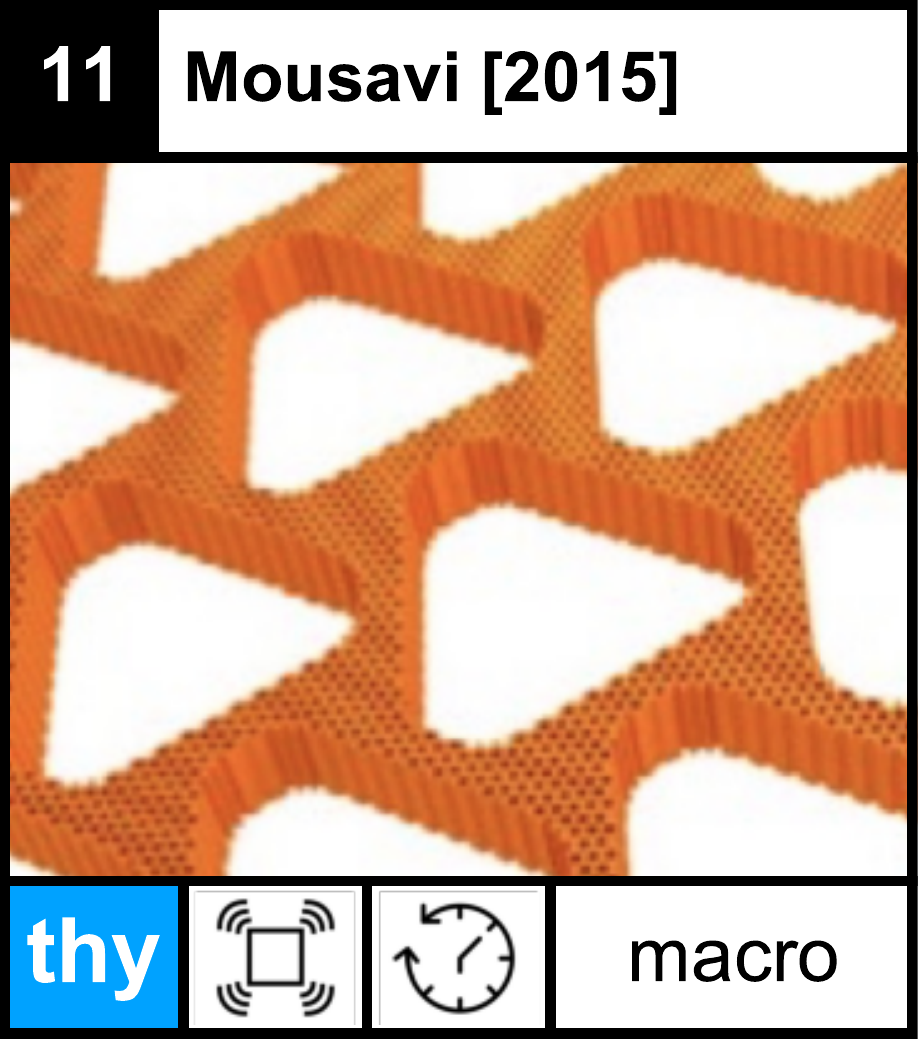}
\\
\includegraphics[width=28mm]{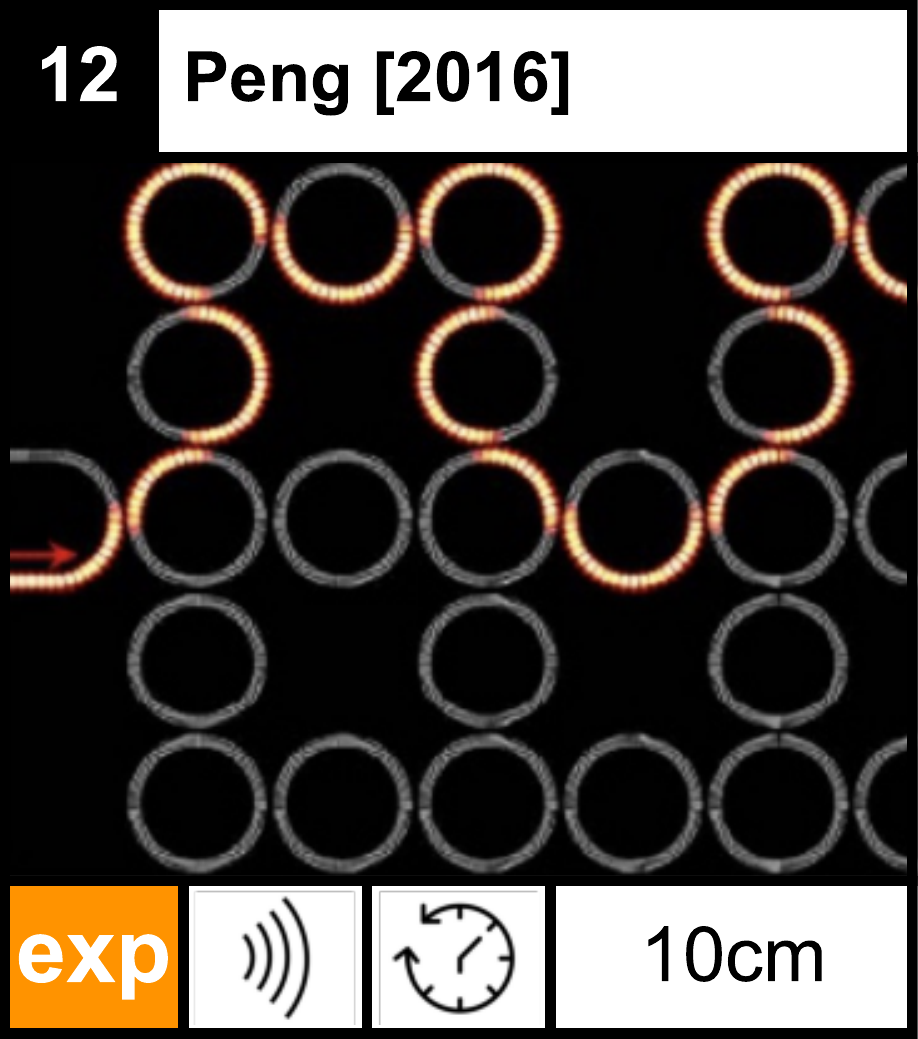}
\includegraphics[width=28mm]{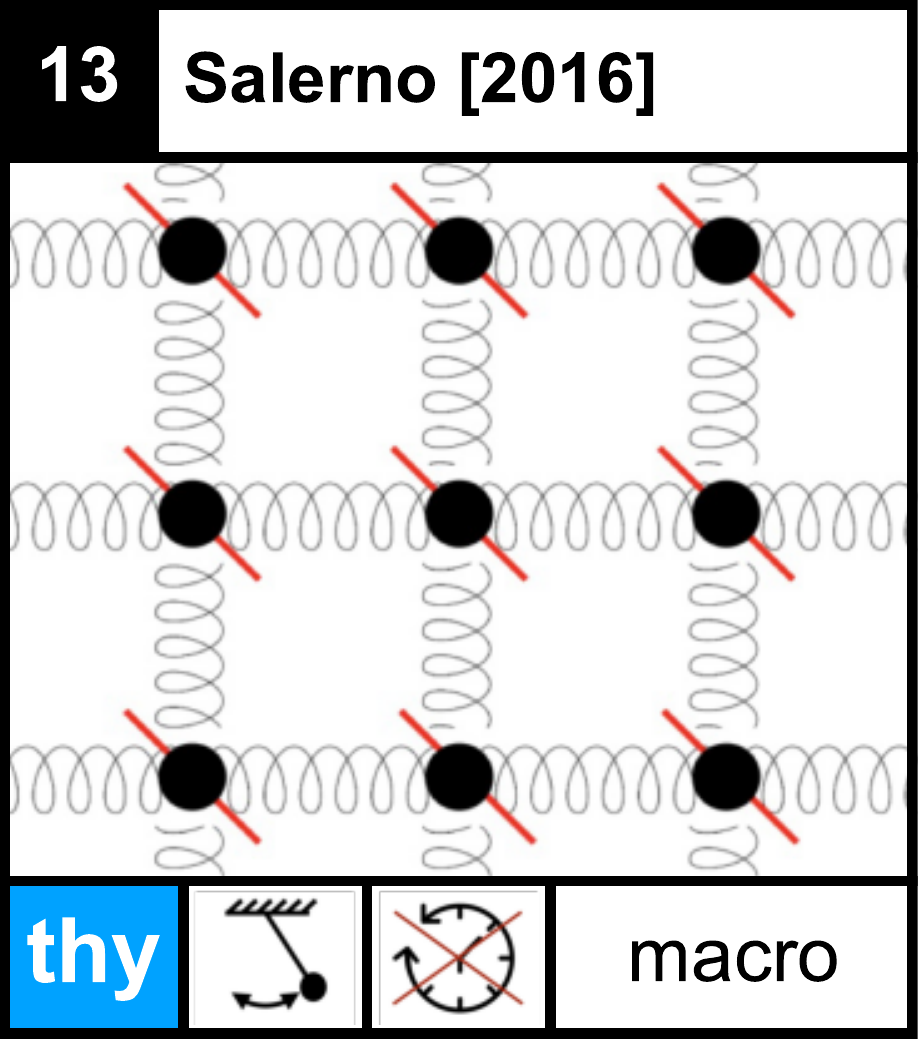}
\includegraphics[width=28mm]{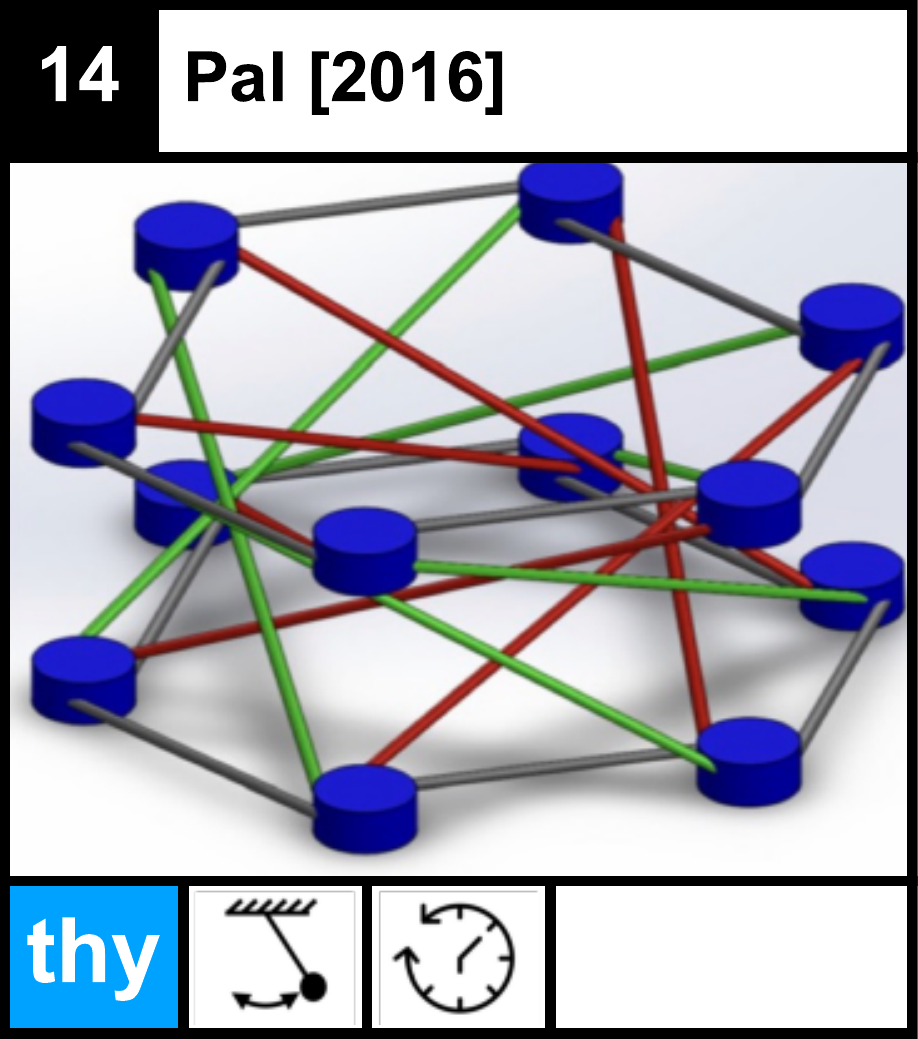}
\includegraphics[width=28mm]{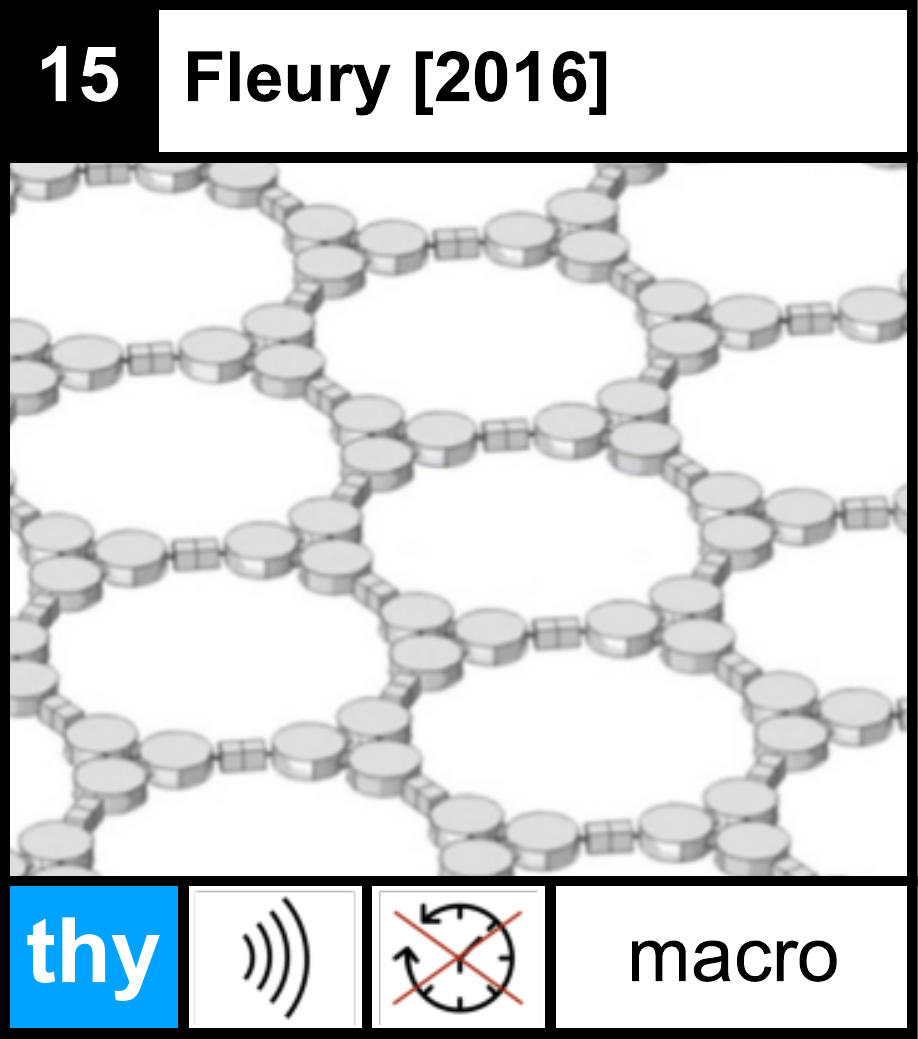}
\includegraphics[width=28mm]{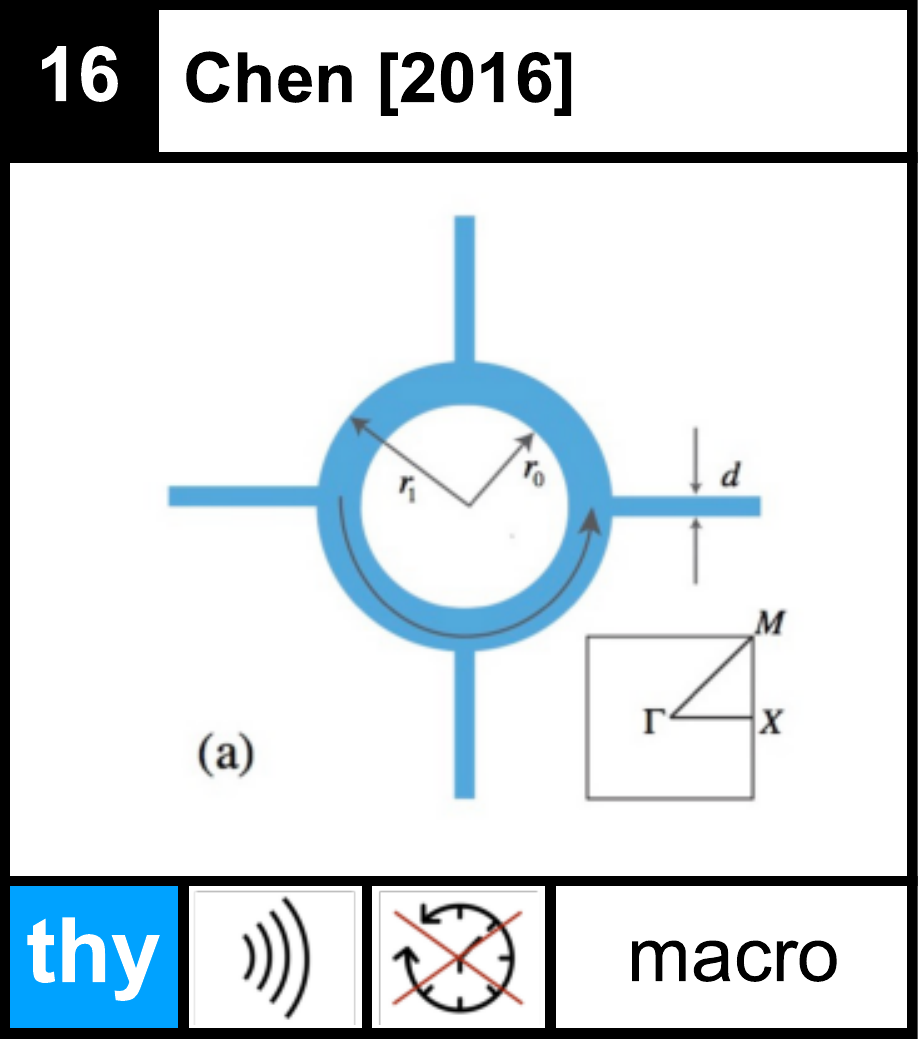}
\includegraphics[width=28mm]{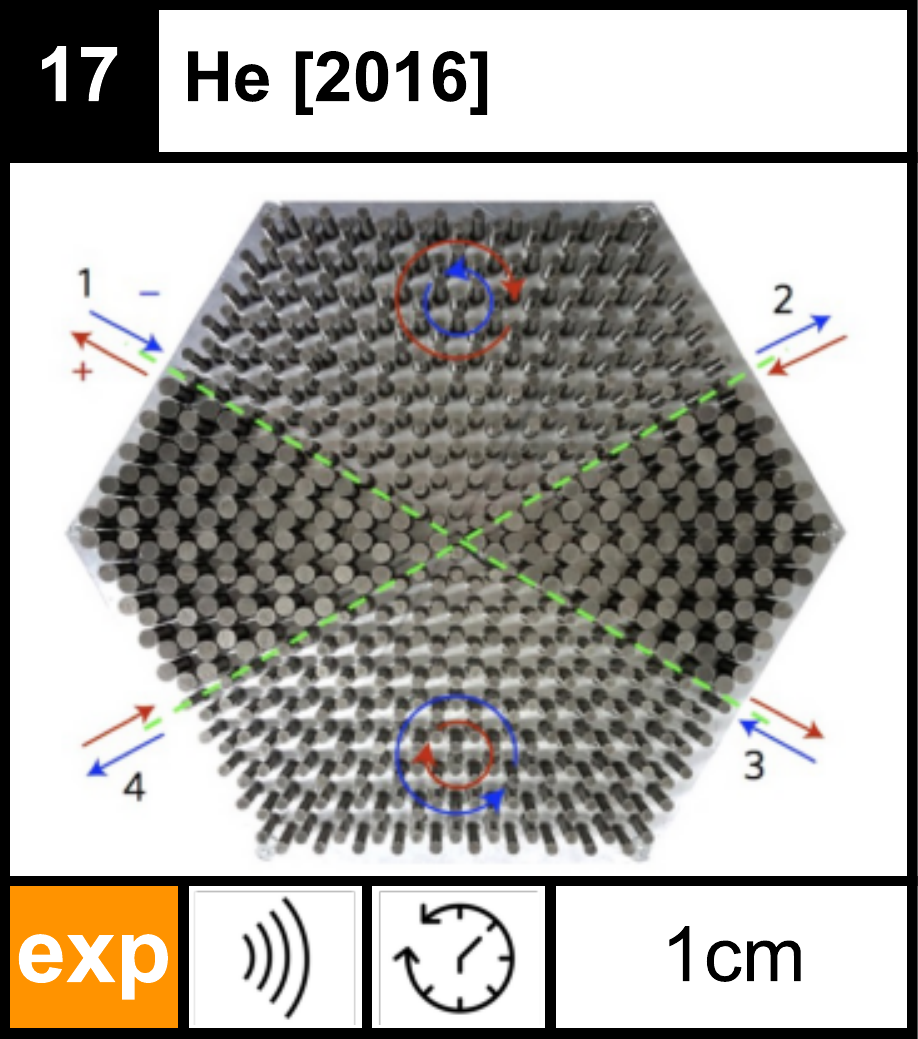}
\\
\includegraphics[width=28mm]{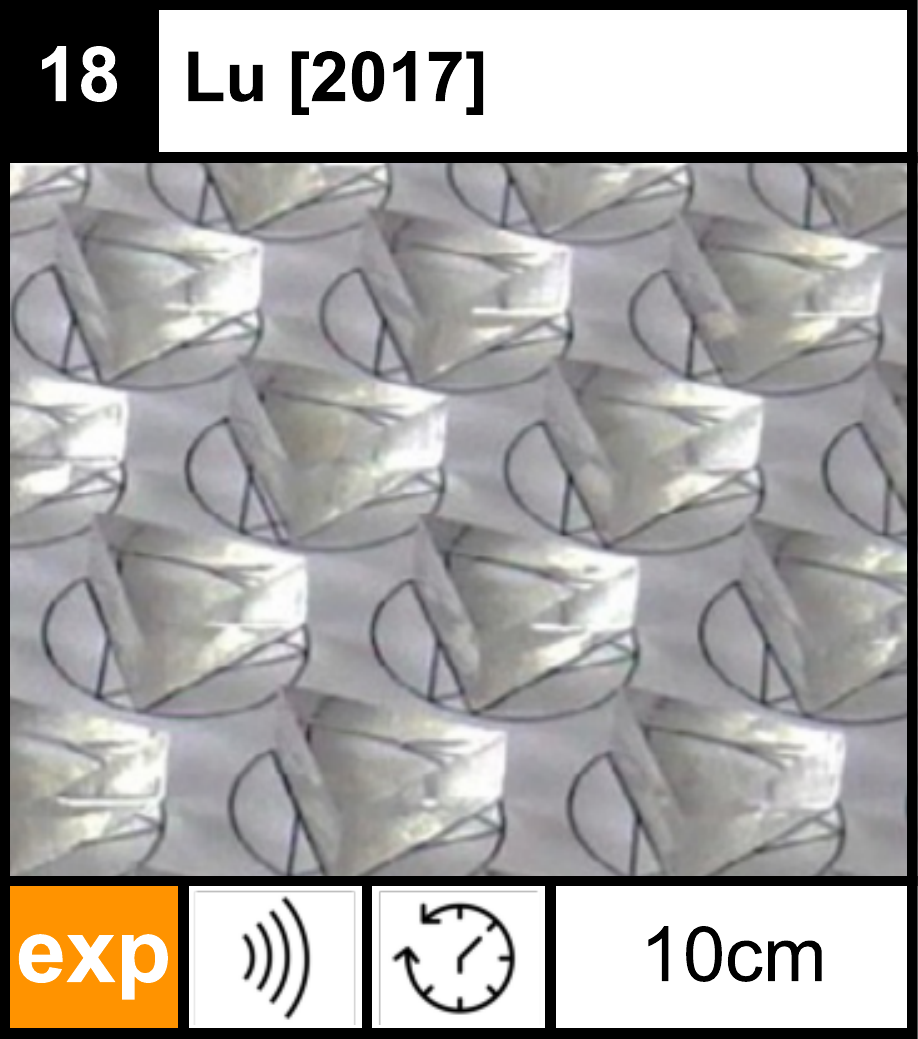}
\includegraphics[width=28mm]{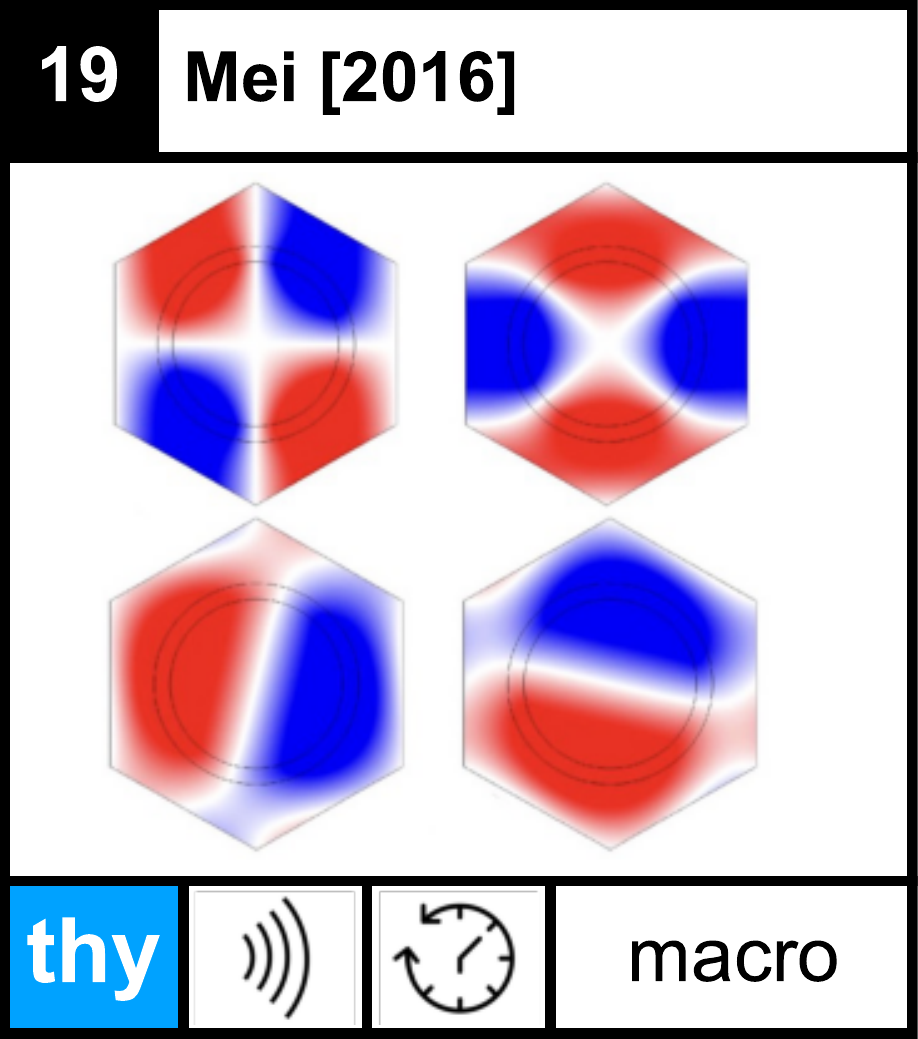}
\includegraphics[width=28mm]{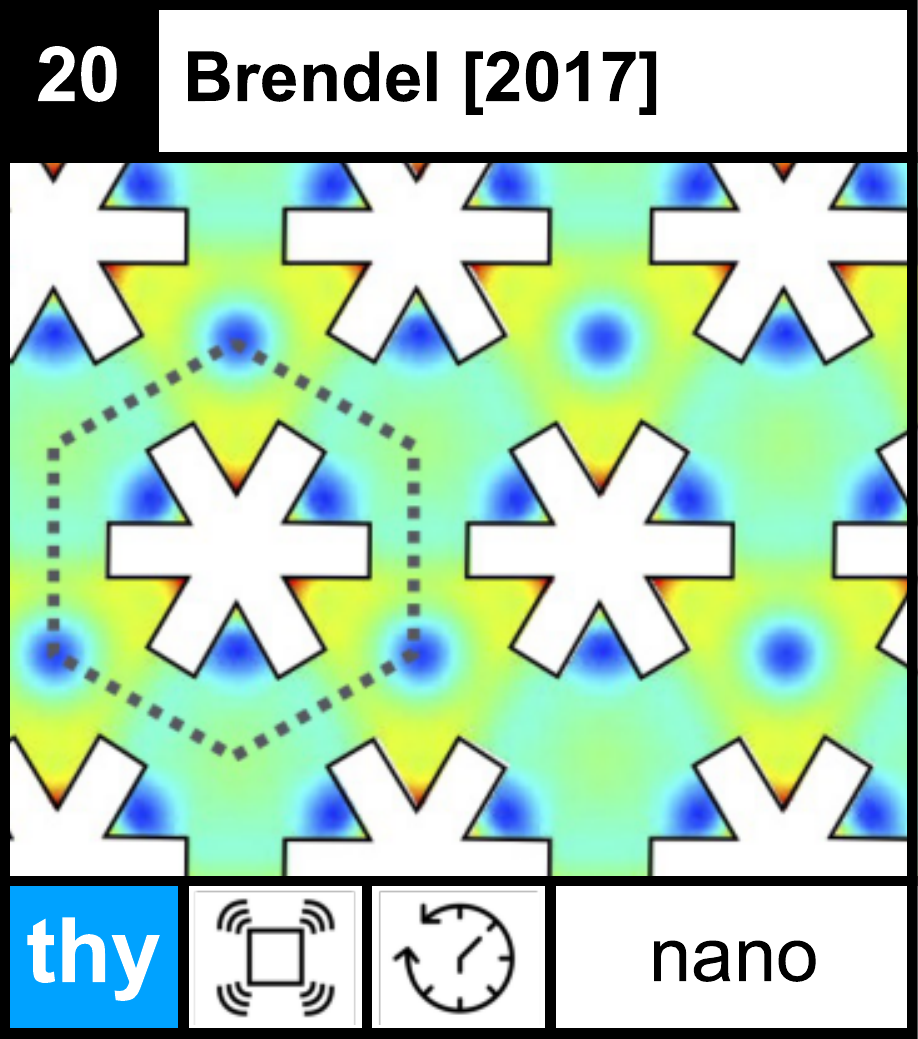}
\includegraphics[width=28mm]{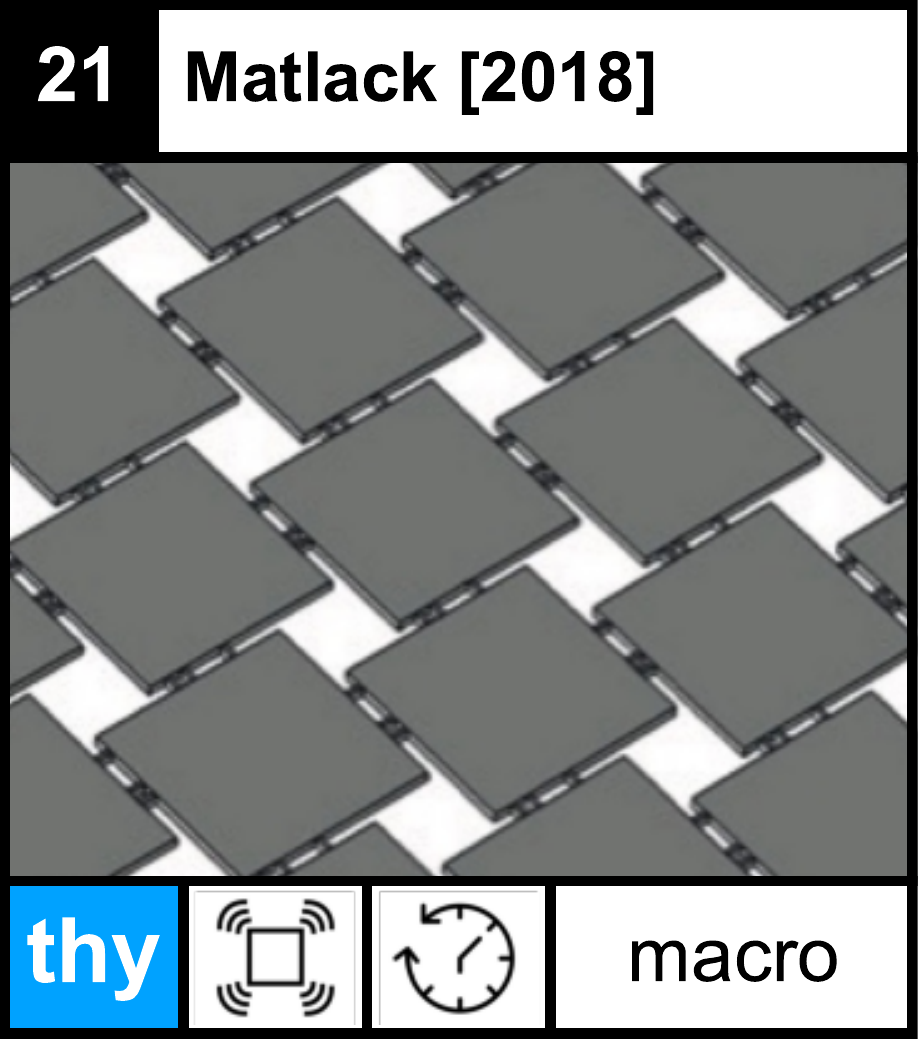}
\includegraphics[width=28mm]{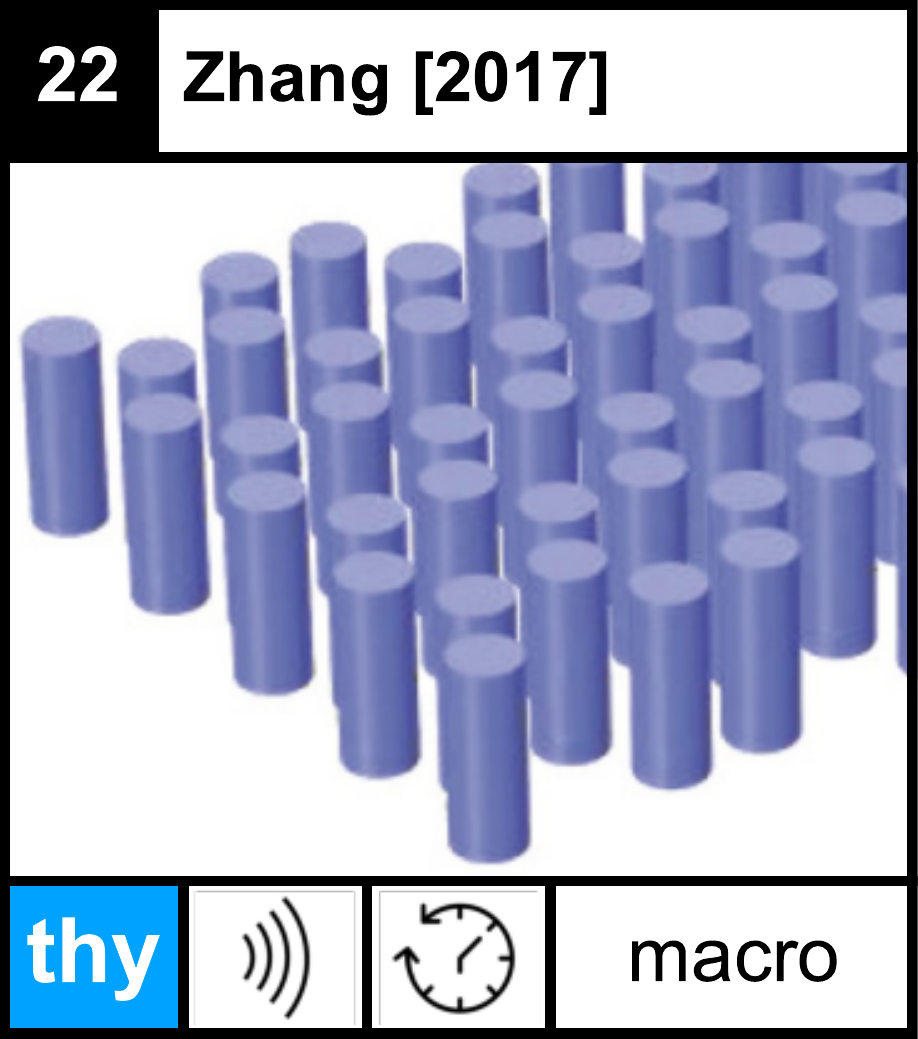}
\includegraphics[width=28mm]{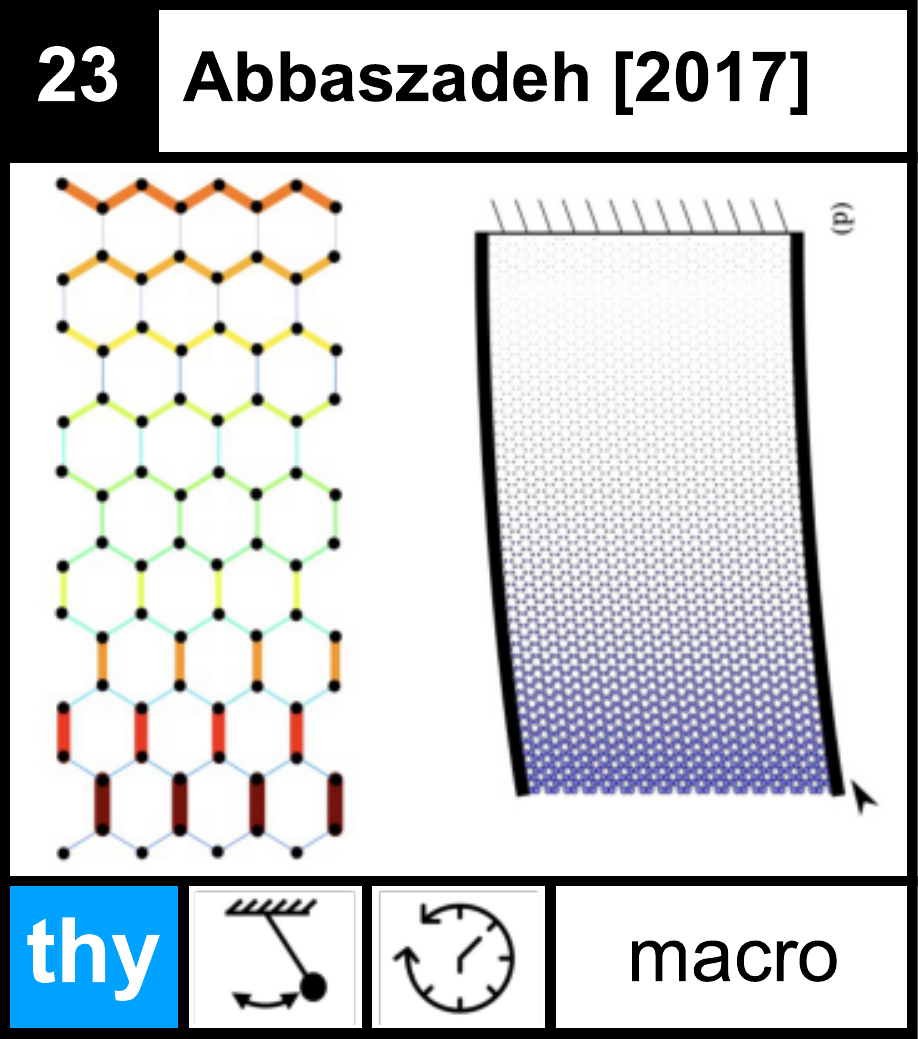}
\\
\includegraphics[width=28mm]{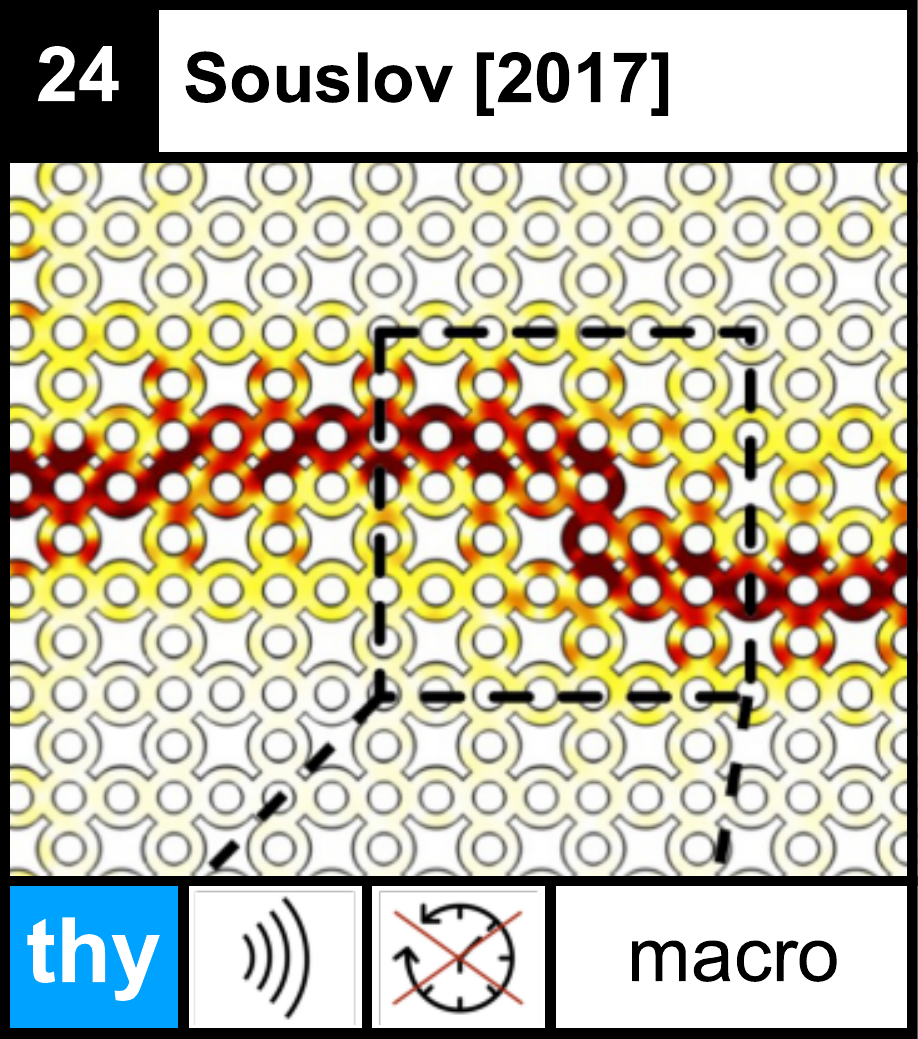}
\includegraphics[width=28mm]{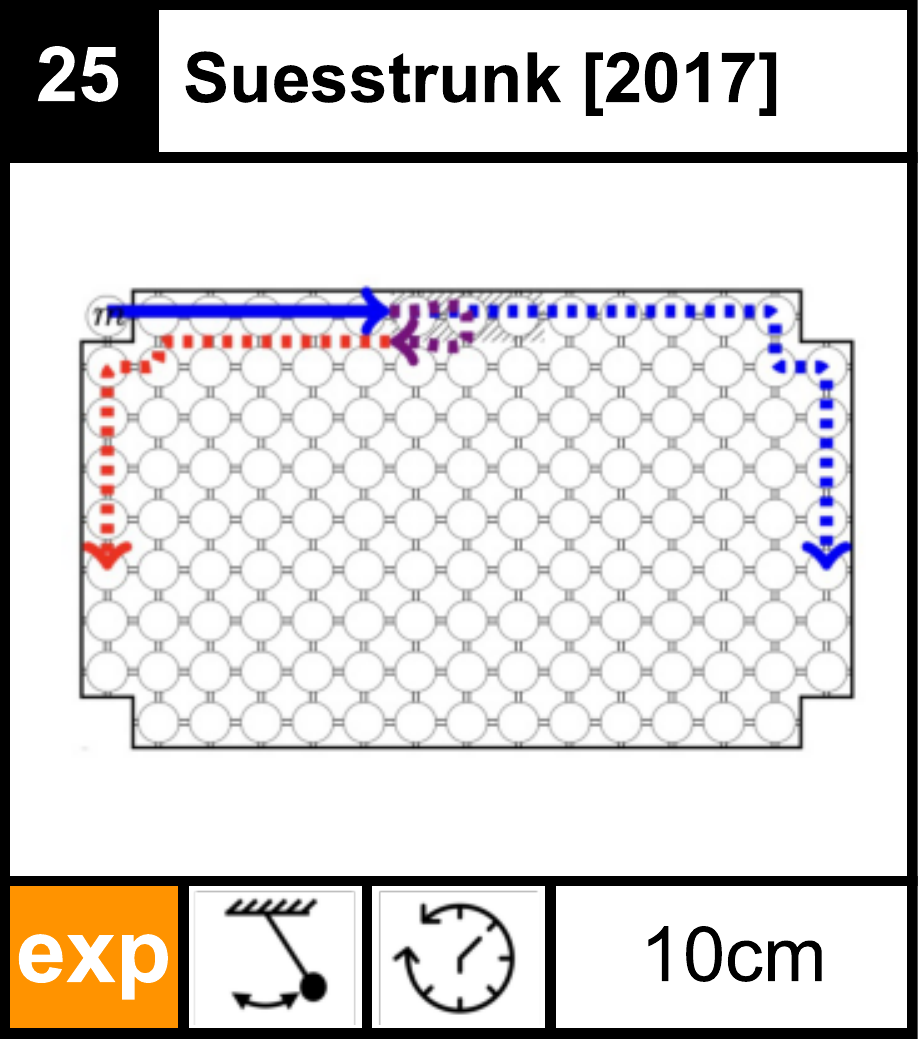}
\includegraphics[width=28mm]{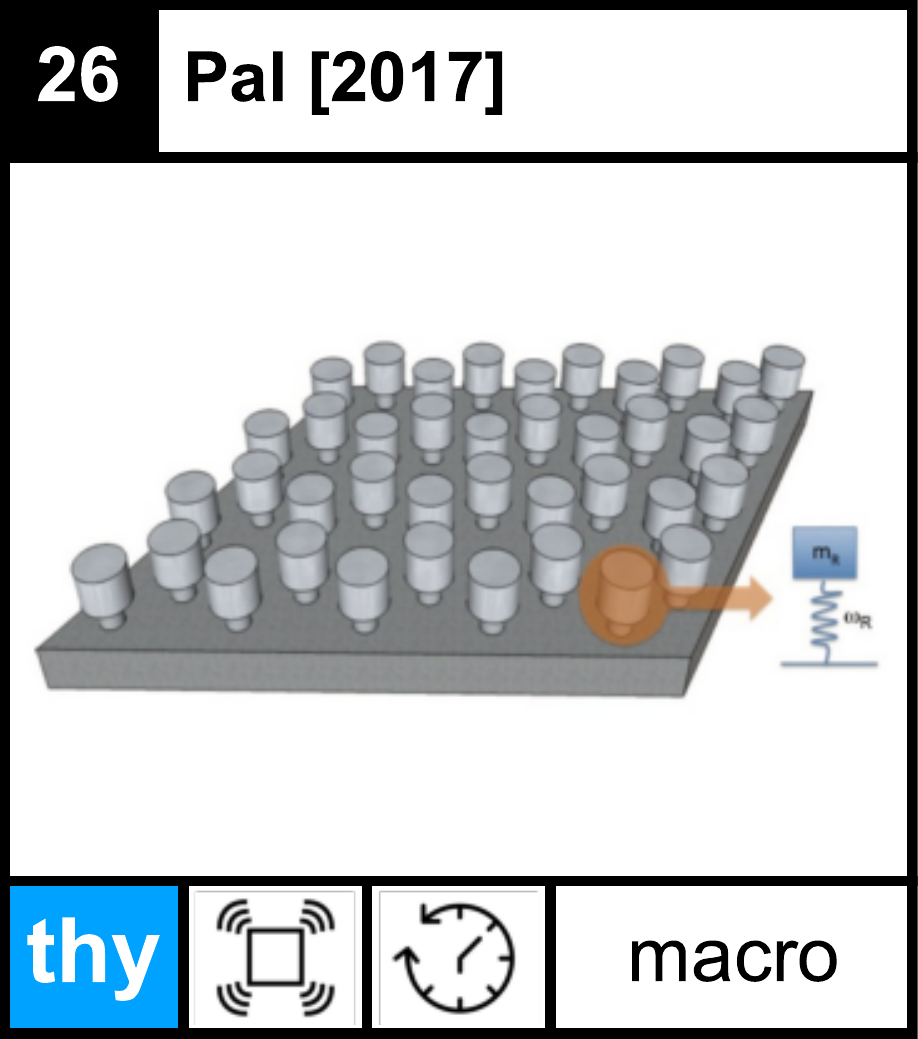}
\includegraphics[width=28mm]{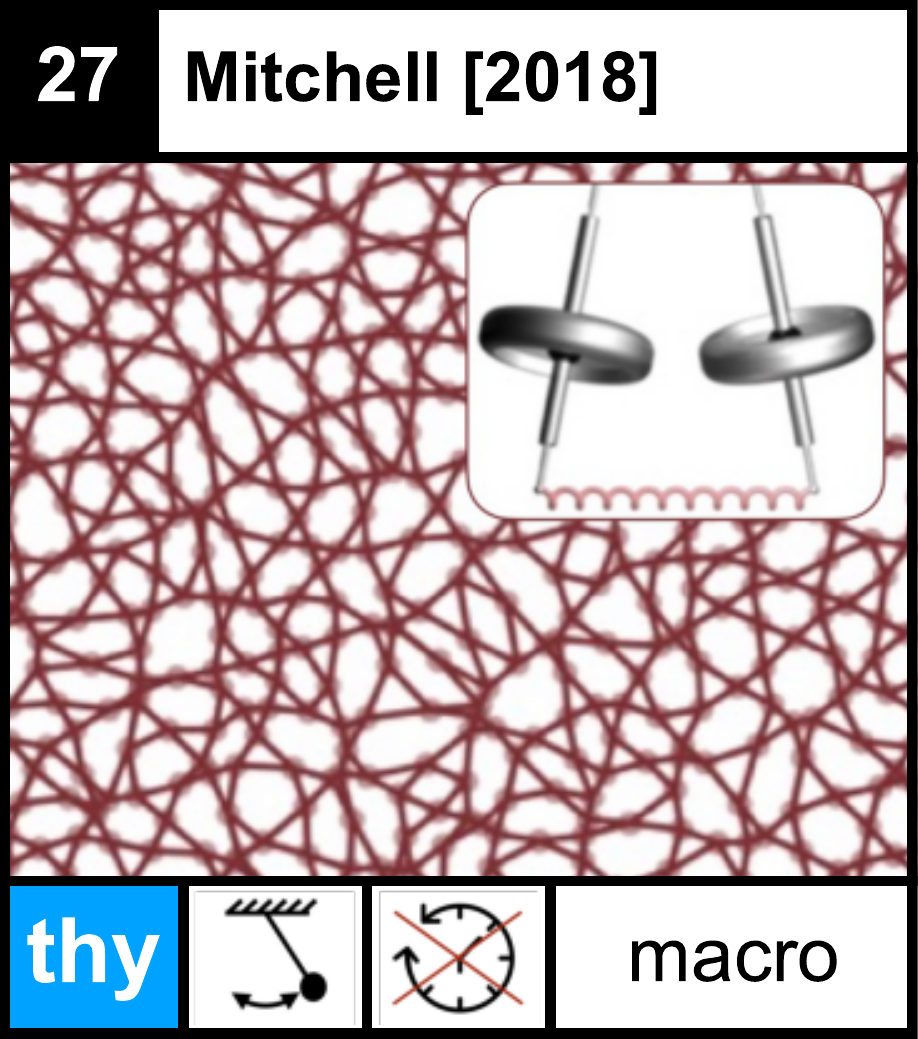}
\includegraphics[width=28mm]{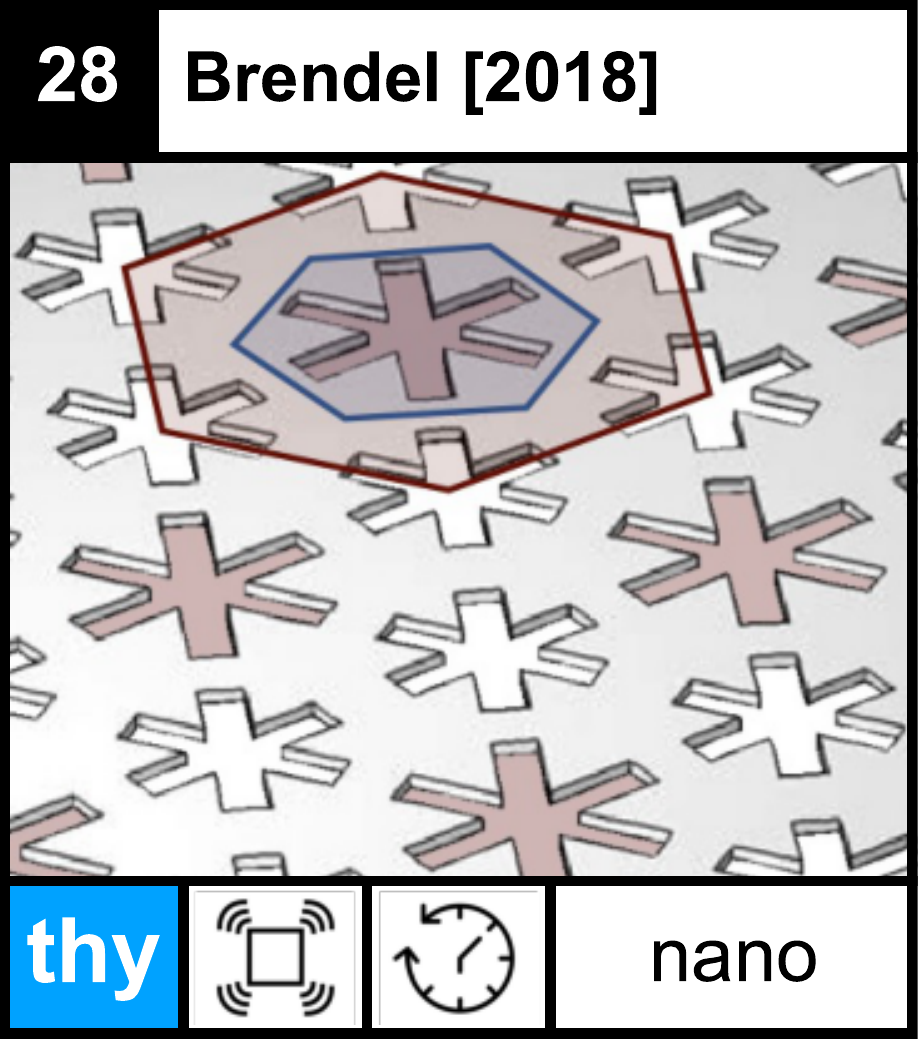}
\includegraphics[width=28mm]{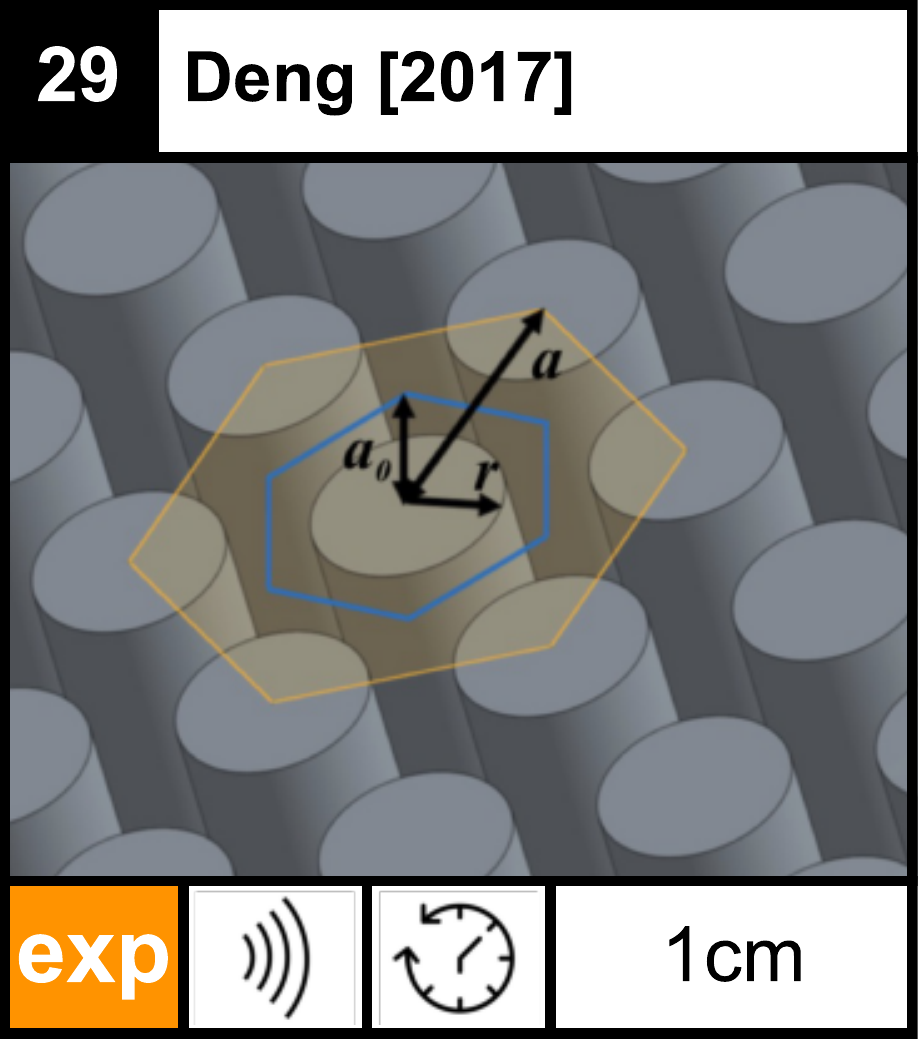}
\\
\includegraphics[width=28mm]{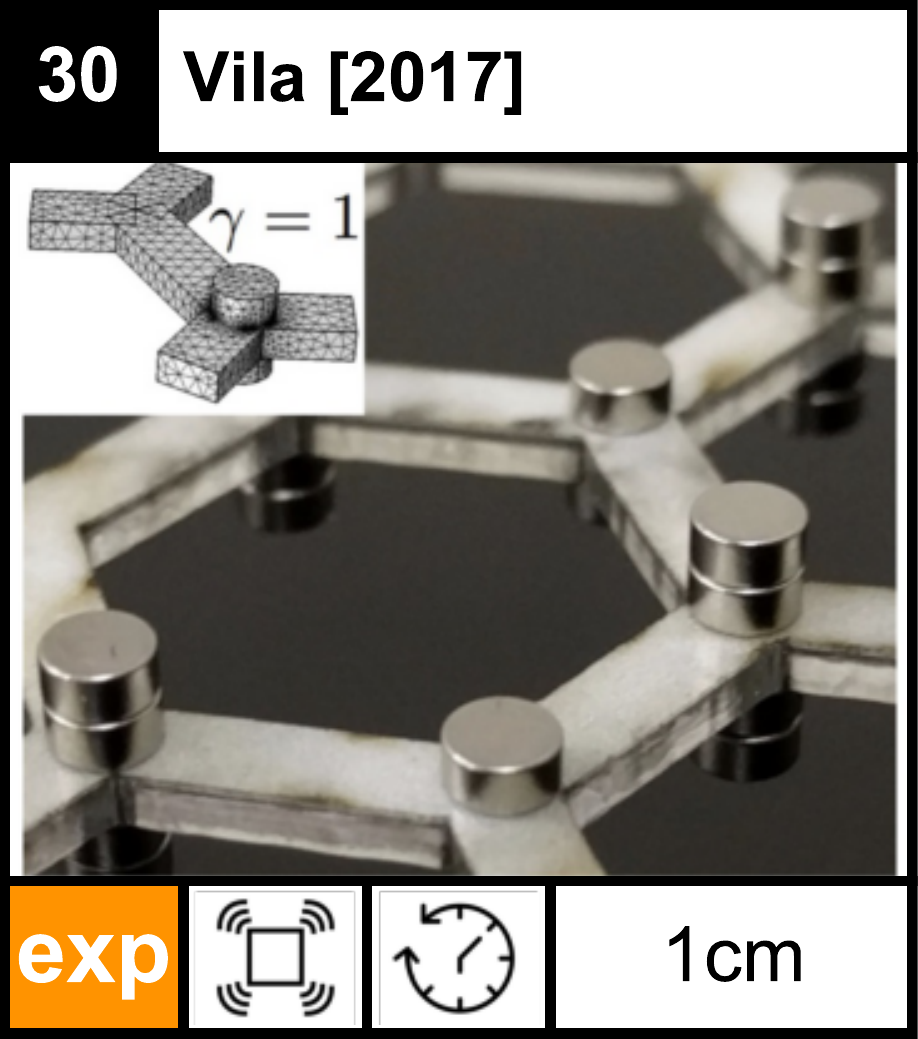}
\includegraphics[width=28mm]{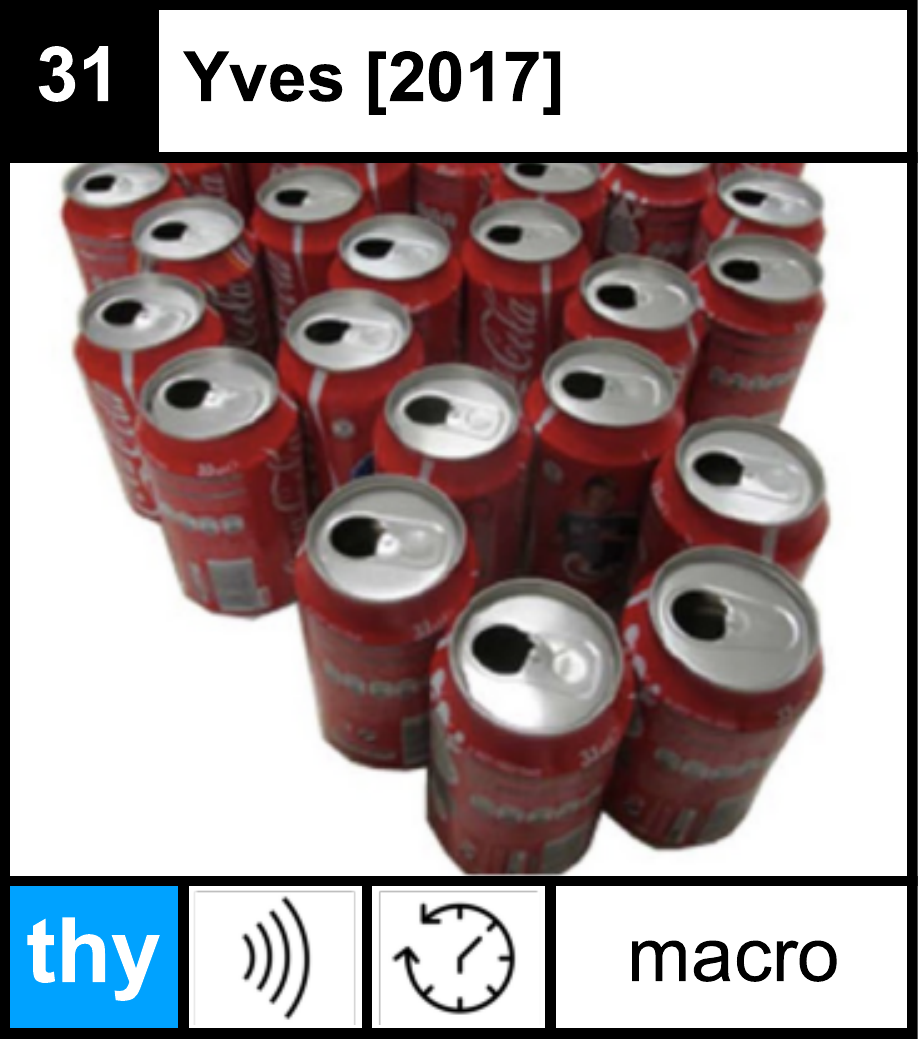}
\includegraphics[width=28mm]{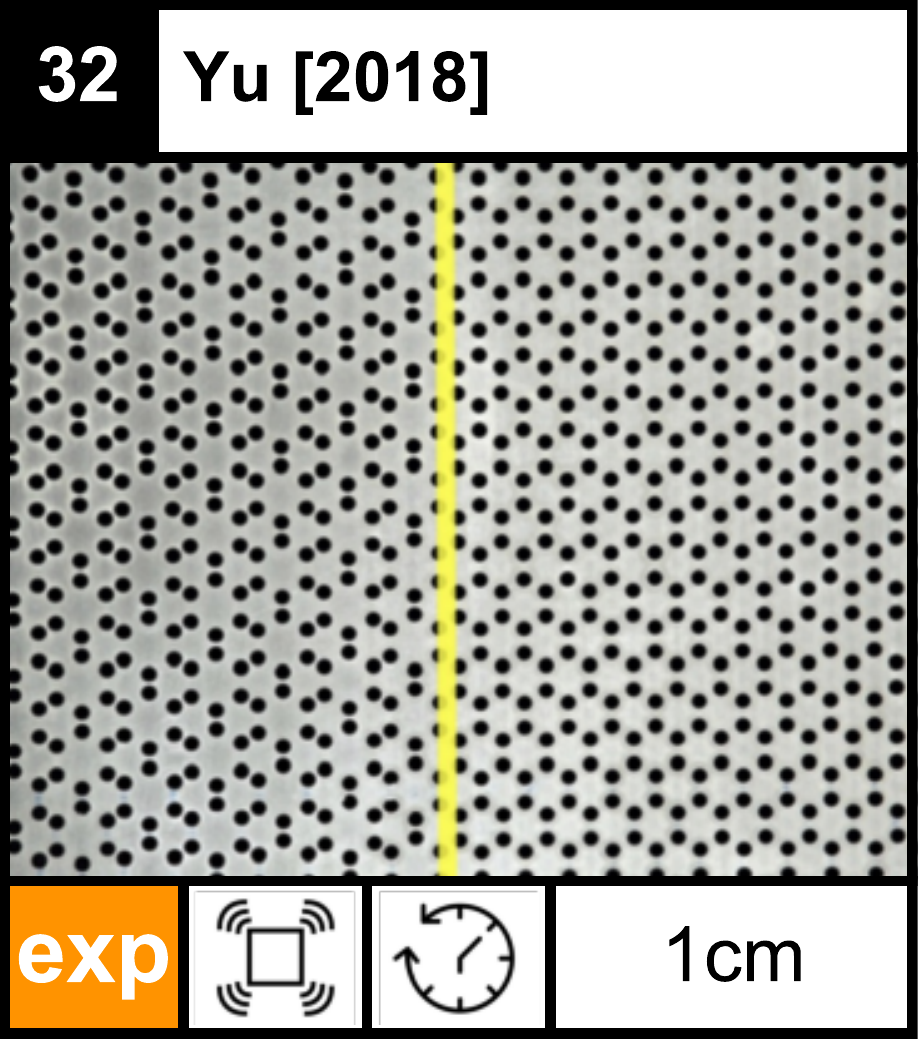}
\includegraphics[width=28mm]{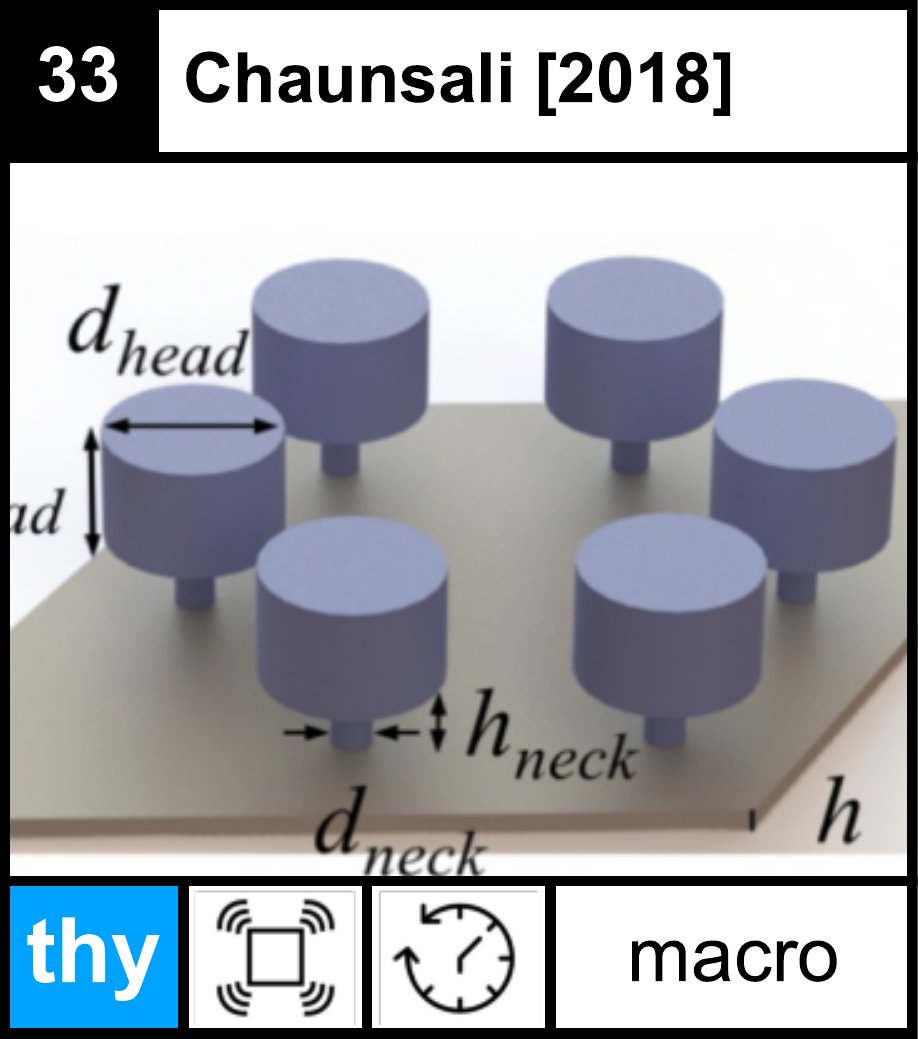}
\includegraphics[width=28mm]{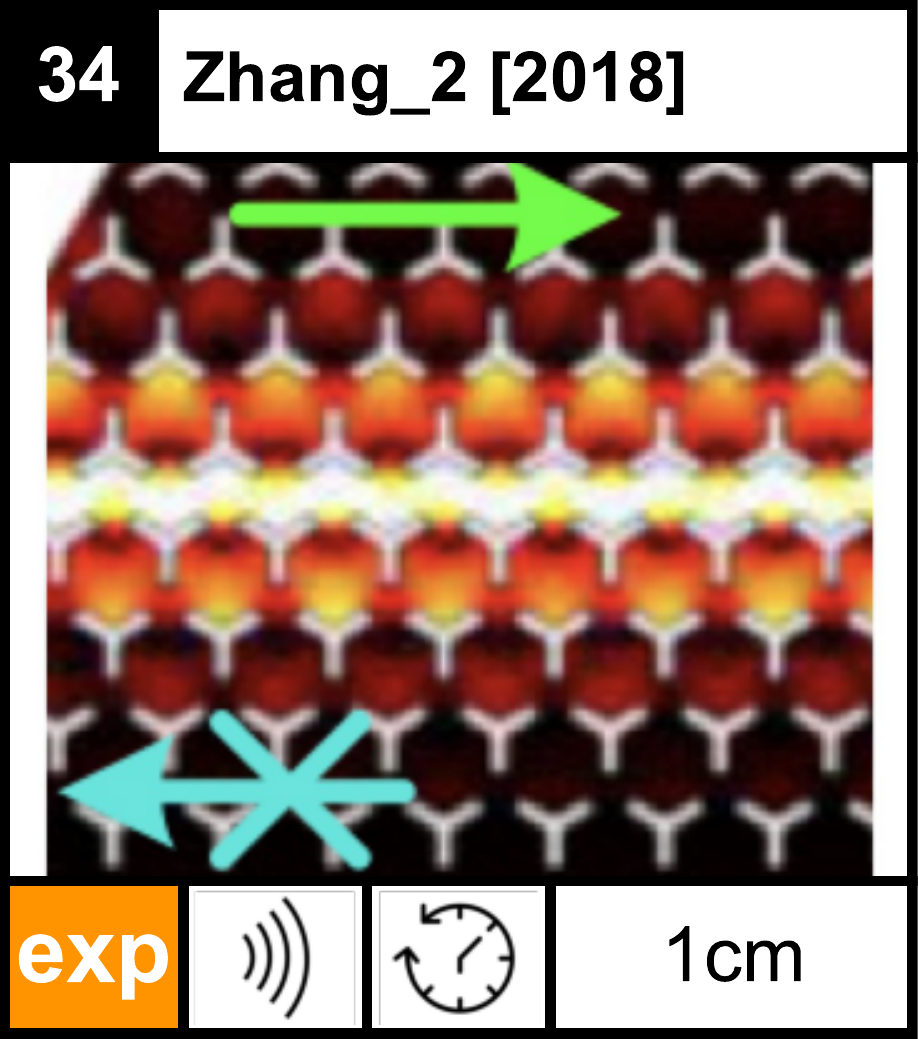}
\includegraphics[width=28mm]{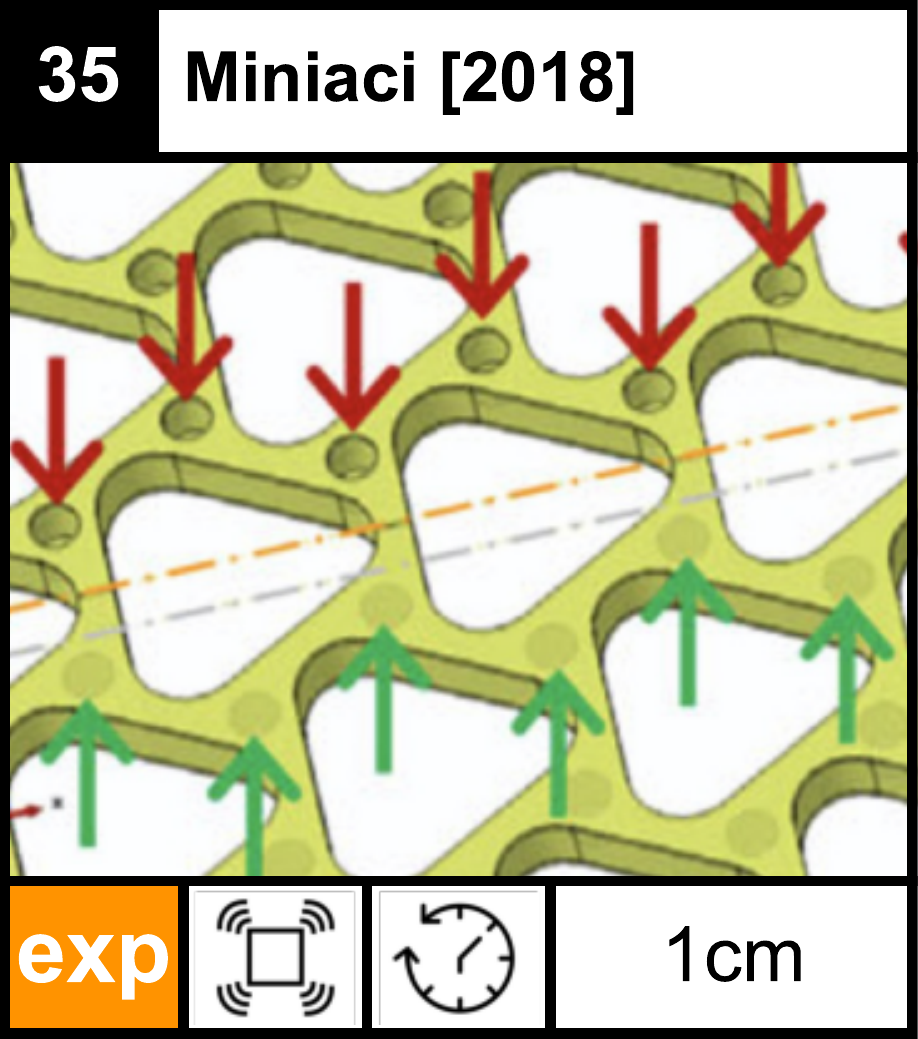}
\end{tabular}
\caption{Graphical timeline of works on topological transport in vibrational systems. We depict both theoretical proposals and experimental works.  
Each panel is labeled with the name of the first author. The icons indicate whether time-reversal symmetry is broken and whether the model employed in the work is based mainly on the coupling of discrete localized modes (pendulum symbol), acoustics in gases or liquids (acoustic wave), or  elastic vibrations in solids (vibrating plate).  
The length-scale is shown for experimental implementations. Whenever theoretical works propose concrete designs that can work at the nanoscale, this is indicated; likewise, we indicate when concepts by their nature are geared towards macro-/mesoscopic scales. References, from top left to bottom right (ordered according to first appearance on the arXiv or submission to a journal): 
\cite{prodan_topological_2009},  \cite{peano_topological_2015}, \cite{wang_coriolis_2015}, \cite{yang_topological_2015},  \cite{ni_topologically_2015},  \cite{susstrunk_observation_2015},  \cite{wang_topological_2015},  \cite{nash_topological_2015},  \cite{kariyado_manipulation_2015},  \cite{khanikaev_topologically_2015},  \cite{mousavi_topologically_2015},  \cite{peng_experimental_2016},  \cite{salerno_floquet_2016}, \cite{pal_helical_2016},  \cite{fleury_floquet_2016},  \cite{chen_tunable_2016},  \cite{he_acoustic_2016},  \cite{lu_observation_2017},  \cite{mei_pseudo-time-reversal_2016},  \cite{brendel_pseudomagnetic_2017},  \cite{matlack_designing_2018},  \cite{zhang_topological_2017}, 
\cite{abbaszadeh_sonic_2017},  \cite{souslov_topological_2017}, \cite{susstrunk_switchable_2017}, \cite{pal_edge_2017},
\cite{mitchell_amorphous_2018},  \cite{brendel_snowflake_2018},  \cite{deng_observation_2017},  \cite{vila_observation_2017},  \cite{yves_topological_2017},  \cite{yu_elastic_2018},  \cite{chaunsali_subwavelength_2018},  \cite{zhang_topological_2018-1},  \cite{miniaci_experimental_2018}. (All the figures are reproduced with permission.)}
\label{fig:timeline_1}
\end{figure*}

\begin{figure*}
\centering
\begin{tabular}{ccccc}
\includegraphics[width=28mm]{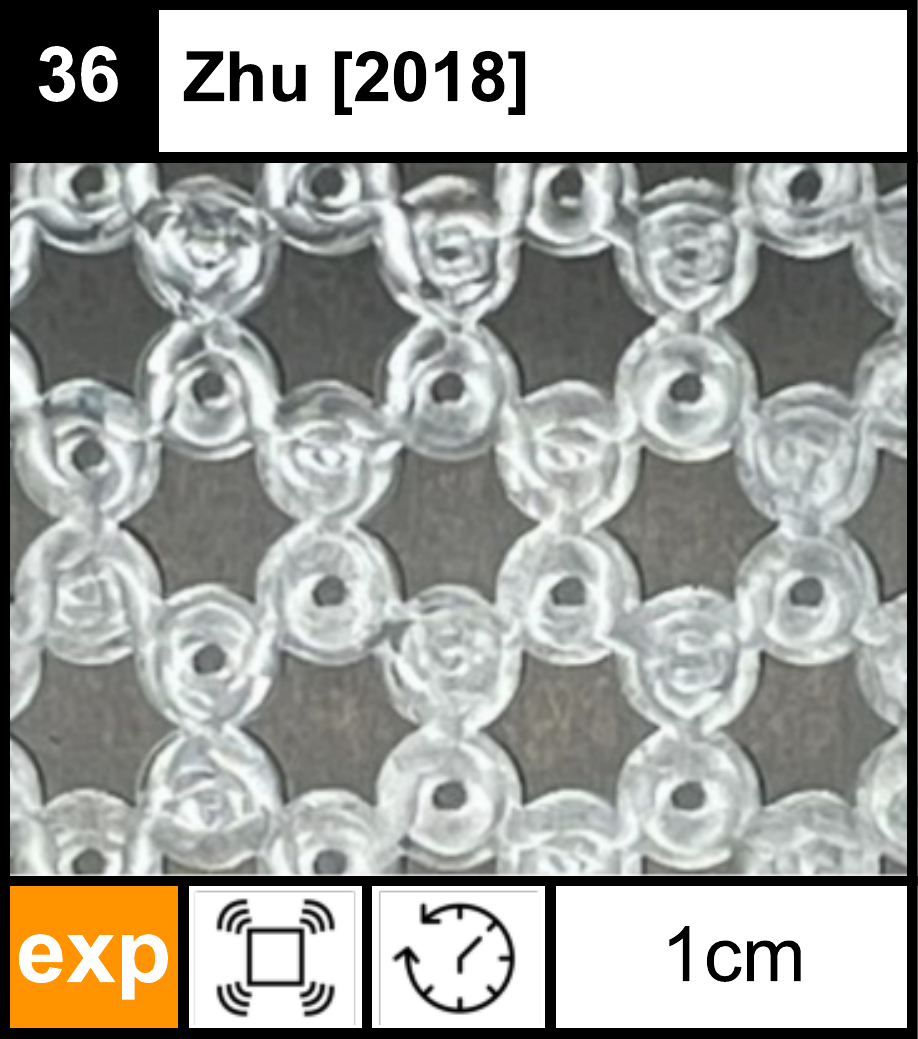}
\includegraphics[width=28mm]{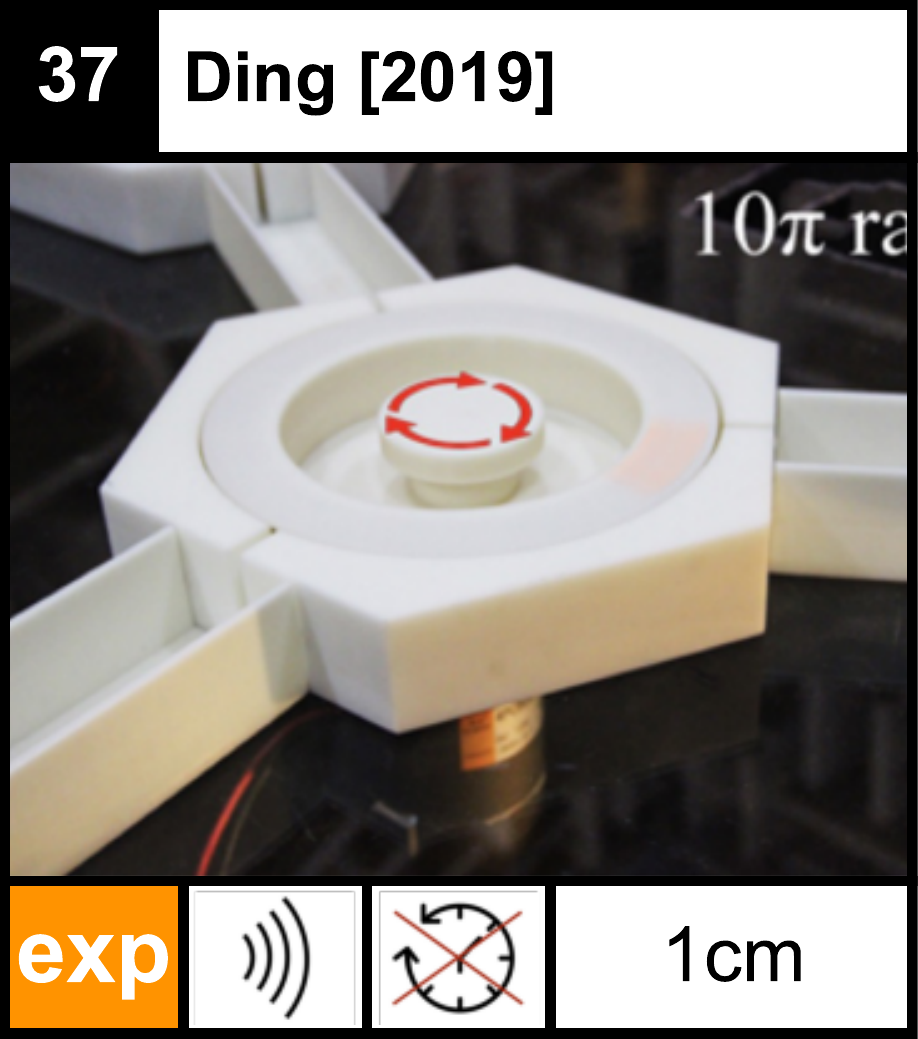}
\includegraphics[width=28mm]{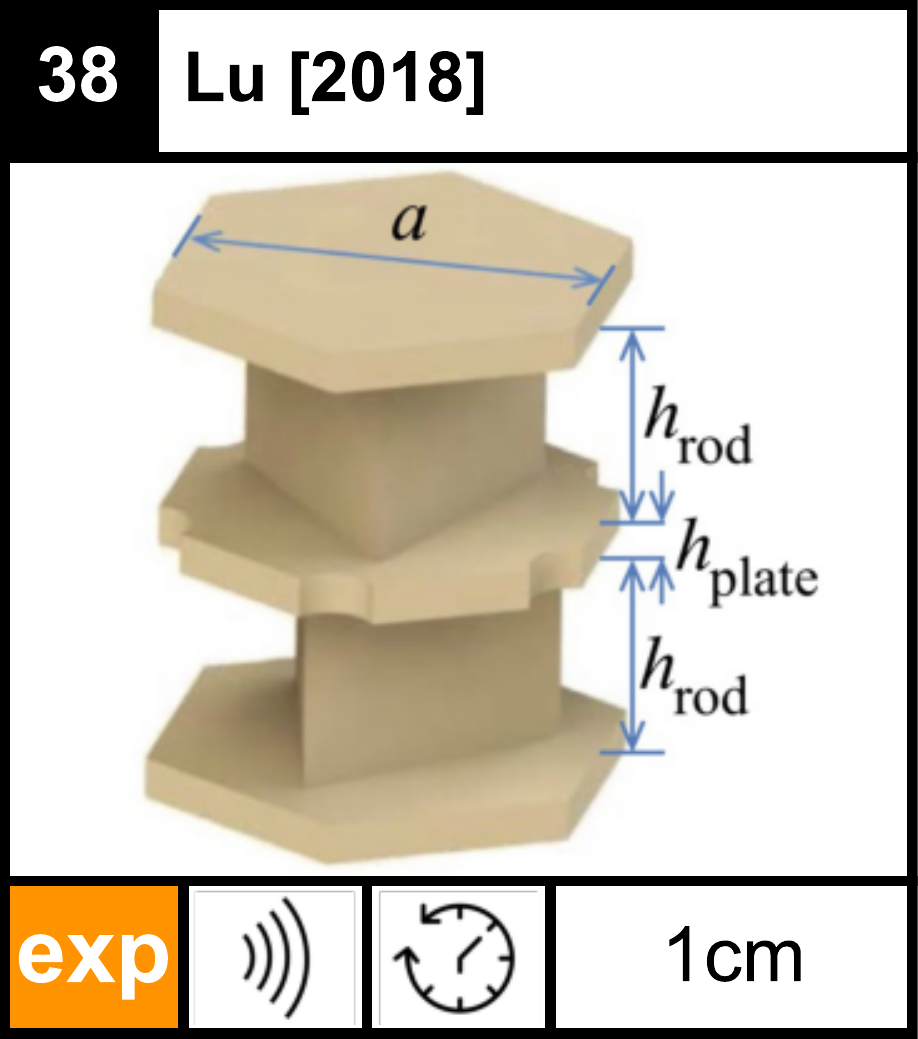}
\includegraphics[width=28mm]{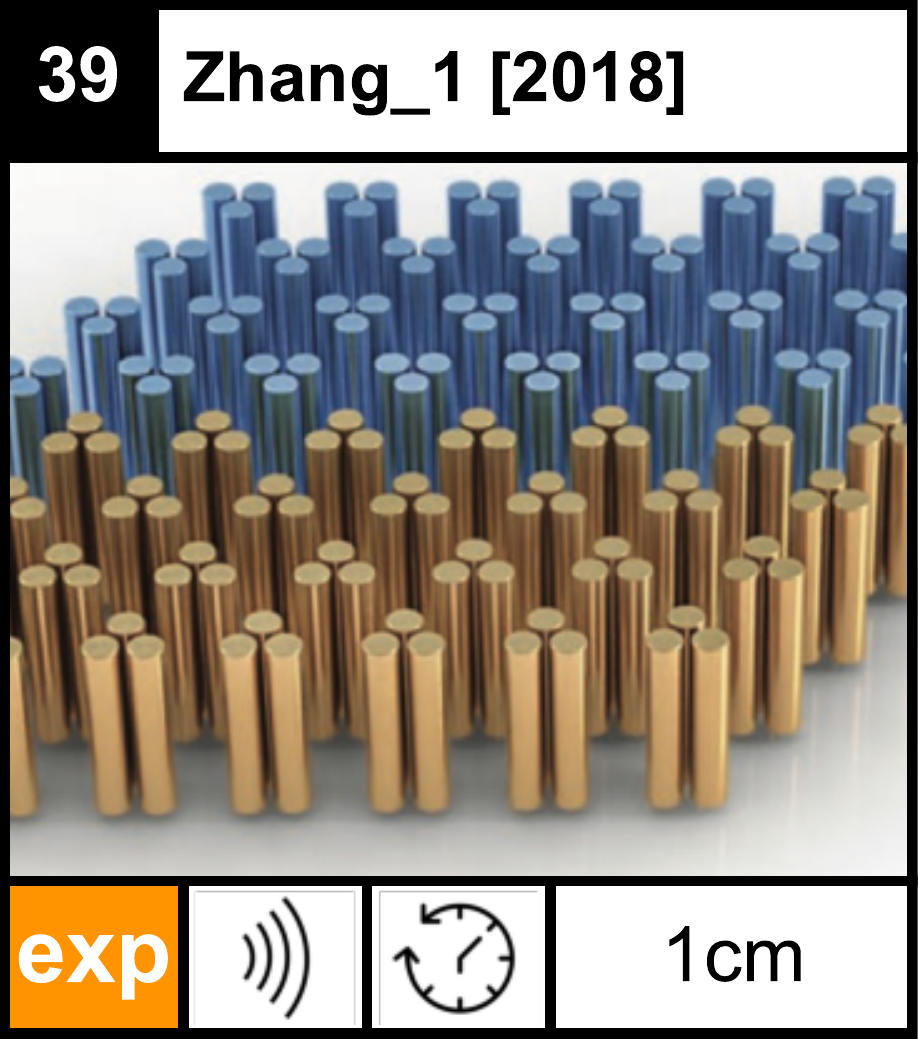}
\includegraphics[width=28mm]{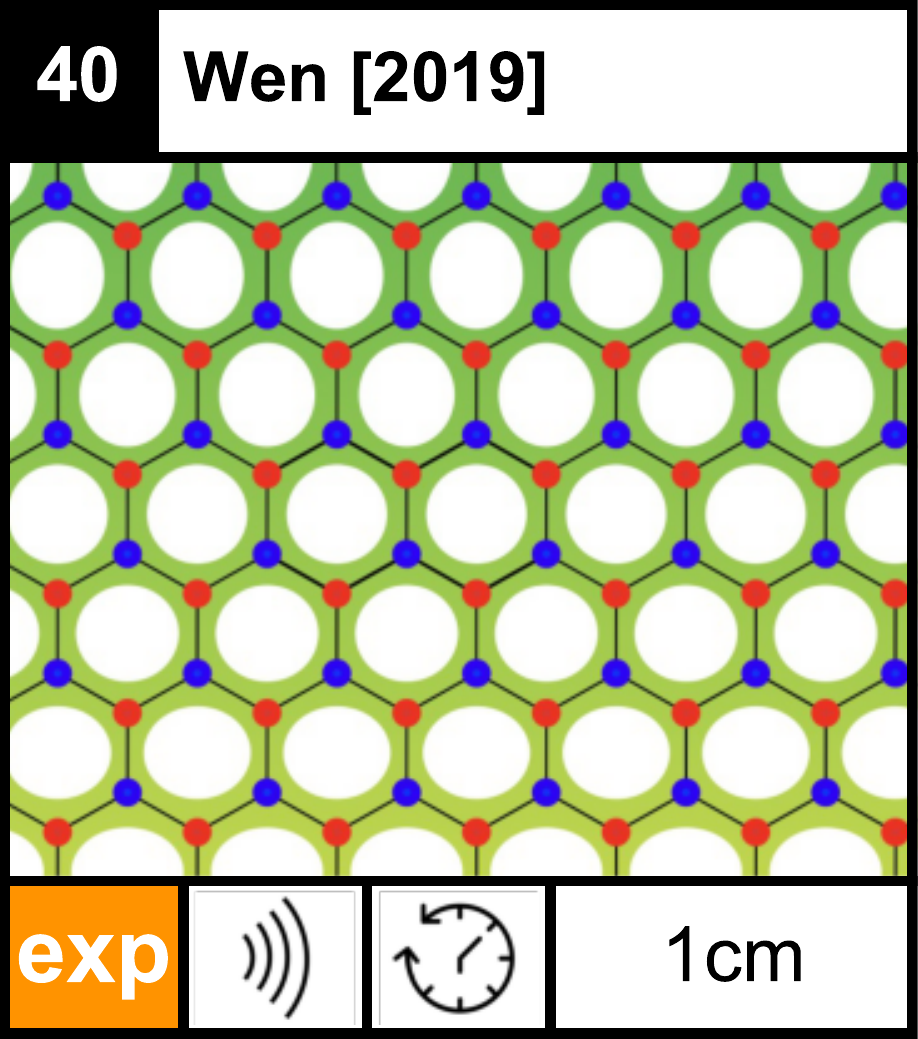}
\includegraphics[width=28mm]{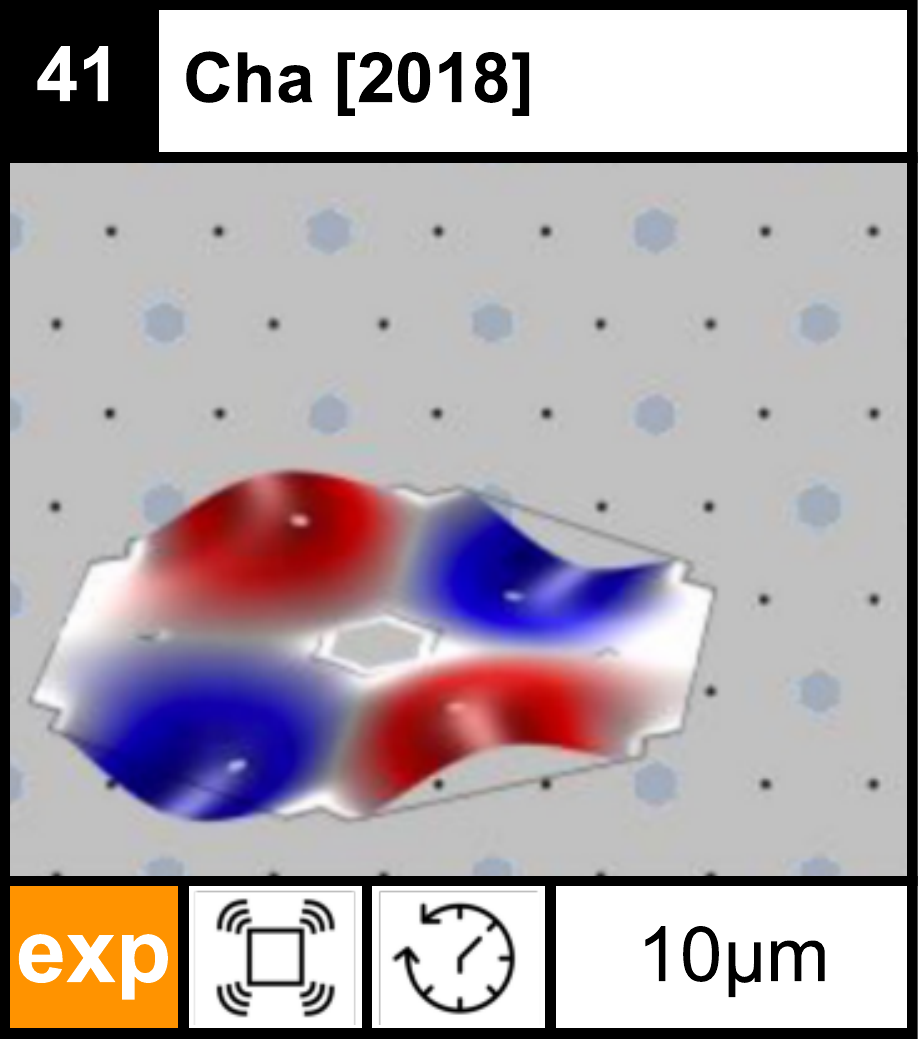}
\\
\includegraphics[width=28mm]{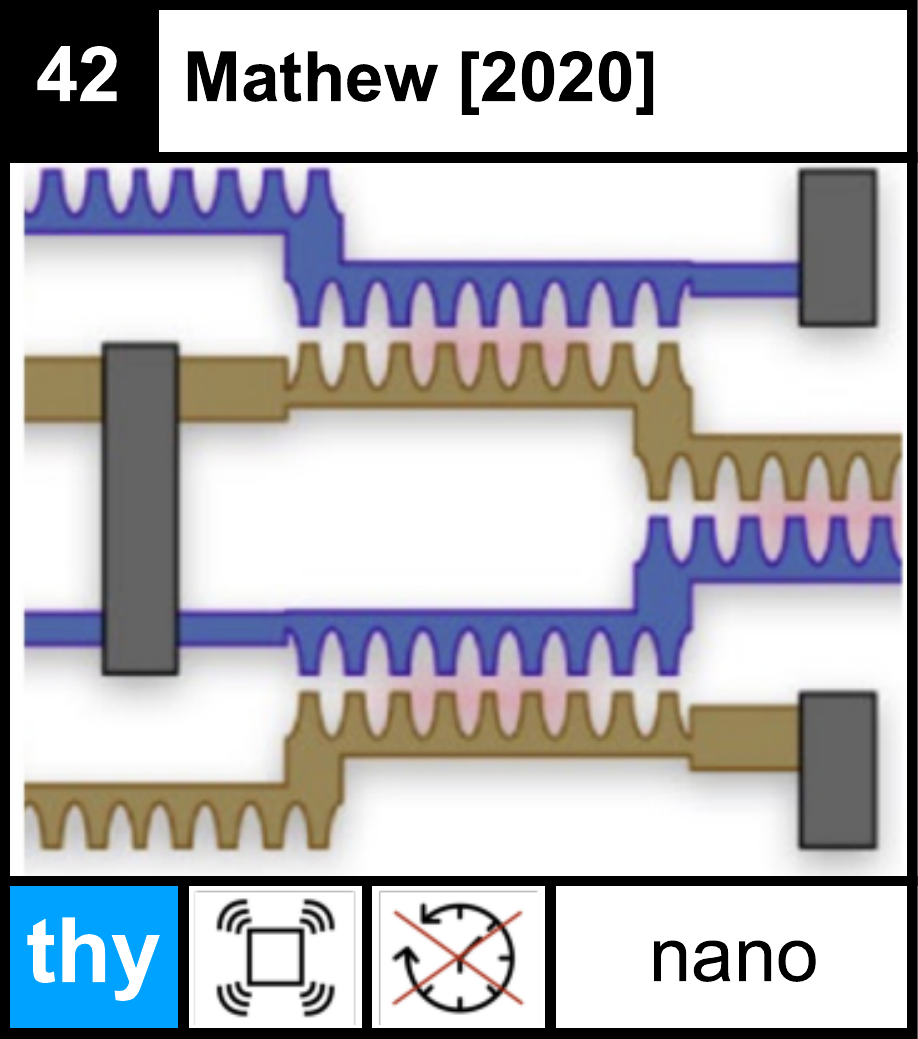}
\includegraphics[width=28mm]{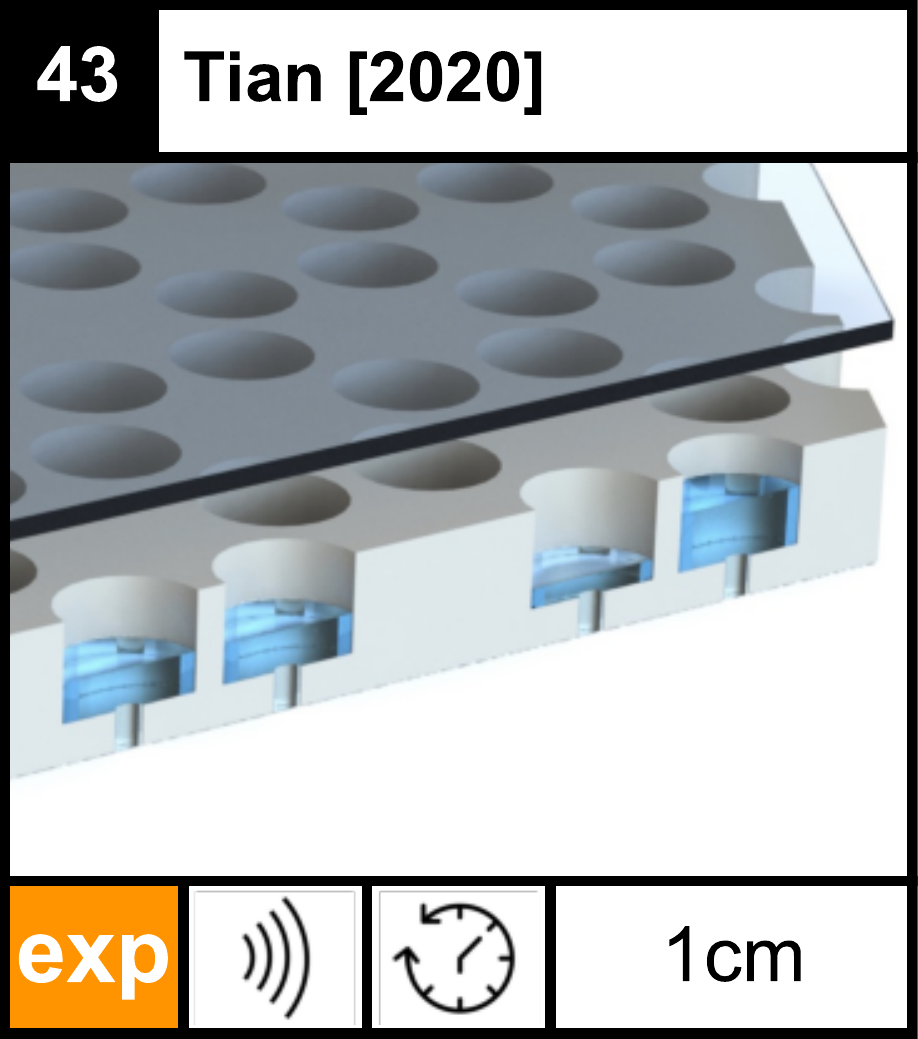}
\includegraphics[width=28mm]{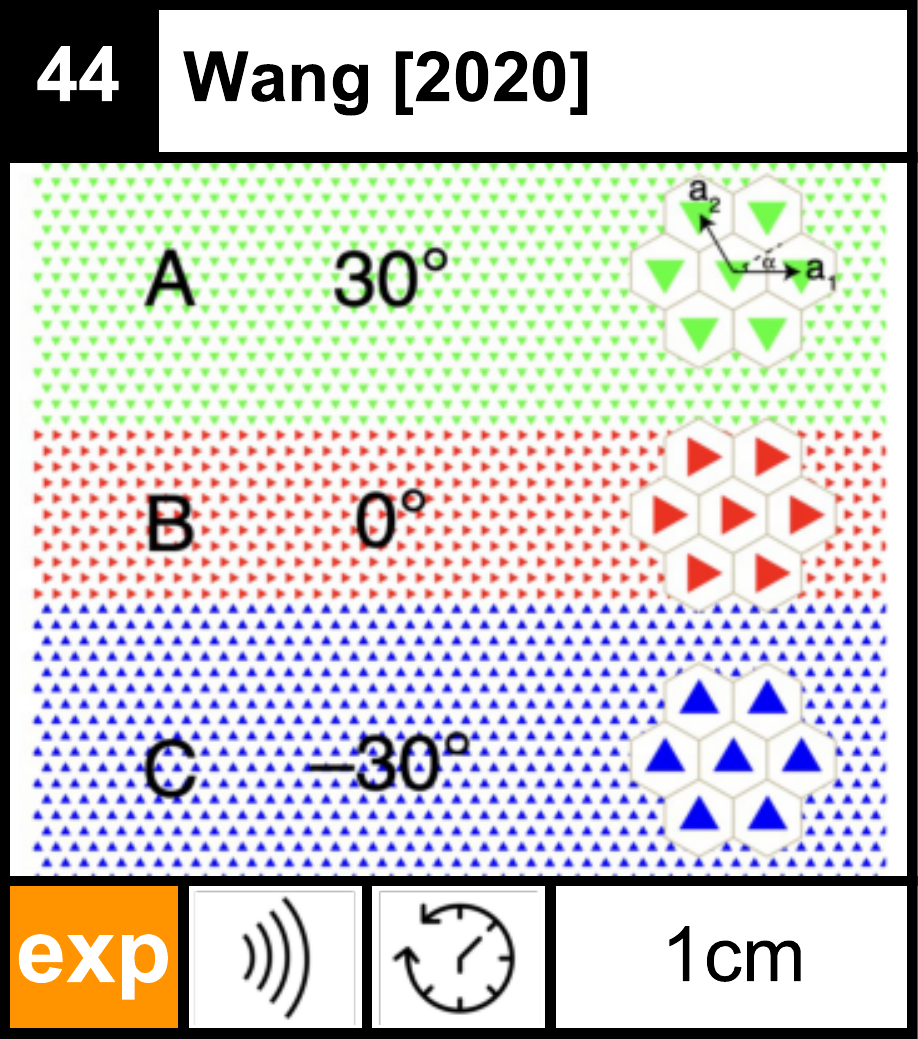}
\includegraphics[width=28mm]{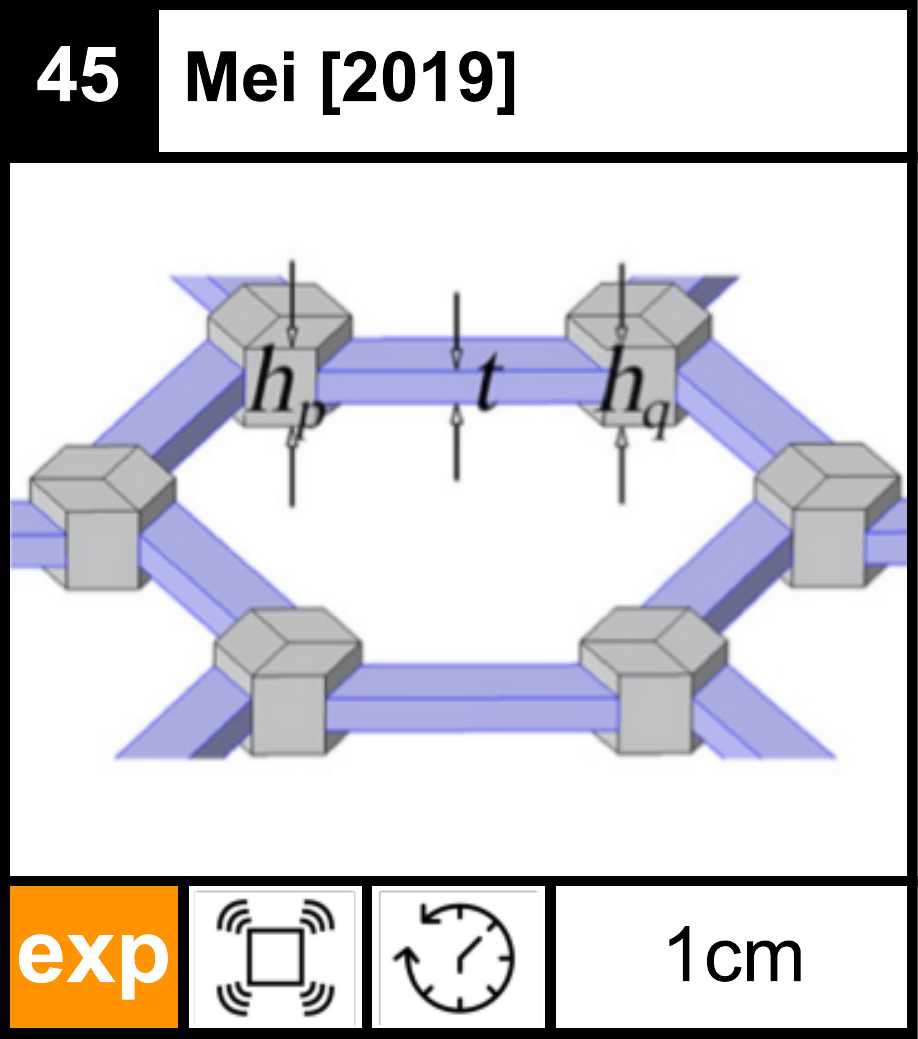}
\includegraphics[width=28mm]{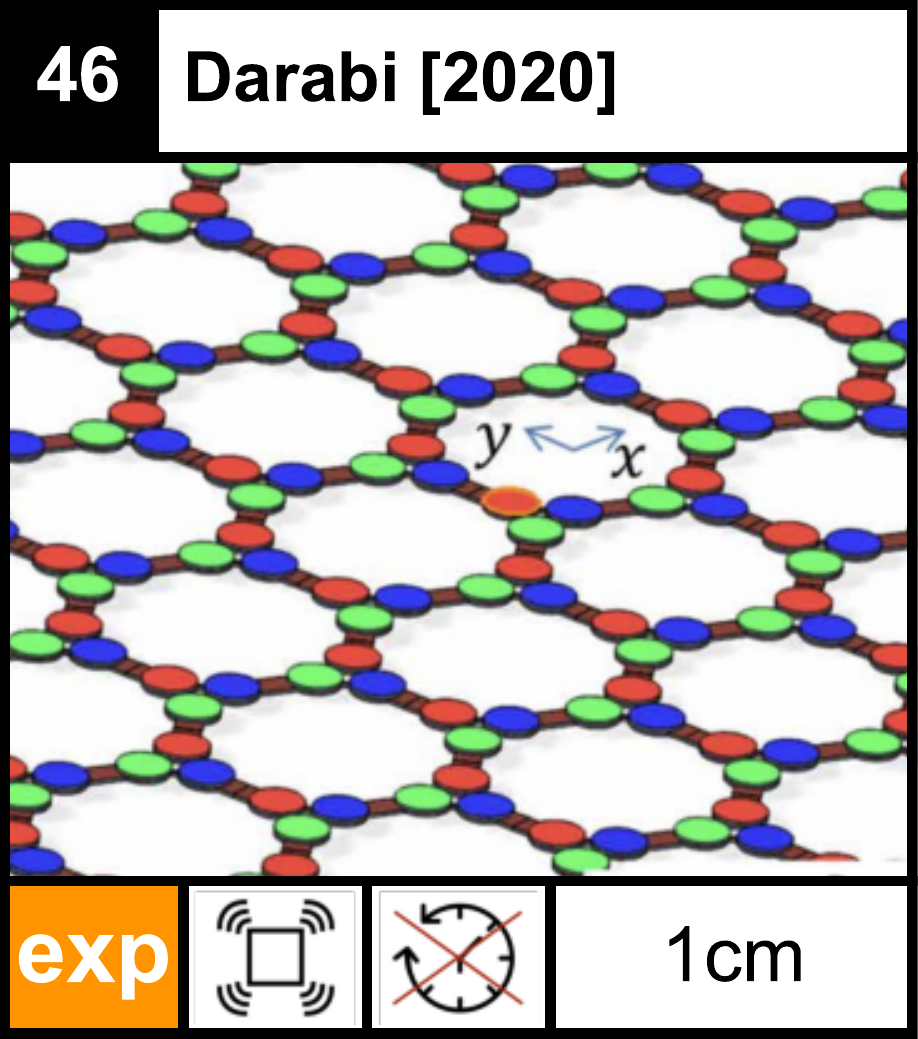}
\includegraphics[width=28mm]{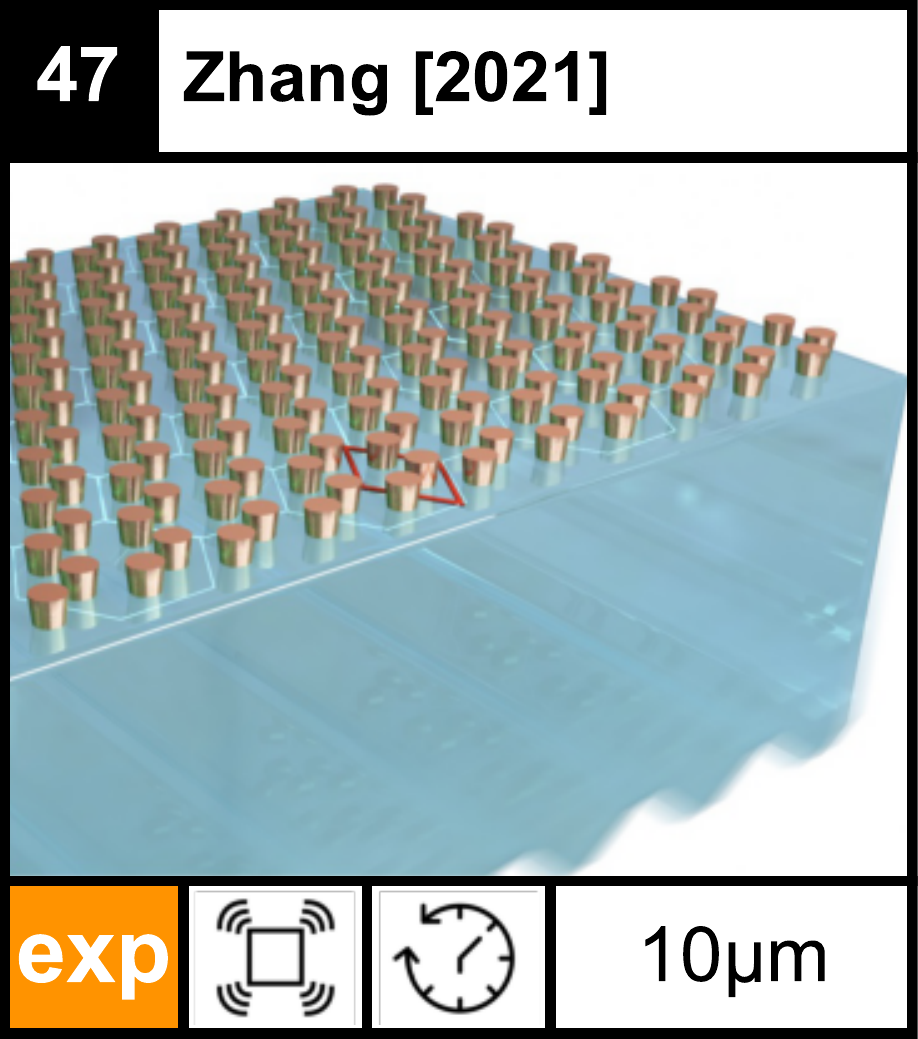}
\\
\includegraphics[width=28mm]{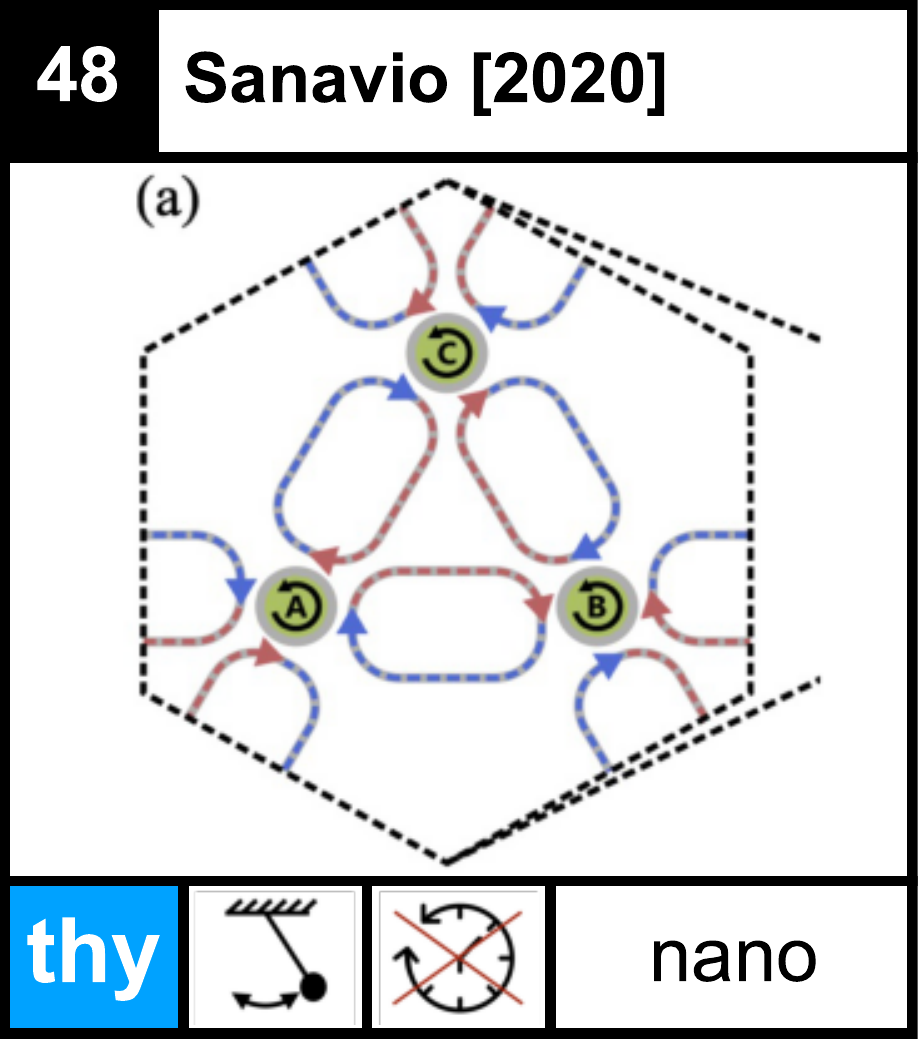}
\includegraphics[width=28mm]{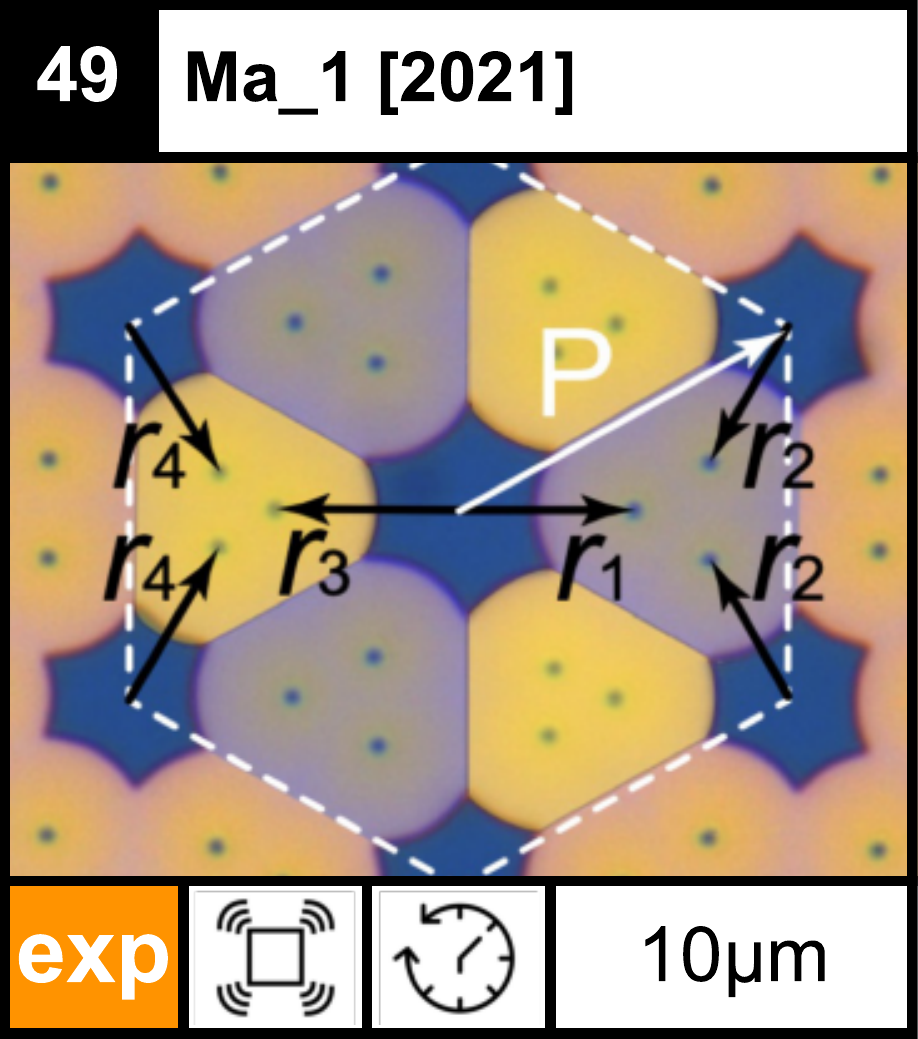}
\includegraphics[width=28mm]{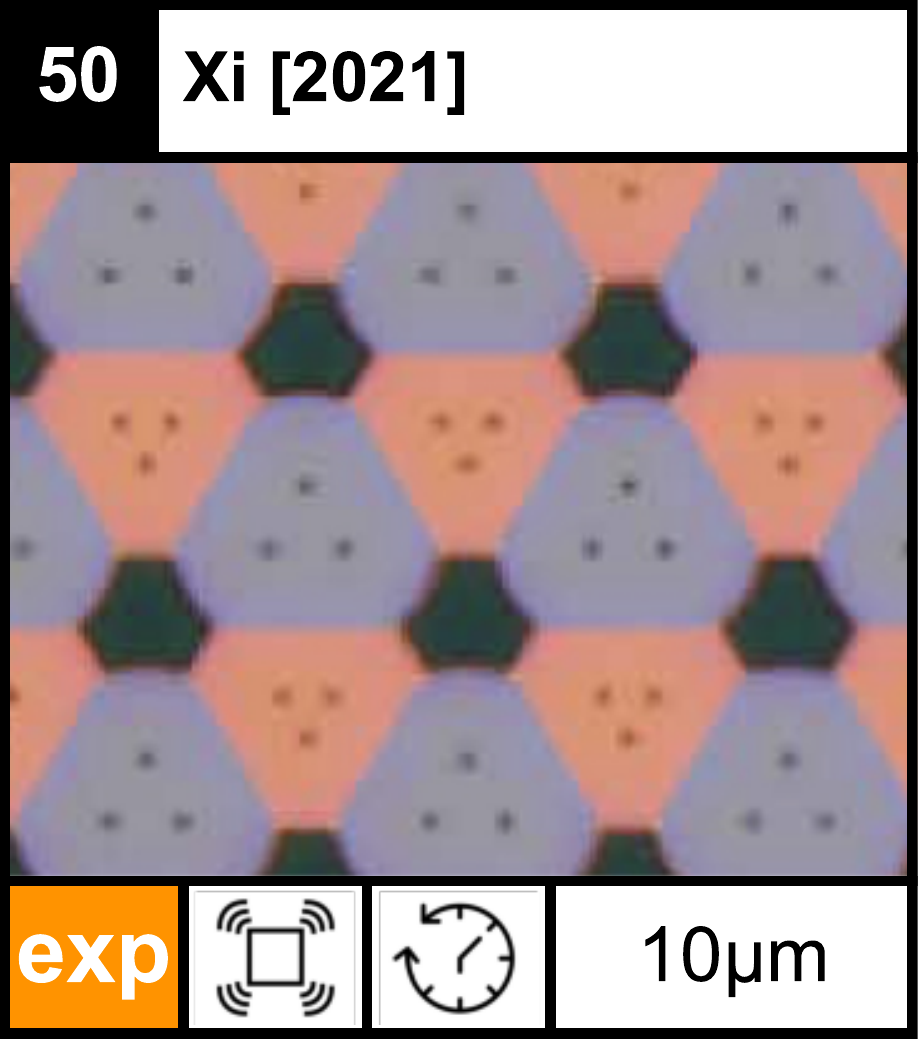}
\includegraphics[width=28mm]{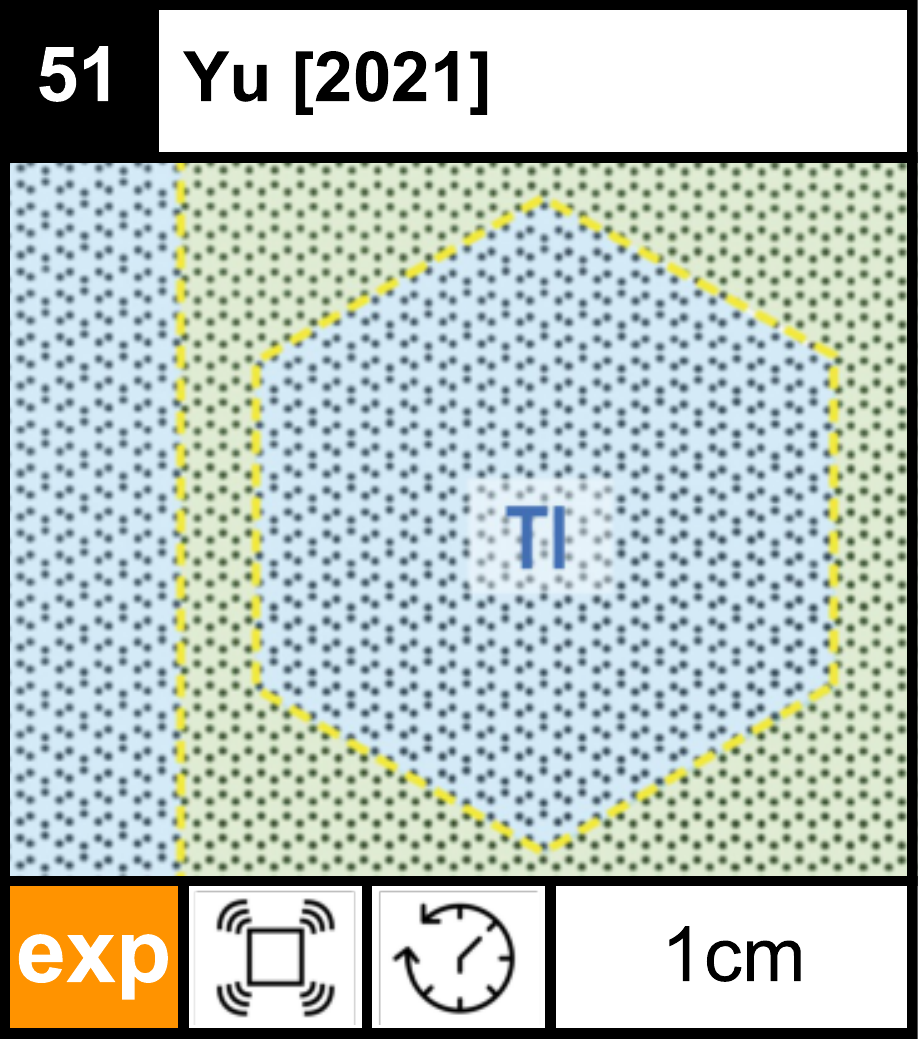}
\includegraphics[width=28mm]{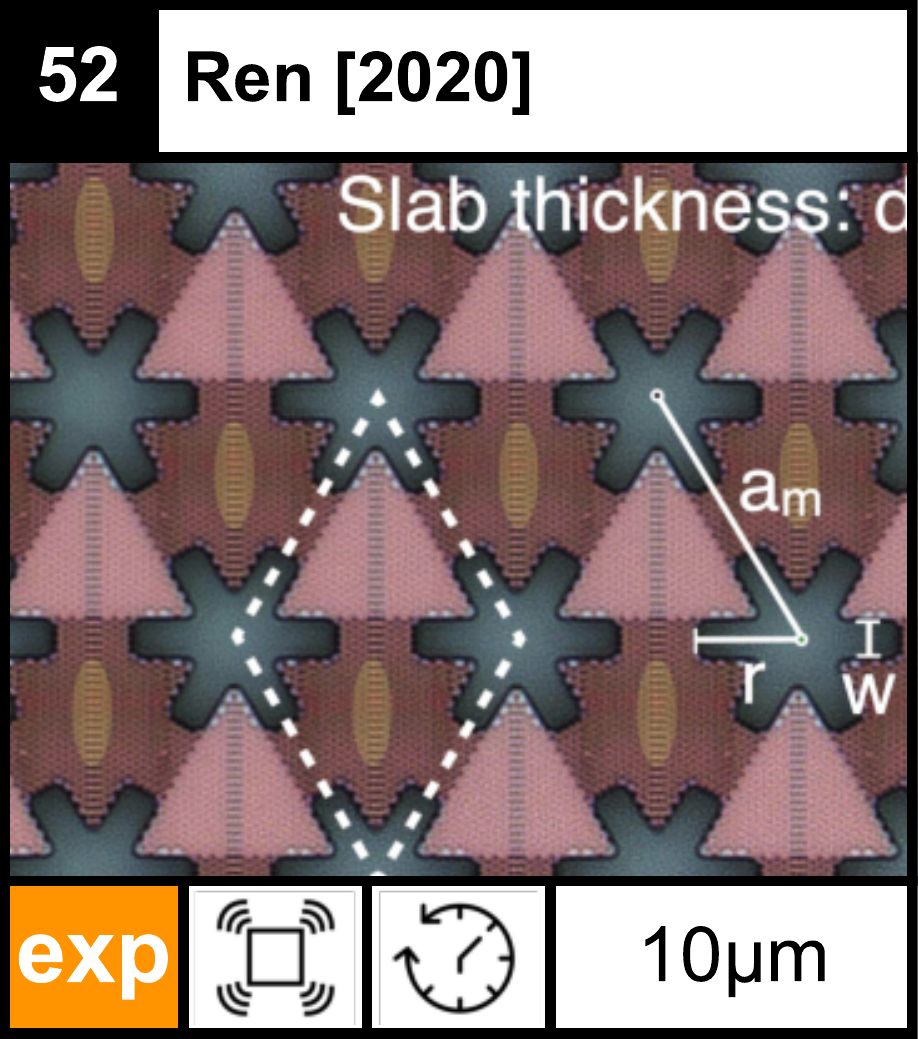}
\includegraphics[width=28mm]{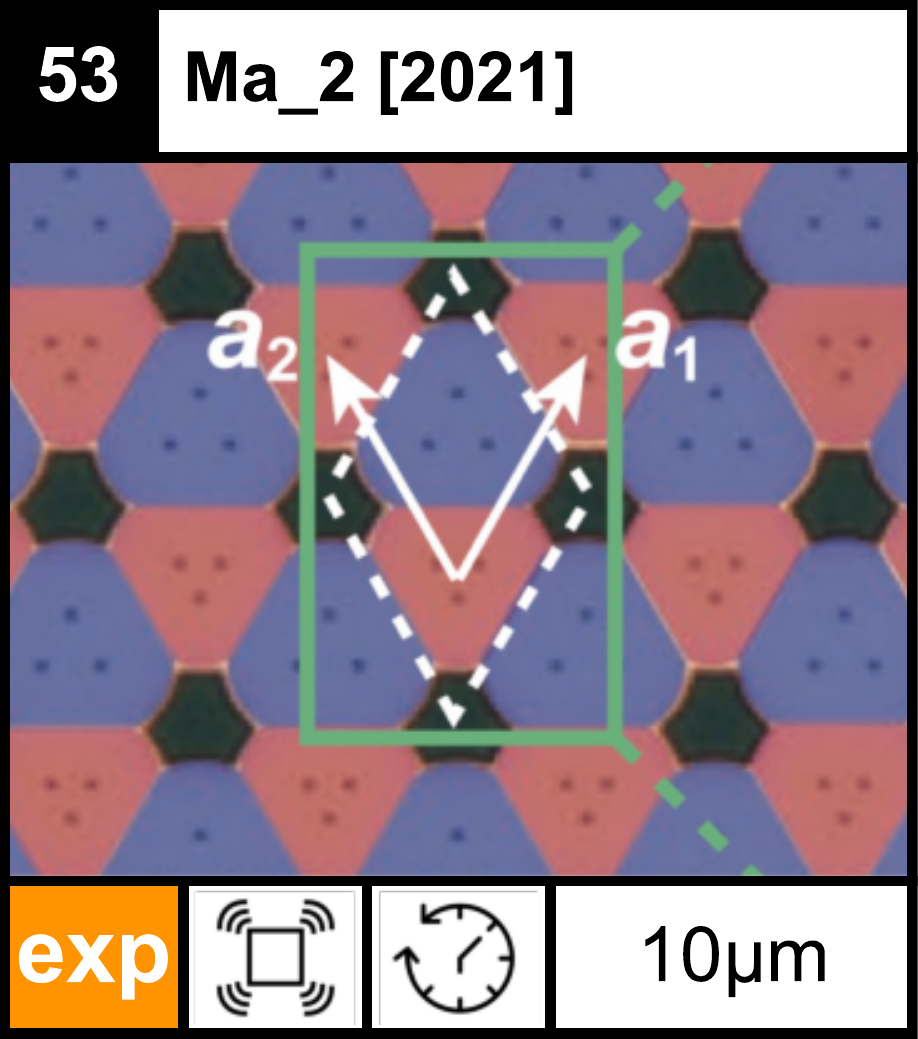}
\\
\includegraphics[width=28mm]{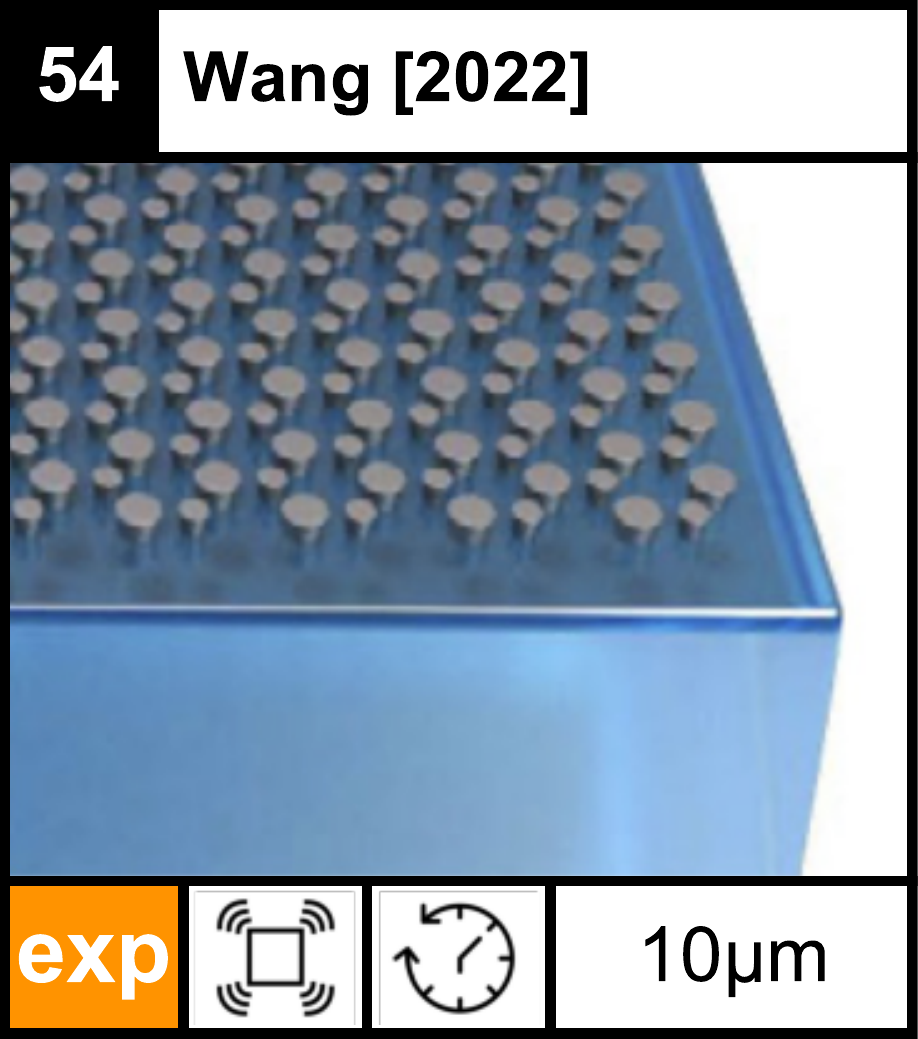}
\includegraphics[width=28mm]{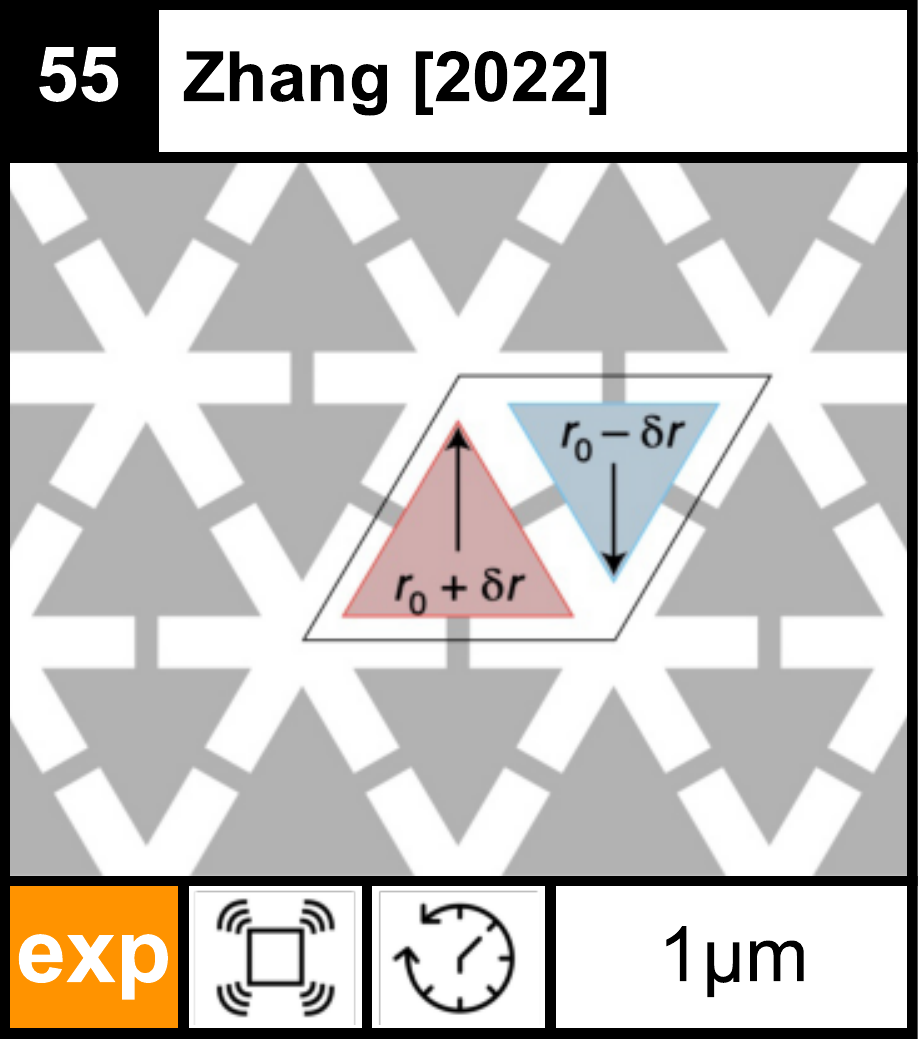}
\end{tabular}
\caption{Graphical timeline of works (continued).
References, from top left to bottom right:
\cite{zhu_design_2018}, \cite{ding_experimental_2019}, \cite{lu_valley_2018},  \cite{zhang_directional_2018}, \cite{wen_acoustic_2019}, \cite{cha_experimental_2018},  \cite{mathew_synthetic_2020},  \cite{tian_dispersion_2020},  \cite{wang_valley-locked_2020},  \cite{mei_robust_2019},  \cite{darabi_reconfigurable_2020},  \cite{zhang_topological_2021},  \cite{sanavio_nonreciprocal_2020},  \cite{ma_nanomechanical_2021}, \cite{xi_observation_2021}, \cite{yu_critical_2021},  \cite{ren_topological_2022},  \cite{ma_experimental_2021}, \cite{wang_extended_2022}, \cite{zhang_gigahertz_2022}. (All the figures are reproduced with permission.)
}
\label{fig:timeline_2}
\end{figure*}

After having understood the general fundamental mechanisms behind topologically protected edge states, we now focus on mechanical vibrations. Adopting the usual terminology in the field, we will often refer to topologically protected phonon transport, even though all of the systems studied experimentally so far are firmly in the classical regime, where many phonons are involved in the high-amplitude classical vibrational states. As long as one stays in the linear regime, the mathematics remains the same regardless of the amplitude and regardless of whether individual phonons can be excited and resolved.

Since the design of geometrical structures is key in this field, we decided to show a pictorial overview of some of the most relevant theoretical as well as experimental works from the field, in chronological order, in Figs.~\ref{fig:timeline_1} and \ref{fig:timeline_2}. 

\subsection{Different classification criteria}

It is useful to systematically classify these works according to several criteria.
We adopt the following categories, which are used to label works in the timeline of Figs.~\ref{fig:timeline_1} and \ref{fig:timeline_2}. Beyond distinguishing experimental implementations from theoretical proposals, we show the typical length scales of the system. Since most possible future applications of interest will most likely be on a micrometer- or nanometer-scale chip, this is a crucial key characteristic. While theoretical proposals may not commit themselves to a certain length scale, they sometimes clearly rely on a macroscopic setting or, alternatively, take care to present a design that would work on the nanoscale, which we try to indicate. 

Furthermore, we can separate the different works based on the type
of topological protection. We distinguish between systems that break time-reversal symmetry (and thus result in a Chern insulator) and setups that do not break the time-reversal symmetry.

Moreover, most works can also be classified according to one of the following three categories: discrete systems, acoustic-wave systems and elastic-wave systems. The first category corresponds to edge states that are formed by vibrational excitations of discrete lattice sites, i.e. coupled modes in a lattice. Exemplary systems are coupled pendula and theoretical works that take tight-binding models as their starting point. On the other hand, both acoustic and elastic wave systems start from a continuum description, although sometimes this might eventually be transformed into an effective coupled-mode theory. Acoustic wave systems feature topological modes that appear in the continuous fields of the pressure distribution within fluids. The air flow in an array of solid steel rods or water waves in a pipe of changing diameters are systems of that category. Finally, systems with topological edge modes appearing in the continuous deformation field of solid materials, e.g. suitably patterned slabs of silicon, are referred to as elastic-wave models. The distinction according to these three categories is of course not always clear-cut, but we found it helpful nonetheless.

\begin{figure}
\includegraphics[width=1\columnwidth]{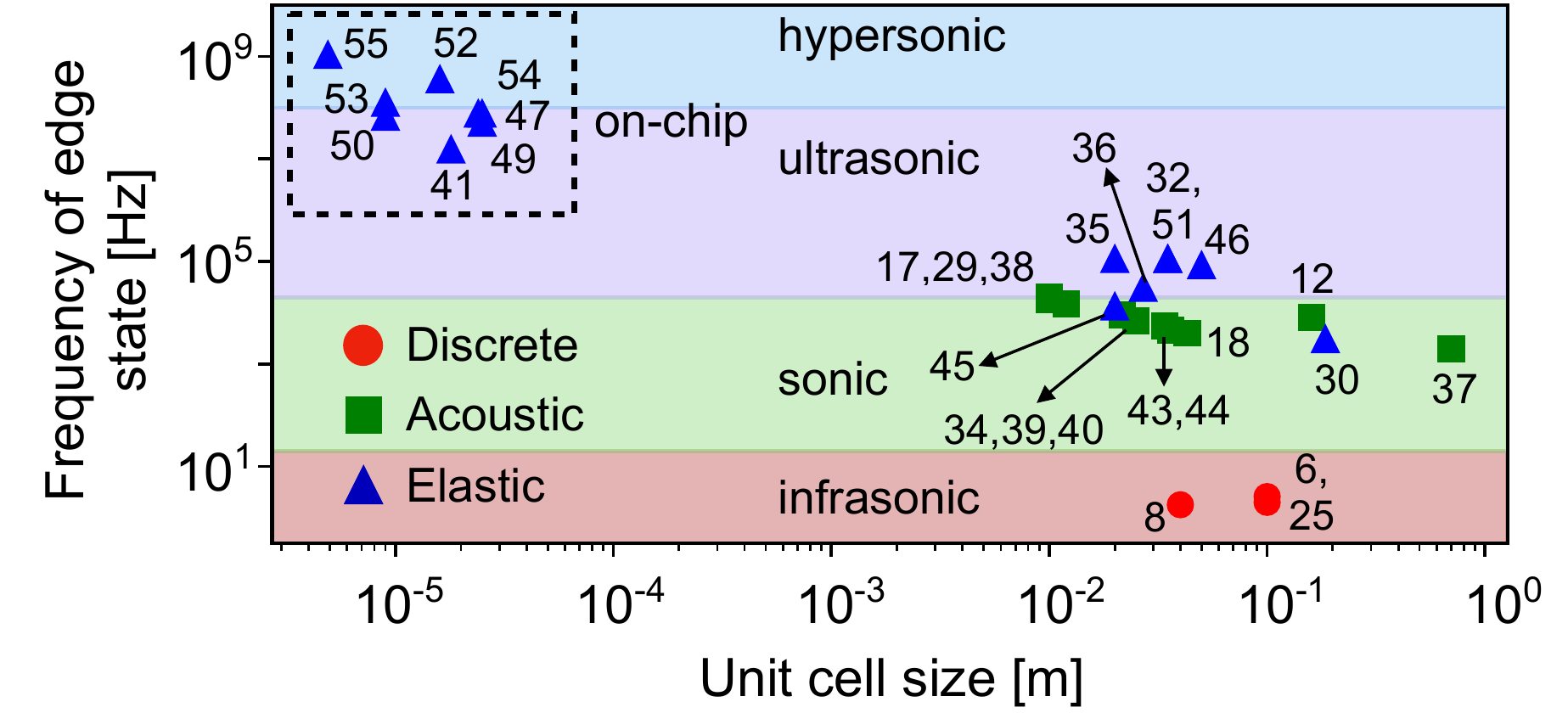}
\caption{
Classification of different experimental works based on the frequency of the phononic edge channel and the size of the unit cell. The different categories of mechanical systems, in increasing order of frequency (decreasing size), are: discrete macroscopic systems, acoustic systems, and elastic systems. The on-chip elastic systems are on the micro- or nanoscale and have frequencies in the MHz regime. The numerical labels refer to the indices of the respective works in the timeline of Figs.~\ref{fig:timeline_1} and \ref{fig:timeline_2}. 
}
\label{fig:refs_freq_size}
\end{figure}

Overall, the topological systems of interest in this review  cover a  broad range of carrier frequencies and unit cell length scales,  
from the infrasonic to the hypersonic regime and  from the micrometer to the meter scale (see the scatter plot Figure \ref{fig:refs_freq_size}).


\subsection{Evolution of research in this field: brief chronological overview}




In this section, we will merely give a first brief overview of the most salient early works in chronological order, to provide some general context for the following, more detailed sections which will aim to be comprehensive. 

The first work to point out that topologically protected transport of phonons is a possibility remarkably stems from the field of biophysics. In their very early pioneering paper, \cite{prodan_topological_2009} conjectured that under certain assumptions, vibrational edge states may form at the ends of long macromolecules that are part of the cell skeleton, the so-called microtubules, provided magnetic elements give rise to time-reversal symmetry breaking. This hypothesis still remains to be tested.

Moving to the domain of deliberately engineered topological phonon transport, the first idea for how this might be realized was introduced in 2014. The approach discussed there \cite{peano_topological_2015} employed the optomechanical coupling between light and mechanical modes to generate time-reversal symmetry breaking and a phononic Chern insulator. It also was the first work to describe a potential nanoscale implementation, based on phononic crystals. 

A few months afterwards, it was pointed out that topological transport can be designed also in macro-scale engineered systems of fluids, where time-reversal symmetry can be broken by circulating flows, which results in a Chern insulator for the acoustic modes propagating inside the fluid \cite{yang_topological_2015}. This was followed by further analyses of the circulating-flow scenario in different settings \cite{ni_topologically_2015,khanikaev_topologically_2015}.
\cite{fleury_floquet_2016} proposed to implement a Chern insulator in an array of coupled acoustic resonators by modulating their resonant frequencies   via piezoelectric actuators. 

Around the same time, many implementations of Chern insulators were proposed for discrete systems. Wang et al. \cite{wang_topological_2015} predicted chiral edge states in an array of gyroscopes coupled with springs. Alternatively, Coriolis force was suggested to break the time-reversal symmetry in a lattice of mass coupled with springs \cite{wang_coriolis_2015,kariyado_manipulation_2015}.

Realizing mechanical topological phases requires a high degree of control in the engineering of the mechanical structures. Not surprisingly, therefore, the earliest successful {\em experimental} implementations of topological mechanical systems  were established in the realm of macroscopic centimeter-scale systems, starting in 2015 from 2D arrays of coupled pendula \cite{susstrunk_observation_2015}, with time-reversal symmetry intact (modeling a topological insulator), and of coupled gyroscopes \cite{nash_topological_2015}, with time-reversal symmetry broken. These systems provided direct and convenient experimental access and served to attract considerable attention to the possibilities of engineered topological transport in mechanical systems, paving the way for the subsequent rapid developments.

Subsequent experimental works expanded into the domains of acoustic waves directed by engineered scatterers \cite{he_acoustic_2016,lu_observation_2017}, and shortly thereafter elastic vibrations in metal plates patterned with holes \cite{vila_observation_2017,yu_elastic_2018,miniaci_experimental_2018}. More recently,  Chern insulator phases have been realized for acoustic waves \cite{ding_experimental_2019} in the presence of circulating a similar setup as the original proposal of \cite{yang_topological_2015}, see above) and for elastic waves \cite{darabi_reconfigurable_2020} (in an array of piezoelectric membranes).

In parallel developments,  theoretical investigations continued to explore the potential for nano- and microscale topological vibrational materials, already mentioned above. The first theory work to present a design in this domain for time-reversal-preserved topological transport  \cite{mousavi_topologically_2015} relied on a free-standing patterned mechanical metamaterial with suitable symmetries. Other works in this direction used honeycomb (graphene-type) lattices, either producing pseudomagnetic fields via engineered distortions \cite{brendel_pseudomagnetic_2017} or adopting enlarged unit cells \cite{brendel_snowflake_2018}.

Building on these proposals as well as recent advances in microfabrication techniques, the first on-chip nanoscale   topological phonon transport was realized in 2018   \cite{cha_experimental_2018}, followed by recent very promising further experimental developments in this direction \cite{mathew_synthetic_2020,ren_topological_2022,ma_experimental_2021,ma_nanomechanical_2021,xi_observation_2021,zhang_gigahertz_2022}. Beyond free-standing devices,  topological waves can be engineered on-chip also in the form of surface acoustic waves, as recently demonstrated in two pioneering experiments  \cite{zhang_topological_2021,wang_extended_2022}. These efforts are part of an ongoing important quest to reduce the footprint and increase  phonon frequencies and bandwidths, as well as to explore more versatile actuation and detection schemes. 

\section{Approaches for engineering topological transport of phonons}
\label{sec:approaches-for-engineering}

\begin{figure}
\includegraphics[width=1\columnwidth]{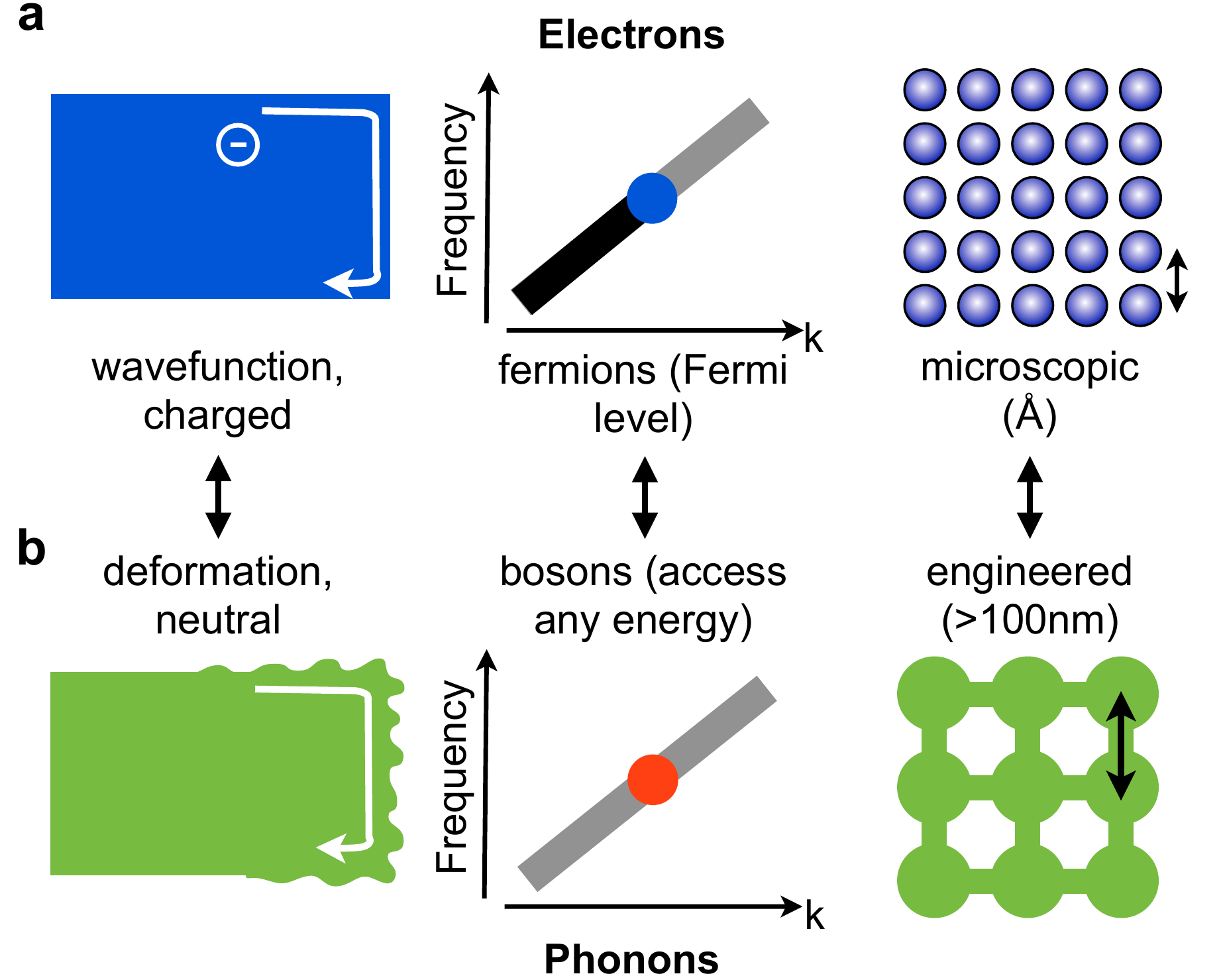}
\caption{
Main differences between electrons (\textbf{a}) and phonons (\textbf{b}), from the perspective of topological transport. Left column: Electronic transport can be affected by magnetic fields or spin-orbit coupling,  while transport of neutral phonons requires different approaches. Middle column: While electronic transport typically involves only states at the Fermi level, vibrations can be excited at any frequency in the band structure. Right column: Electronic topological transport may exploit the microscopic crystal structure, while phononic transport is engineered on larger length scales (nanoscopic up to macroscopic).
}
\label{fig:electrons_vs_phonons}
\end{figure}


\subsection{Electrons vs. electromagnetic waves vs. vibrations}

The underlying mathematics of topological protection is equivalent for any kind of waves, be they electronic matter waves, electromagnetic waves, or vibrational waves, as long as interactions can be neglected. 

Nevertheless, there are fundamental physical differences -- distinguishing fermionic and bosonic systems, but also with respect to other characteristics like the presence of a charge and the typical length- and frequency scales that need to be manipulated in engineered structures. These differences lead to distinct, platform-specific approaches for implementing and exploiting topologically protected waveguides, cf Fig.~\ref{fig:electrons_vs_phonons}. 

Electrons are charged and can therefore be manipulated with the help of electromagnetic fields. Breaking the time-reversal symmetry is straightforward by using a static magnetic field, which produces the Quantum Hall Effect, a Chern insulator, with its characteristic robust protection. The fermionic nature of electrons implies that transport properties are only affected by the electronic matter waves near the Fermi energy, up to which all the levels are occupied. 

By contrast, electromagnetic waves and vibrational waves are neutral, so breaking the time-reversal symmetry needs to be engineered with some effort, e.g. via some forms of time-dependent driving, as detailed in the next section. Alternatively, one can rely on geometrical engineering to produce time-reversal-preserved systems that are the counterpart of the electronic topological insulators (which, in the electronic domain, arise from spin-orbit coupling). Both types of waves, electromagnetic and vibrational, are also bosons. There is no Fermi energy, and waves can be injected and probed at any frequency. One can view this as a unique advantage over electrons, since it allows easy access to the full band structure and all edge states. At the same time, this also means that interactions (i.e. nonlinearities in the wave equation) can more easily scatter waves into other states, while these effects are suppressed for electrons due to the Pauli principle. 

There are differences regarding scales that are important for engineering. Electronic wave functions live in materials with Angstrom lattice periodicity, and 'engineering' takes place via chemistry, apart from the comparatively rare cases where one relies on the design of superlattices. By comparison, both photonic and phononic wave functions are being engineered on a wide range of length-scales, depending on the frequencies, ranging from macroscopic cm-scale setups down to 100 nm dimensions. The methods employed encompass tools as diverse as 3D printing and lithography. 

So far, we have highlighted the similarities between sound waves and electromagnetic waves, contrasting them with electronic matter waves. However, in a number of aspects important for topological transport sound and light differ considerably. Differences include the possibility to engineer fluid flows for time-reversal-symmetry breaking of sound waves, the fact that vibrational waves are necessarily confined to the material and cannot radiate away energy into free space nearly as easily as electromagnetic waves, and their very compact footprint (small wavelength) at a given frequency.

We now turn to the core of this review: a detailed discussion of the different engineering approaches employed in the literature to implement topological transport in mechanical systems.

\subsection{Broken time-reversal symmetry: Chern insulators}

\begin{figure}
\includegraphics[width=1\columnwidth]{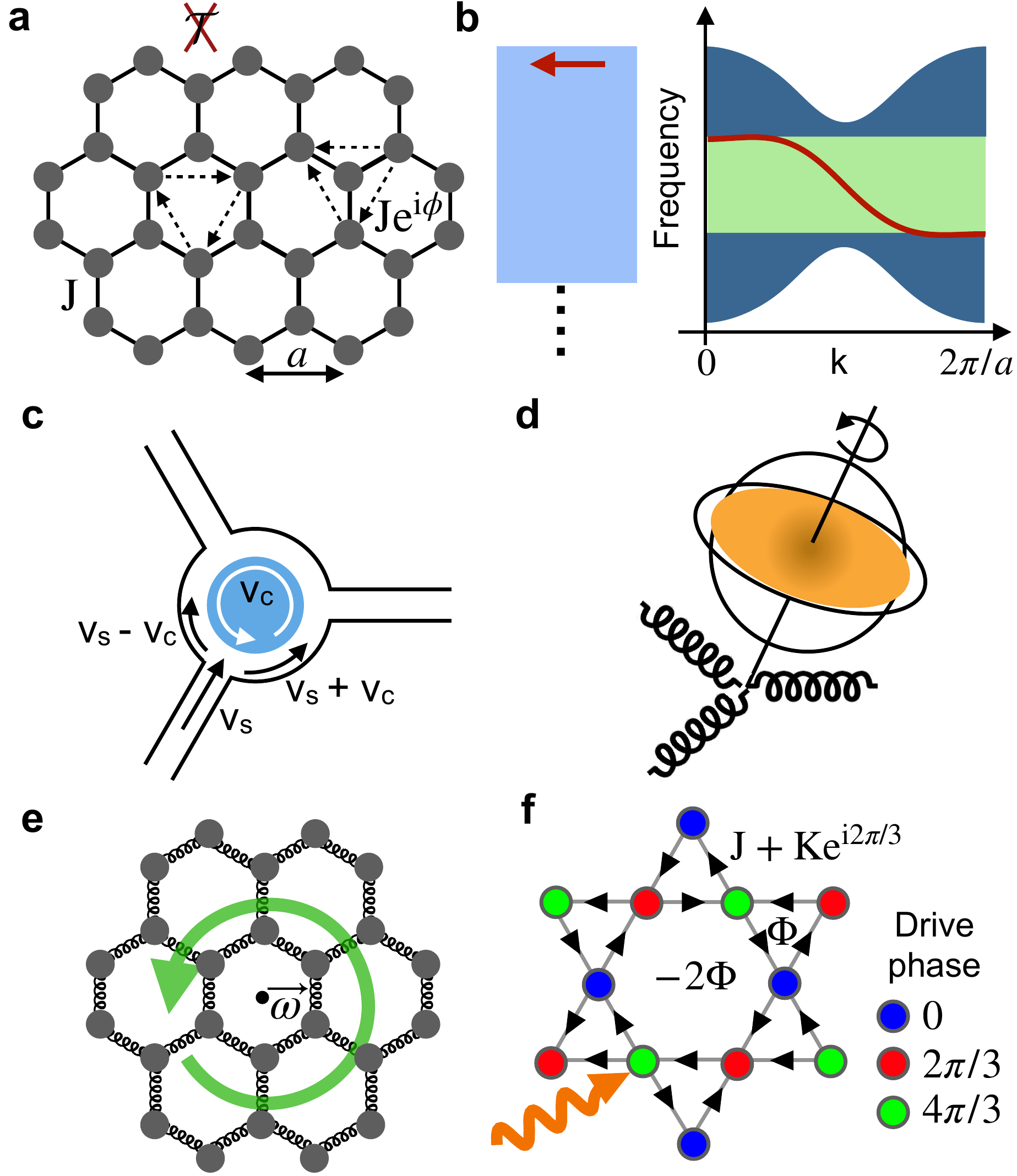}
\caption{
Chern insulators. 
\textbf{a}, Tight-binding description of the Haldane model, consisting of sites arranged in a honeycomb lattice with couplings between nearest (solid lines) and next-to-nearest (dashed lines) neighbors, the latter breaking time-reversal symmetry. 
\textbf{b}, Resulting band structure (right) of the semi-infinite translationally invariant strip configuration (shown on the left; infinitely extended in the horizontal direction). The topological edge mode is shown in red. 
\textbf{c-f}, Different mechanisms to break the time-reversal symmetry in topologically protected phononic systems: 
\textbf{c}, A rotating cylinder (blue) generates a fluid flow that results in a different sound speed along the two circulation directions of the acoustic waves, breaking time-reversal symmetry \cite{yang_topological_2015, ni_topologically_2015, khanikaev_topologically_2015, chen_tunable_2016}. 
\textbf{d}, Precession of a spinning gyroscope, part of a lattice connected with springs \cite{nash_topological_2015, wang_topological_2015, mitchell_amorphous_2018}. 
\textbf{e}, Rotation of a lattice to create a Coriolis force acting on spring-coupled masses \cite{wang_coriolis_2015,kariyado_manipulation_2015}. 
\textbf{f}, Time-dependent modulation of lattice sites in a chiral fashion, e.g. using optomechanical interactions and suitable illumination or other mechanisms \cite{peano_topological_2015, mathew_synthetic_2020, fleury_floquet_2016, darabi_reconfigurable_2020}.
}
\label{fig:chern_insulators}
\end{figure}











As discussed in Section~\ref{subsec:mechanism_chern_insulator}, Chern numbers can be non-zero only for systems that break time-reversal symmetry. For fermions, this can be achieved via an external magnetic field. However, this route is precluded for neutral vibrations. Alternative ideas that have been useful for electromagnetic topological transport, like employing magneto-optical effects, cannot be translated in any obvious way to vibrations. This poses a major obstacle to building a Chern insulator for phonons.

The solution arises from inspecting the example of a particle in a magnetic field and understanding it in terms of the Aharonov-Bohm effect. If a particle of charge $q$ travels around a loop containing a magnetic flux $\Phi$, then it will pick up a phase $e^{iq\Phi /\hbar}$. Importantly, the sign of this phase depends on the direction in which the loop is traversed. As a consequence, any physical mechanism that can give rise to phases of this kind will eventually produce the physics of a particle in a magnetic field, including robust edge channels. 

Taking advantage of this way of thinking, it is possible to build Chern insulators for phonons by engineering complex hopping amplitudes between the lattice sites, so that the total phase picked up by a phonon around a closed loop is non-zero.  We remark in passing that these complex amplitudes are, in most cases, matrix elements of  effective Hamiltonians defined in a rotating frame after a rotating wave approximation.

Incidentally, there is no need to mimic a constant magnetic field in order to obtain edge channels. This was demonstrated by the Haldane model, which is the prototypical example of a tight-binding model with broken time-reversal symmetry but zero average magnetic field (See Fig.~\ref{fig:chern_insulators}a). The model includes contribution from both the nearest-neighbor and next-to-nearest-neighbor hoppings on a honeycomb lattice. The net magnetic flux is zero. Nevertheless, the model features a topological edge state in the bulk band gap, cf Fig.~\ref{fig:chern_insulators}b. This edge state is localized at the physical edge of the strip as illustrated in Fig.~\ref{fig:chern_insulators}b, in the same way that would be generated by a simple non-zero magnetic field in the conventional Quantum Hall Effect.

As mentioned before, the first theoretical analysis of phononic topologically protected edge-states in any system is due to Prodan and Prodan in 2009 \cite{prodan_topological_2009}, and it is an example of a Chern insulator. They presented the hypothesis that topological effects could play a role in the vibrational modes of microtubules, i.e. the self-assembled hollow protein tubes that are part of the cytoskeleton of living cells. A microtubule can be seen as a tube that is produced by wrapping around the underlying two-dimensional membrane-like protein assembly. Prodan and Prodan pointed out that the phonon spectrum of this two-dimensional sheet contains Dirac points. Under the additional (speculative) assumption that time-reversal symmetry may be broken by some magnetic properties of the microtubules or the surrounding medium, they discovered a splitting of the Dirac cones and the emergence of phonon bands with non-trivial Chern-numbers. The corresponding topological vibrational modes would be localized at the rims forming the ends of the microtubules.

Moving on from this example of potentially naturally occuring topological transport of sound waves, we now turn to ideas for designing artificial structures. 

The first such idea \cite{peano_topological_2015} proposed to exploit the coupling between light and sound to generate the required time-reversal symmetry breaking. The starting point would be a suitably engineered phononic crystal with a micrometer-scale unit size, in this case consisting of a Kagome pattern of coupled mechanical sub-units. Using the principles of cavity optomechanics, the interaction between light and sound could be boosted by embedding photonic-crystal defect cavities, creating an optomechanical array. When the whole structure is illuminated by a superposition of three laser beams, optical vortices are formed which impose their nonvanishing orbital angular momentum on the sound waves via the optomechanical interaction. A typical choice of the driving laser phase, shown in Fig.~\ref{fig:chern_insulators}f, is to have a phase difference of $2\pi/3$ between neighboring sublattices. Eventually, this creates a Chern insulator for phonons in the underlying 2D nanomechanical structure, with topologically protected edge channels around the boundaries of the sample. The possibilities offered by this approach include in-situ time-dependent switching of topological domains via the light. Depending on parameters, one may choose to have only a small admixture of photonic excitations or deliberately enter a regime of strong hybridization and topological transport of combined photon-phonon excitations. This proposal pointed the way towards nanomechanical topological phonon transport, and experiments in recent years are moving closer to eventually realizing such ideas, with \cite{ren_topological_2022} already showing 800 unit cells in an optomechanical array.

Other proposals based on optomechanical arrays were recently  put forward \cite{sanavio_nonreciprocal_2020,mathew_synthetic_2020}.  \cite{sanavio_nonreciprocal_2020} proposed an implementation based on optomechanical microtoroids. In their approach, the array is driven only from the perimeter and neighboring microtoroids  are coupled  via the evanescent radiation without any direct mechanical coupling.  If the optical modes are engineered to have a spin-orbit coupling \cite{hafezi_imaging_2013}, the optomechanical interaction induces a non-reciprocal mechanical coupling between the breathing modes. The approach of \cite{mathew_synthetic_2020}, with time-dependent modulation of the laser intensity generating coupling between mechanical modes of different frequencies, requires less precise control of the optical modes, and they experimentally demonstrated a building block of the envisaged lattice. Similar schemes, involving the parametric modulation of an array of mechanical oscillators with a periodic pattern of phase delays, have also been proposed  for arrays of pendula \cite{salerno_floquet_2016}, and of acoustic cavities \cite{fleury_floquet_2016}. Both works also investigate the full time-dependent dynamics beyond the rotating-wave approximation, using the Floquet formalism. More recently, the driving protocol and geometry of \cite{fleury_floquet_2016} (involving  oscillator trimers arranged on a honeycomb lattice) has  been experimentally  demonstrated for an array of piezoelectrically-modulated elastic membranes  \cite{darabi_reconfigurable_2020}, representing  the first experimental realization of a Chern insulator for elastic waves.


A number of schemes employing time-reversal symmetry breaking are based on acoustic wave propagation in fluids. The unit-cell lengths for these systems are in the centimeter scale. The initial theoretical proposal \cite{yang_topological_2015}, followed by \cite{ni_topologically_2015, khanikaev_topologically_2015}, and experimentally demonstrated by \cite{ding_experimental_2019} involved breaking the time-reversal symmetry by running a circulating air flow around cylinders. Due to this external air flow, the acoustic wave speed in the direction along the flow and opposite to it is different. In this way, the wave picks up a different propagation phase in the clockwise and the counter-clockwise directions (See Fig.~\ref{fig:chern_insulators}c), effectively leading to an acoustic Aharonov-Bohm effect. Chen et al. \cite{chen_tunable_2016} used this scheme to propose a tunable topological phononic crystal. In a conceptually similar setting, \cite{souslov_topological_2017} proposed and theoretically analyzed the topological density waves appearing in a liquid composed of self-propelled particles.

Chronologically, the first  experimental implementation of a Chern insulator for mechanical vibrations was presented in early 2015, for a macroscopic system: an array of coupled spinning motor-driven gyroscopes on a honeycomb lattice
\cite{nash_topological_2015}.    The precession of the gyroscopes around the suspension point breaks the time-reversal symmetry. To provide the coupling, a small magnet is placed in each gyroscope, leading to a magnetic repulsion between neighboring gyroscopes which can be modelled as a spring (See Fig.~\ref{fig:chern_insulators}d). This array of about 50 gyroscopes, with a cm-scale unit size, was then imaged directly in real time, revealing the chiral propagation of a wave packet along the boundary of the array. 
Also in 2015, a similar gyroscope-based platform had been independently proposed by \cite{wang_topological_2015}. Subsequently, following up on their own experimental work presented above, Mitchell et al. \cite{mitchell_amorphous_2018} studied (both theoretically and experimentally) Chern insulators in an amorphous lattice of gyroscopes, analyzing the interplay with strong disorder. 

An alternative  mechanism  to  implement  a Chern insulator phase for discrete coupled mechanical systems is to take advantage of the Coriolis force by rotating a lattice of masses coupled with springs, as illustrated in Fig. 9e and proposed in \cite{wang_coriolis_2015,kariyado_manipulation_2015}.



In summary, Chern insulators for vibrations have been proposed, and in a few cases already experimentally implemented, based on time-reversal symmetry breaking via magnetic interactions, optomechanical interactions, time-dependent piezoelectric modulation, circulating air flows, gyroscopic motion, and Coriolis forces. 

\subsection{Preserved time-reversal symmetry}
\label{sec:preserved-time-reversal-systems}


While time-reversal symmetry breaking gives rise to robust edge states it can sometimes be easier if no such measures are required. This is especially true for nanoscale systems, where purely passive geometric designs are easiest to implement. 

In Section \ref{sec:Time-reversal-mechanism}, we have identified two complementary approaches to implement helical edge states in bosonic systems: i) implement an  effective  Spin-Hall Hamiltonian (block-diagonal across the BZ and supporting an engineered time-reversal symmetry ${\cal T_{\rm en}}$), cf \ref{sec:Time-reversal-mechanism_kramers}.   ii)  via  effective Dirac Hamiltonians describing selected bands in a limited quasimomenum region, cf \ref{sec:Time-reversal-mechanism_dirac_engineering}. Here, we discuss their corresponding implementations in phononic systems starting from the first approach.

\subsubsection{Implementing Spin-Hall Hamiltonians}
\label{sec:kramers_degeneracy_implementation}
\begin{figure}
\includegraphics[width=1\columnwidth]{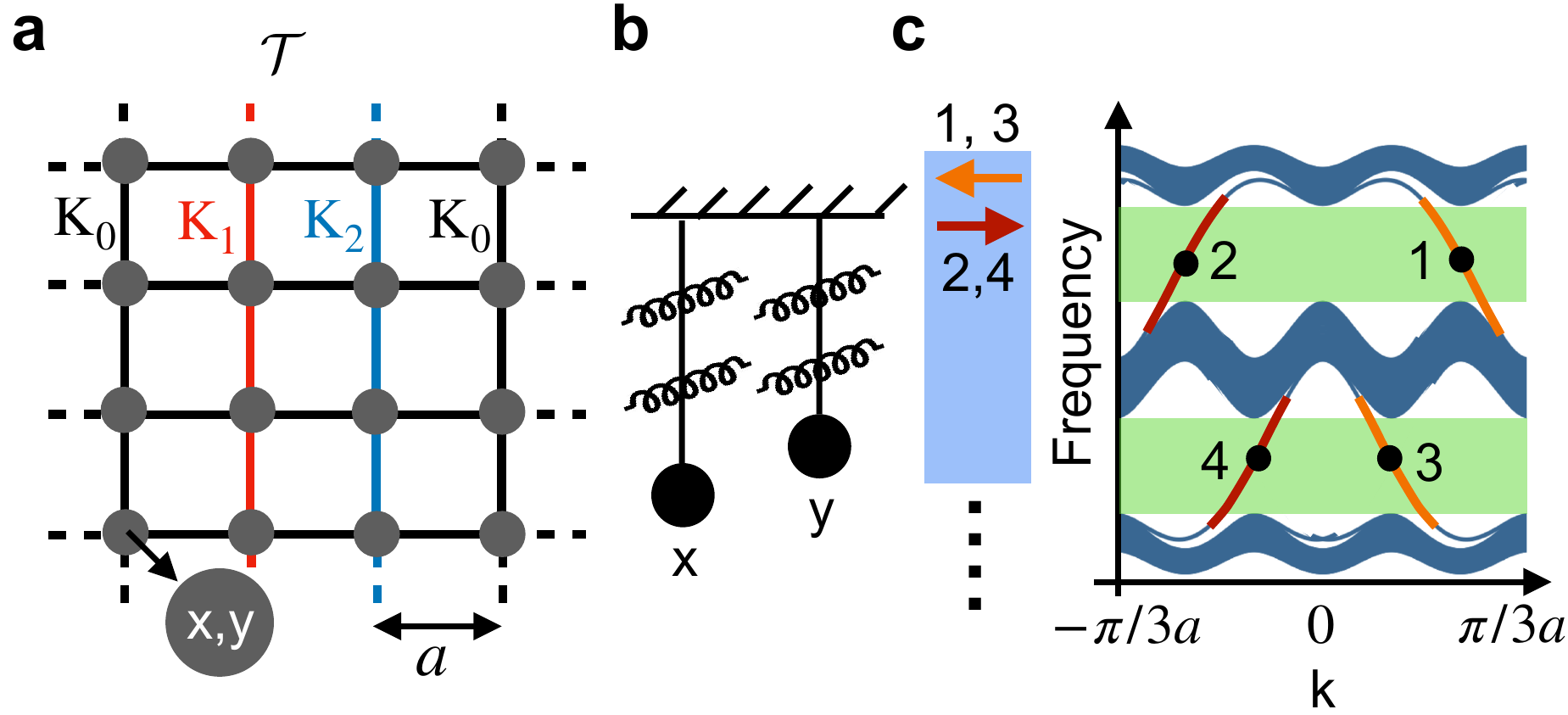}
\caption{
Phononic Spin-Hall Hamiltonians. 
\textbf{a}, Schematic illustration of the two-copy version of the Hofstadter model. Each lattice site is composed of two modes $x,y$. The colored lines indicate the couplings (colors encode the different type of couplings) between the two modes of the neighboring sites (see main text). 
\textbf{b}, The two modes $x,y$ are realised in \cite{susstrunk_observation_2015} as two one-dimensional pendula, whereas the intricate couplings in the model are implemented by connecting the pendula with springs. 
\textbf{c}, Band structure (right) of the semi-infinite strip configuration (left) featuring topological edge states (red and orange).
}
\label{fig:engineered_kramers_degeneracy}
\end{figure}


How can the classical dynamics of coupled mechanical harmonic oscillators, governed by Newton's laws $m \Ddot{r}_i=-\mathcal{D}_{ij}r_j$ ($\hat{\mathcal{D}}$ is the dynamical matrix), imitate the dynamics of the Spin-Hall Hamiltonian Eq.~(\ref{eq:Ham_Top_Ins})? 

This question was addressed by Suesstrunk and Huber in their work \cite{susstrunk_observation_2015}  which constitutes the first experimental realization of a two-dimensional mechanical phononic topological insulator (as opposed to a Chern insulator, see above). They employed an array of cm-scale macroscopic pendula, each of them with only one motional degree of freedom, but coupled pairwise in a suitably designed way with several mechanical springs. In this way, they created a whole array imitating the required dynamical matrix $\mathcal{D}$ (see Fig.~\ref{fig:engineered_kramers_degeneracy}a).

More specifically, \cite{susstrunk_observation_2015}  considered two copies of the Hofstadter model \cite{hofstadter_energy_1976} with flux $\Phi=2\pi/3$ per plaquette i.e. $\hat{H}_{\uparrow} = \hat{H}_{\uparrow, \Phi}$ and $\hat{H}_{\downarrow} = \hat{H}_{\downarrow,- \Phi}$ (Fig.~\ref{fig:engineered_kramers_degeneracy}a). The Hamiltonian $\hat{H}$ is complex-valued and cannot be easily realized with coupled harmonic oscillators. Therefore, a local unitary transformation $\hat{U}=\hat{u}\otimes 1\!\!1_{\rm lattice}$ is performed $\hat{U}^{\dagger}\hat{H}\hat{U}=\hat{\mathcal{D}}$, such that the dynamical matrix $\hat{\mathcal{D}}$ becomes real-valued and symmetric. More explicitly, at each lattice site, the pseudo-spin states ($\psi_{\uparrow}, \psi_{\downarrow}$) of the double-Hofstadter model are related to the vibrations $x$ and $y$ of two pendula (Fig.~\ref{fig:engineered_kramers_degeneracy}b) via the unitary operation
\begin{equation}
\begin{pmatrix}
\psi_{\uparrow} \\ \psi_{\downarrow}
\end{pmatrix}
=\hat{u}\begin{pmatrix}
x \\ y
\end{pmatrix},\quad 
\hat{u}=\frac{1}{\sqrt{2}}
\begin{pmatrix}
1 & -i \\
1 & i
\end{pmatrix}
.
\end{equation}
Thus, the two pseudo-spin eigenstates of $\hat{H}$ can be seen as certain relative motion patterns of the two pendula at each lattice site, which would correspond to left- and right-hand circular polarisation in electromagnetism. The effective spring constants between pendula pairs at neighboring sites are described by a matrix, \begin{eqnarray} K_n=J\hat{u}^{\dagger}\begin{pmatrix}
e^{i\theta_n} & 0 \\
0 & e^{-i\theta_n}
\end{pmatrix}\hat{u}=J\begin{pmatrix}
\cos\theta_n & \sin\theta_n \\
-\sin\theta_n & \cos\theta_n
\end{pmatrix},
\end{eqnarray}
with $\theta_n=2\pi n/3$ and the different versions of these couplings (n=0,1,2) are arranged in a suitable fashion (Fig.\,\ref{fig:engineered_kramers_degeneracy}). Effective negative spring constants are engineered by coupling two pendula via a lever arm.

In a strip configuration, this particular system gives rise to two topological edge-states in each of the two bulk band-gaps, as illustrated in Fig.~\ref{fig:engineered_kramers_degeneracy}c. The experiment of \cite{susstrunk_observation_2015}  demonstrated the existence of the edge states and tested their robustness against various boundary deformations. As a sequel to this work, the authors demonstrated the idea of switchable topological phonon channels in the same kind of platform \cite{susstrunk_switchable_2017}. 

The edge states demonstrated by Suesstrunk et al. are protected against any perturbation that preserves the engineered time-reversal symmetry
$\cal{T}_{\rm en}=\cal{U}\cal{T}$. Here, $\cal{T}$ is the physical time-reversal symmetry and  $\cal{U}$ is the local unitary symmetry $(x,y)\to (y,-x)$. We note that  arbitrary disorder in the spring constants would  preserve $\cal{T}$  but not $\cal{T}_{\rm en}$ and can, thus,  introduce backscattering. In other words, generic time-symmetric disorder coupling the two pseudo-spins plays the same role  as magnetic disorder for an electronic topological insulator. This is generally true for any bosonic implementation of topological insulators.   Although this seems like a huge disadvantage, experiments and numerical simulations show that the resulting edge states may still be surprisingly robust, since the typical geometrical disorder arises from fabrication errors that are often distributed smoothly.   

Apart from arrays of pendula,  the implementation of Spin-Hall Hamiltonians has also been proposed for other phononic platforms, including bilayer lattices of masses coupled via springs  \cite{pal_helical_2016}  and elastic metamaterials \cite{matlack_designing_2018} comprising perforated plates coupled by beams. In addition, \cite{peng_experimental_2016} have proposed and experimentally demonstrated a topological insulator based on coupled-resonator acoustic waveguides. Their platform is the acoustic analogue of the well-known topological photonics platform based on optical resonators coupled via coupler waveguides, see e.g. \cite{hafezi_imaging_2013}. The relevant pseudo-spin is the chirality of the sound propagating within a lattice resonator. It is conserved because the propagation direction does not change when the sound is transmitted between a lattice resonator and a coupling resonator.  \cite{peng_experimental_2016} observed the topological phase transition predicted by \cite{liang_optical_2013} for symmetric couplers in the strong coupling regime. A similar platform was proposed by \cite{he_topological_2016} for the spin-polarized transport of underwater sound.

Beyond Spin-Hall Hamiltonians, the connection between mass spring models  
and single-particle quantum mechanical Hamiltonians  has been analyzed 
by  \cite{susstrunk_classification_2016},  who have developed a classification of the topological  phases of such models based on  the symmetry class and spatial dimension (in $1$, $2$, and $3$ dimensions).

We now describe several schemes in which robust helical edge states are engineered via an effective Dirac Hamiltonian, see Section \ref{sec:Time-reversal-mechanism_dirac_engineering}. In these methods, the bulk normal modes need not to be engineered across the BZ but only near some special high-symmetry points, based on robust symmetry-based designing principles.  This family of approaches are particularly suitable for elastic vibrations and acoustic waves whose dynamics can not be easily reduced to an effective tight-binding model.

\subsubsection{Valley Hall}
\label{sec:valley-Hall}
\begin{figure}
\includegraphics[width=1\columnwidth]{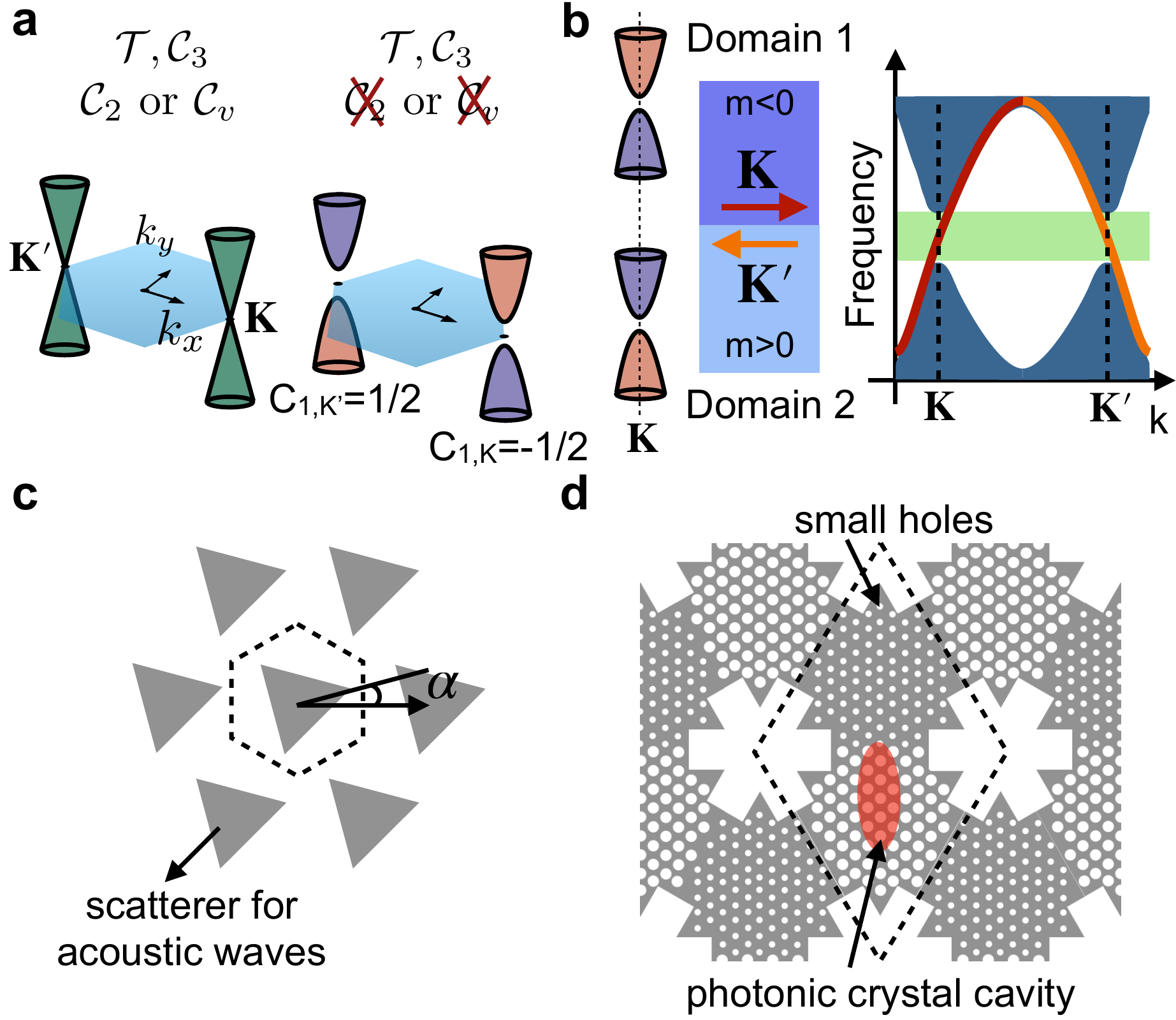}
\caption{
Valley Hall effect. 
\textbf{a}, Effect of different symmetries on the Dirac cones. The degeneracy splits on breaking the $\mathcal{C}_2$ or $\mathcal{C}_v$ symmetry of the original $\mathcal{C}_6$ or $\mathcal{C}_{3v}$ crystal. The color of the resulting hyperbolic bands encodes the valley Chern number.
\textbf{b}, (left) Translationally invariant strip including two domains of opposite mass, with a helical edge channel arising at the domain interface (propagation direction depending on valley index). Note the different valley Chern number $C_{\rm K}$ of the bands for each domain. (right) 
\textbf{c}, Array of triangular scatters for acoustic waves in \cite{lu_observation_2017}. The scatters are rotated, to break the mirror symmetry $C_v$, which opens a gap in the Dirac cones. 
\textbf{d}, Implementation of valley Hall effect in an optomechanical crystal \cite{ren_topological_2022}. The silicon slab consists of patterned holes of two different sizes, which are used to manipulate the Dirac cones. The vibrations are measured optically via the photonic-crystal cavity mode.}
\label{fig:valley_hall}
\end{figure}

How can we most easily implement the Dirac Hamiltonian Eq.~(\ref{eq:Dirac_Hamiltonian}) in an engineered structure like a patterned slab or an array of scatterers? 

We start by observing that
Dirac cones are ubiquitous for band structures in crystals in the wallpaper groups $p6$ (point group ${\cal C}_6$ formed by the 6-fold rotations)  or $p31m$  (point group $\mathcal{C}_{3\nu}$,  three mirror planes, each including a primitive lattice vector), when  time-reversal symmetry $\mathcal{T}$ is respected. A prominent example is of course graphene. In this scenario,  each Dirac cone stems from an essential degeneracy of two Bloch waves with opposite quasi-angular momentum at the high-symmetry points $\mathbf{K}$ or $\mathbf{K'}$  of the Brillouin zone. The Dirac cones become gapped when the point group is reduced to  $\mathcal{C}_{3}$,  
 Fig.~\ref{fig:valley_hall}(a). We, thus, arrive at Eq.~(\ref{eq:Dirac_TRS}) where $\tau_z=1$ ($\tau_z=-1$)  and $\mathbf{q}=\mathbf{k}-\mathbf{K}$ ($\mathbf{q}=\mathbf{k}-\mathbf{K}'$) in the valley around the high symmetry point $\mathbf{K}$  ($\mathbf{K}'$). At a domain wall for the mass parameter, this leads to the appearance of a pair of counter-propagating edge channels, with  each channel  localized about a different valley, cf Fig.~\ref{fig:valley_hall}(b). This is analogous to the Quantum Spin Hall effect, with the two valleys playing the role of  effective pseudo-spin directions, and it is, for this reason, known as  the valley Hall effect.

The   valley Hall effect has been originally predicted for bilayer graphene \cite{martin_topological_2008}, where it has been experimentally demonstrated by \cite{ju_topological_2015}. However, it is especially straightforward to implement in engineered crystal structures. In that setting, it is very easy to design a mass domain wall because from a geometry with weakly broken symmetry ($\mathcal{C}_{2}$  or $\mathcal{C}_{\nu}$) and  mass parameter $m= m_{\rm bk}$, one can  obtain a geometry with  mass parameter $m= -m_{\rm bk}$ by applying the broken symmetry transformation to the unit cell. 

The first experimental implementation of acoustic topological edge states utilizing the valley Hall effect relied on acoustic waves propagating in a 2D lattice of triangular rod-like scatterers \cite{lu_observation_2017} (See Fig.~\ref{fig:valley_hall}c). As the triangles are not aligned with the hexagonal lattice, a mismatch of mirror symmetries between the lattice and scatterers arises, breaking the $\mathcal{C}_{3v}$ symmetry, producing the valley Hall effect and eventually edge channels.

A number of publications based on valley-polarized acoustic waves followed this initial work \cite{wang_valley_2018,xia_observation_2018,lu_valley_2018,zhang_directional_2018,zhang_topological_2018-1,wen_acoustic_2018,yang_acoustic_2018,han_experimental_2019,shen_valley-projected_2019,tian_dispersion_2020,wang_valley-locked_2020,fan_tracking_2022}. Lu et al. \cite{lu_valley_2018} proposed a bilayer design of scatterers for air waves. Xia et al. \cite{xia_observation_2018} demonstrated valley-polarized acoustic edge channels between a square lattice of scatterers, thereby going beyond the conventional triangular lattice employed in valley Hall effect. The concept of topological acoustic valley transport has been leveraged to build acoustic antennas \cite{zhang_directional_2018} and acoustic delay lines \cite{zhang_topological_2018-1}, cf Sec.~\ref{sec:applications}. The acoustic antennas are used to out-couple air waves from the lattice of rod-shaped scatterers to the surroundings in a desired direction, and to perform the reverse process of receiving sound only from a source in the desired direction. The delay line is used to control the time it takes for the energy to travel between two different points on a lattice. Zhang et al. \cite{zhang_topological_2018-1} implemented acoustic delay lines using three-legged rod scatterers, where these rods can be rotated with computer-controlled motors to build reconfigurable topological edge-states. An alternative recent implementation of tunable lattices, this time based on coupled acoustic cavities, is presented in \cite{tian_dispersion_2020}.  With the goal of  providing a more flexible interface to other acoustic devices, \cite{wang_valley-locked_2020} proposed and experimentally demonstrated a setup in which  two topologically distinct domains are separated by a massless domain.


Approaches based on the valley Hall effect for elastic vibrations in macroscopic solids soon followed the first ideas for  airflows discussed above \cite{pal_edge_2017,vila_observation_2017,huo_simultaneous_2017}. Pal et al. \cite{pal_edge_2017} proposed an array of resonators arranged in a triangular lattice on an elastic plate. The same group implemented it experimentally by building a hexagonal elastic lattice out of an acrylic panel \cite{vila_observation_2017}. 

The first experimental demonstration of nanoscale topological phonon transport based on the valley Hall effect was presented recently. It utilizes an optomechanical array \cite{ren_topological_2022} which is designed starting from a patterned Si slab phononic crystal with snowflake-shaped holes \cite{brendel_pseudomagnetic_2017}. The Dirac cones are gapped by breaking the $\mathcal{C}_v$ symmetry via engineering different properties of the A/B sublattice units. Furthermore, a photonic crystal is embedded inside the larger phononic unit cell (see Fig.~\ref{fig:valley_hall}d). The purpose of producing an optical photonic-crystal defect cavity at each site of this array is to enhance the  sensitivity of light-based detection of mechanical motion (by a factor of the finesse of the cavity). This allowed, for the first time in any system, to measure thermal topological vibrations, of amplitudes around 10 femtometer, and to do so in a spatially resolved way, scanning along the domain wall. Valley-locked edge states have been demonstrated in other recent nanoscale experiments, for surface acoustic waves \cite{zhang_topological_2021} and in an array of suspended Silicon Nitride membranes \cite{ma_experimental_2021,xi_observation_2021}  (similar to the setup used in \cite{cha_experimental_2018} but with a different symmetry of the etching pattern, cf Figure \ref{fig:zone_folding}d below).  \cite{zhang_gigahertz_2022}  reached for the first time the hypersonic regime using an Aluminiun-Nitride snowflake phononic crystal actuated piezoelectrically. 

Next, we highlight two important theoretical contributions to the general understanding of the Valley Hall physics.  \cite{fan_tracking_2022} drew an interesting connection between valley Hall edge states and  the topological edge states of 3D Weyl semimetals. They focused on setups supporting a  geometrical angle parameter $\alpha$ which controls the value of the mass parameter (cf Fig.~\ref{fig:valley_hall}c for an example). Viewing this angle as the quasi-momentum for a third synthetic dimension allows to promote the Dirac cones to 3D Weyl cones. 
In the presence of a straight boundary or domain wall, the valley edge states for varying values of the angle parameter $\alpha$ and longitudinal quasi-momentum $k$ are, thus, mapped onto the topological edge states of a Weyl semimetal.
In addition, the synthetic quasi-momenta ($\alpha,k$) for the subset of valley edge states with eigenfrequency exactly in the middle of the bulk band gap  form  arcs in the $\alpha k$-plane. Each arc connects the projections of a pair of artificial Weyl points onto this plane, analogous to Fermi arcs in Weyl semimetals \cite{Wan_2011_Topological_semimetal}. \cite{shah_tunneling_2021} analysed the  residual  backscattering induced by large quasi-momentum transfer providing an interpretation of these transitions  as tunneling processes in quasimomentum space, more on this in Section \ref{sec:Challenges}.





\subsubsection{Zone folding}

\label{sec:zone-folding-scheme}


\begin{figure}
\includegraphics[width=1\columnwidth]{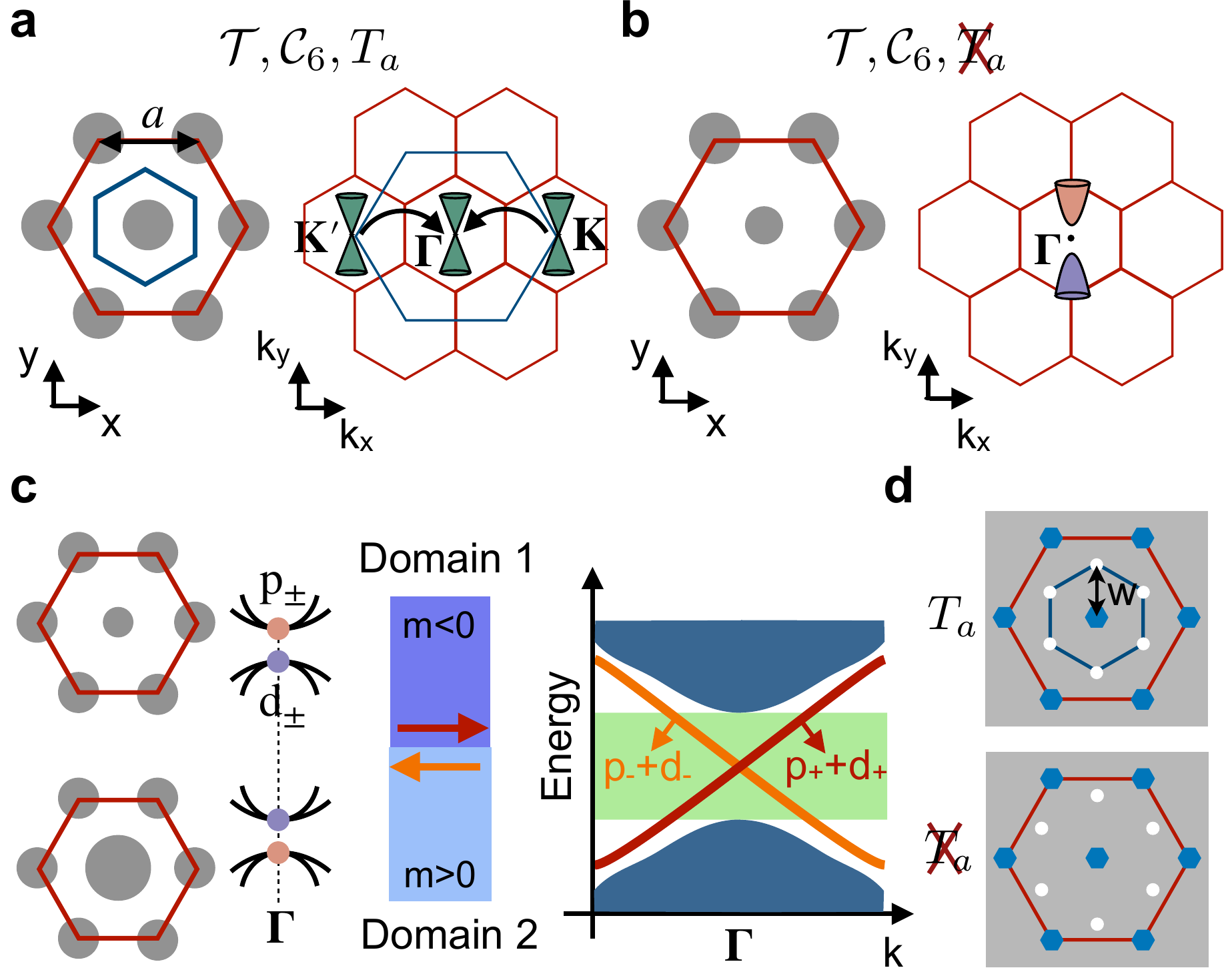}
\caption{Zone folding
\textbf{a}, (left) Unit cell of an array of rod-shaped scatters for acoustic waves used in \cite{deng_observation_2017}. (right) Description of the Dirac cone in k-space for both the primitive (blue) and a larger (red) real-space unit cell, indicating the folding back to the $\boldsymbol{\Gamma}$ point. 
\textbf{b}, Breaking the original discrete translational symmetry $\mathcal{T}_a$ by making the central rod smaller induces a gap in the Dirac cone 
\textbf{c}, Domain wall configuration, consisting of two domains with inverted band arrangement. The radius of the central rod relative to that of its surroundings is varied in the two domains. The helical edge states appear in the bulk band gap of the strip band structure. 
\textbf{d}, Schematic implementation of the zone folding scheme for elastic waves on a patterned plate, as used in the experiment of \cite{cha_experimental_2018}. The distance (w) of the etched holes (white dots) from the center unetched region (blue) is varied to break the translational symmetry. 
}
\label{fig:zone_folding}
\end{figure}

The zone-folding scheme is an alternative symmetry-based approach to engineer an effective Dirac Hamiltonian with a tuneable mass.  As for the valley Hall effect, the Hamiltonian Eq.~(\ref{eq:Dirac_TRS}) is obtained by modifying an initial design with ${\cal C}_6$ point-group symmetry that supports a pair of Dirac cones. In this scheme, the mass term is obtained by breaking the original translational symmetry without breaking the ${\cal C}_2$ symmetry. This generates a smaller Brillouin zone, into which the original bands are folded back, hence the name. This scheme has been originally proposed by \cite{wu_scheme_2015} for photonic crystals.

The effective Hamiltonian for the scenario described above is intimately connected to the valley Hall Hamiltonian. In both scenarios, one starts  from the same assumptions leading to the  effective Dirac equation, Eq.~(\ref{eq:Dirac_TRS}) with $m(\mathbf{x})=0$, before introducing the symmetry-breaking perturbation. We note  that $\mathbf{q}$ is the quasimomentum of the enlarged unit cell. Thus, the band structure can be viewed as supporting a double Dirac cone (i.e. a doubly degenerate Dirac cone) centered at the $\boldsymbol{\Gamma}$-point of the reduced Brillouin zone, cf Fig.~\ref{fig:zone_folding}a. We further note that the Hamiltonian is invariant under any (valley-admixing) rotation  of the pseudospin degree of freedom. In particular, one can combine the Bloch waves in the two valleys to obtain four ${\cal C}_6$-symmetric Bloch waves at the $\boldsymbol{\Gamma}$-point. It can be shown that, under rotations, they behave as $p_\pm$ and $d_\pm$ atomic orbitals (with quasi-angular momentum $\pm 1$ and $\pm 2$, respectively). A perturbation that breaks the (original) translational symmetry $T_a$ without breaking the ${\cal C}_6$-symmetry, splits the degeneracy between the $p$- and the $d$-Bloch waves, leading to a two-fold degenerate  gapped cone band structure, cf Fig.~\ref{fig:zone_folding}b. Just as for the valley Hall effect, the underlying effective Hamiltonian has the form Eq.~(\ref{eq:Dirac_TRS}). However,  now, the (rotated) pseudo-spin $\tau_z$ has a different interpretation: $\tau_z=1$ ($\tau_z=-1$) if the carrier  waves are $p_+$ or $d_+$ ($p_-$ or $d_-$) Boch waves. This reflects that up to leading order in $\mathbf{k}$ for the $\mathbf{k}\cdot\mathbf{p}$ perturbation theory only states that differ by one unit of quasi-angular momentum are coupled. We note that  $\tau_z\sigma_z=1$ $(-1)$ for the $d$ ($p$) Bloch waves. In other words, the sign of the mass parameter $m$ is set by the order of the $p$ and $d$-bands at the $\boldsymbol{\Gamma}$ points, Fig.~\ref{fig:zone_folding}c. Thus, a band inversion at the interface of two adjacent domains (here labeled type 1 and 2) gives rise to helical edge states, cf Fig.~\ref{fig:zone_folding}c.

The zone-folding idea was first transcribed from the photonic world to the phononic domain in \cite{zhang_topological_2017}, where the authors proposed and analyzed theoretically the propagation of airborne sound in the presence of an array of cylindrical scatterers, analogous to the arrangement of the photonic-crystal holes in \cite{wu_scheme_2015}. Shortly thereafter,  \cite{yves_topological_2017} proposed to use the same arrangement for an array sub-wavelength Helmholtz resonators \cite{yves_topological_2017}, in the form of soda cans,  to engineer topological polaritonic waves.
Chaunsali et al. \cite{chaunsali_subwavelength_2018} employed the zone-folding idea to propose topological transport of flexural waves on a thin plate with resonators mounted on its top.

The first experimental implementation of the zone-folding mechanism for mechanical topological transport  \cite{deng_observation_2017} employed an array of rod-like scatterers on a hexagonal lattice (Fig.~\ref{fig:zone_folding}a), shaping the acoustic vibrations of air between those rods. 

Due to its simplicity, zone-folding was recognized as a promising approach for the nanoscale, with the first theory proposal based on elastic waves in a phononic crystal with snowflake-shaped patterned holes \cite{brendel_snowflake_2018}.  

Shortly thereafter, Cha et al. \cite{cha_experimental_2018}  published the first experimental realization of an on-chip nanoscale topological metamaterial. Their experiment is based on a zone-folding design, and their on-chip phononic crystal is made out of a piezoelectric material (SiN), to be able to transduce electrical signals to mechanical motion. The schematic of the geometry is shown in Fig.~\ref{fig:zone_folding}d, with a unit cell size of $18 \mu m$ and a frequency around $15 {\rm MHz}$. The authors were able to detect in a space-time resolved way the propagation of wave packets along domain boundaries. This was achieved using an excitation electrode to inject flexural waves and a Michelson-interferometer, which can be scanned, for optical detection. 

The zone-folding trick has also been used for the first implementation of topological transport for surface acoustic waves  \cite{zhang_topological_2021}. 

\cite{ma_nanomechanical_2021}  proposed and experimentally demonstrated  an interesting extension of the zone-folding trick. Their scheme features  two  different perturbations:
the first breaks the translational symmetry  $\mathcal{T}_a$  and the second two-fold rotational-symmetry.  The strengths of the two perturbations could be tuned independently changing appropriate geometrical parameters in the same platform used in \cite{ma_experimental_2021},  allowing to interpolate between the valley-Hall and the (six-fold symmetric) zone-folding Hamiltonians while always keeping  the same form of the effective Dirac Hamiltonian  Eq.~(\ref{eq:Dirac_TRS}).

\subsubsection{Accidental degeneracy of Dirac cones}
\label{Section:accidental}
\begin{figure}
\includegraphics[width=1\columnwidth]{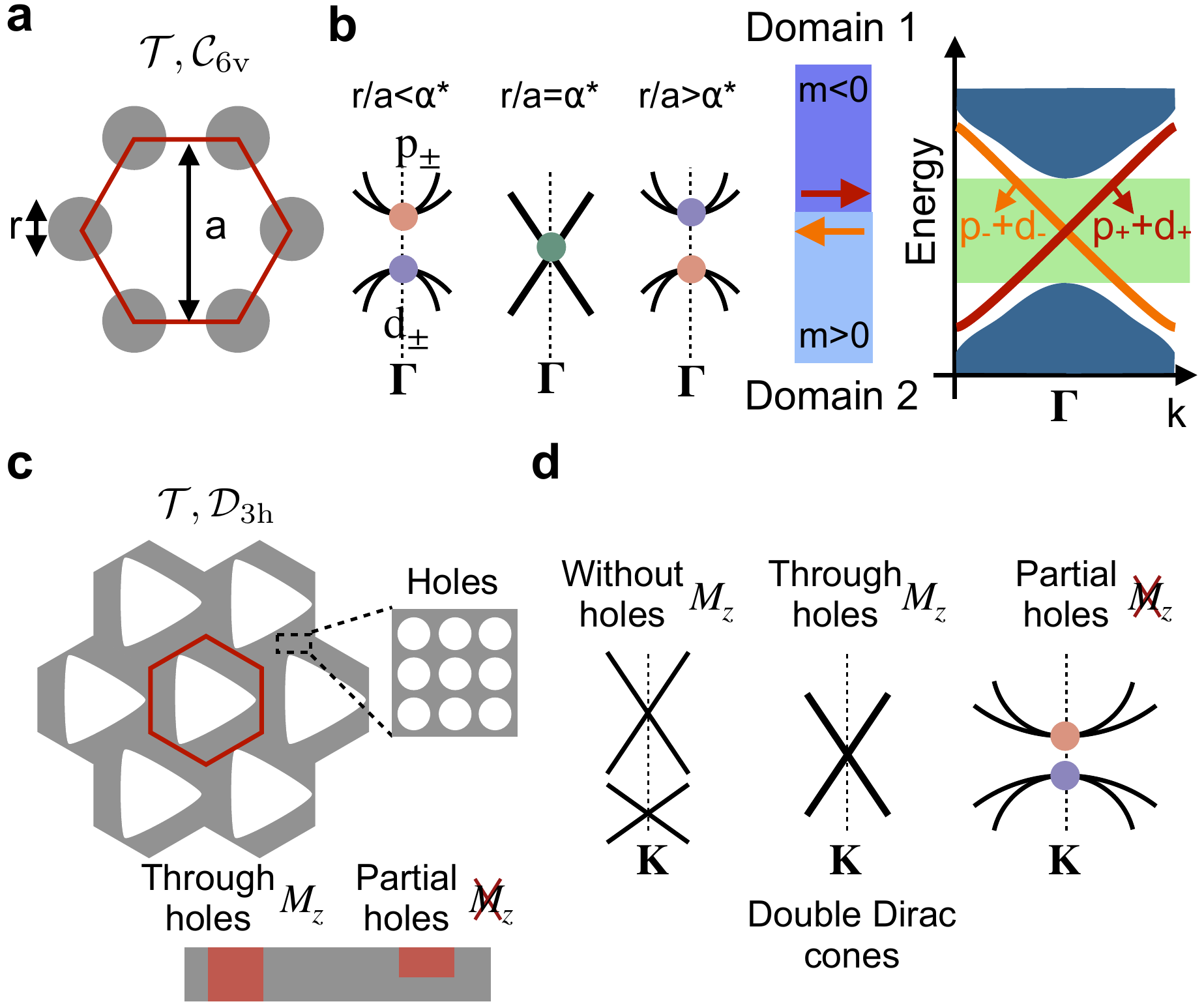}
\caption{Accidental degeneracy of Dirac cones. 
\textbf{a}, Unit cell of an array of rods with $\mathcal{C}_{6v}$ symmetry, acting as scatterers for acoustic waves \cite{he_acoustic_2016}. 
\textbf{b}, (left) The ratio of the diameter to the periodicity is varied to observe the band inversion via a double Dirac cone at the $\boldsymbol{\Gamma}$ point. Similar to the zone folding scheme,  helical edge states appear in the bulk band gap. 
\textbf{c-d}, Metal plates patterned with holes of two sizes: large holes are used to obtain the Dirac cones at the $\mathbf{K}$ point for modes that are symmetric and anti-symmetric about the x-y plane, as suggested in \cite{mousavi_topologically_2015}. The smaller holes (inset) are used to engineer the dispersion for making the two cones identical. One can pattern partial holes to break the mirror symmetry $M_z$, and thereby gap the double Dirac cones \cite{miniaci_experimental_2018}.
}
\label{fig:accidental_degeneracy}
\end{figure}

The zone-folding scheme described above provides a systematic engineering strategy to obtain double Dirac cones in a crystal: consider the band structure of the larger unit cell of a $\mathcal{C}_6$-symmetric crystal. In contrast, as the name itself suggests, the "accidental degeneracy" approach relies on fine tuning. However, there is a logic behind the scheme, which is to take advantage of the fact that a $\mathcal{C}_{6}$-symmetric crystal automatically features  two-fold degenerate symmetric states, $p_{\pm}$ and $d_{\pm}$. If one can tune the parameters of the geometry so as to bring a pair of $p$ and $d$ states to the same frequency, then this would form a double Dirac cone at the $\boldsymbol{\Gamma}$ point (see the schematic bands in Fig.~\ref{fig:accidental_degeneracy}b). Afterwards, one may follow the recipe that we have already described for the zone-folding scheme, to create counter-propagating topological edge-states out of a double Dirac cone.

The first experimental demonstration of phononic edge states produced via the accidental degeneracy mechanism was in the context of acoustic waves scattered by a lattice of steel rods \cite{he_acoustic_2016}. By varying the ratio of the diameter of steel rods to the lattice periodicity across a threshold value, the order of a pair of $p$ and $d$ Bloch waves could be exchanged, cf Fig.~\ref{fig:accidental_degeneracy}a,b. Close to the $\boldsymbol{\Gamma}$ point and for ratios in the vicinity of the threshold value, the group of four bands is described by the Hamiltonian Eq.~(\ref{eq:Dirac_TRS}), allowing to engineer helical edge states at a domain wall for the mass parameter. We note in passing that  essentially the same scheme has been independently proposed by \cite{mei_pseudo-time-reversal_2016}. Going back to \cite{he_acoustic_2016},  their setup featured a new way of arranging the two domains in a beam-splitter like arrangement, that has also been very useful in later helical edge-state implementations to deduce to what extent the helical transport is spin-polarized. The approach based on the accidental degeneracy of pairs of $p$ and $d$ Bloch waves has been also leveraged to implement topological elastic vibrations in a patterned plate, see   \cite{yu_elastic_2018,yu_critical_2021}.

There exists an alternative approach to obtain degenerate Dirac cones accidentally,  which again involves not only fine tuning but also a clever use of the symmetries. 
As for the valley Hall effect, one starts from a geometry with a wallpaper group  supporting symmetry-protected Dirac cones at the $\mathbf{K}$ and $\mathbf{K'}$ points. Typically, planar geometries (such as phononic crystals or flexible plates etc) used for engineered 2D transport have an out-of-plane mirror symmetry $M_z$, for reflection about the 2D plane. This prevents the mixing of modes that are symmetric and anti-symmetric with respect to this symmetry. Therefore, if one can tune the geometry to bring the two Dirac cones corresponding to these two modes to have the same Dirac speed and degeneracy point, then they form a double Dirac cone at the $\mathbf{K}$ and $\mathbf{K'}$ points (see the schematic bands in Fig.~\ref{fig:accidental_degeneracy}d). Subsequently, the mirror symmetry is broken, to produce a tuneable mass term and to eventually create  a domain wall between two topologically distinct  domains. This leads to a pair of counter-propagating edge states at each Dirac point. The underlying Hamiltonian can be viewed as comprising  two copies of Eq.~(\ref{eq:Dirac_TRS}). 

This idea has been presented early on by Mousavi and coworkers \cite{mousavi_topologically_2015}. In their design, they proposed plates patterned with holes arranged in a triangular lattice. The phononic crystal comprises two different feature sizes: the macro-lattice is designed to create a pair of Dirac cones, whereas the smaller features on the micro-lattice are engineered to make the pairs of Dirac cones coincide not only in frequency but also in velocity. Inspired by this theoretical work, Miniaci et al. \cite{miniaci_experimental_2018} later experimentally demonstrated the accidental degeneracy of two Dirac cones in macroscopic patterned elastic plates. Another implementation based on an elastic rod lattice has been demonstrated by \cite{mei_robust_2019}.

\subsubsection{Pseudomagnetic fields}
\begin{figure}
\includegraphics[width=1\columnwidth]{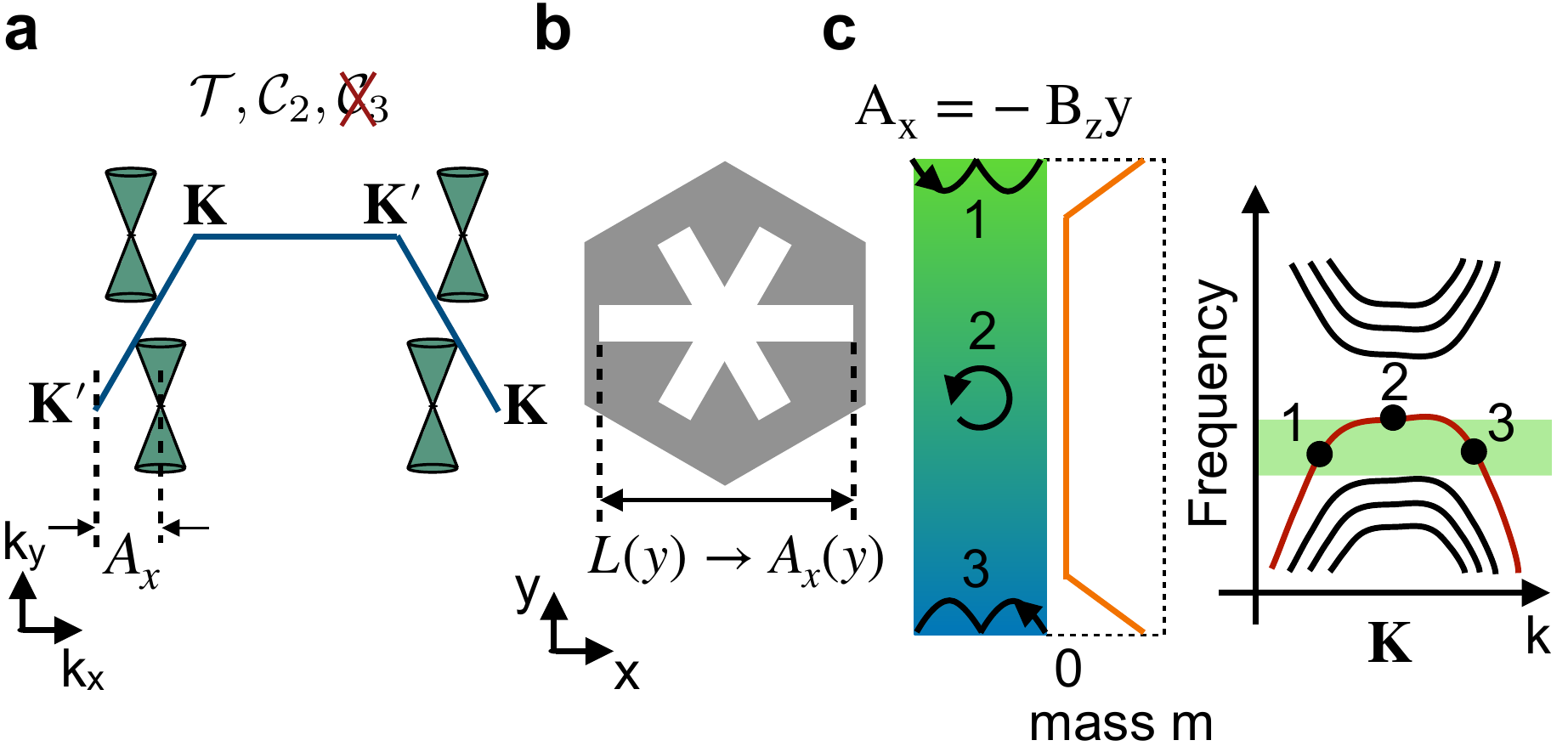}
\caption{Pseudomagnetic fields. 
\textbf{a}, Upon breaking the $\mathcal{C}_3$ point-group symmetry, the Dirac cones shift in the Brillouin zone.  The displacement from the valleys is given by $\mathbf{A}=(A_x,A_y)$ (here $A_y=0$).
\textbf{b}, Scheme proposed in \cite{brendel_pseudomagnetic_2017} to displace the Dirac cones in a phononic crystal. The horizontal arm-length of the snowflake-shaped hole is modified in a spatially-dependent way $L(y)$,  inducing a spatial dependence $A_x (y)$. 
    \textbf{c}, Translationally invariant strip  with smooth boundaries (left). The colors encode  the vector potential $A_x$ (for an out-of-plane constant magnetic field $\mathbf{B}=B_z\mathbf{e}_z$). The position dependence of the mass parameter $m$ (orange) defines the smooth boundaries.  Band structure (right) in the vicinity of the $\mathbf{K}$ point. The bulk band gap between the $n=0$ and $n=-1$ Landau levels is highlighted in green. Away from the $\mathbf{K}$-point the Landau levels turn into gapless edge states. The  semi-classical orbits for three modes are displayed in \textbf{c}.
}
\label{fig:pseudo_magnetic_field}
\end{figure}


All schemes for engineering Dirac Hamiltonians presented so far have in common that they make use of the mass term to generate topologically distinct domains  and, thus, helical edge states at a smooth domain wall.  An alternative to this approach is to use so-called {\em pseudo magnetic fields}.   Pseudomagnetic fields are indistinguishable from magnetic fields if one is allowed to focus on a single pseudo-spin direction, but have opposite sign for two opposite pseudo-spins directions. This ensures that the time-reversal symmetry is preserved, in contrast to  approaches in which the action of a magnetic field  is mimicked across the Brillouin zone (discussed above, for constructing Chern insulators). In graphene-based materials pseudo-magnetic fields are induced by  position-dependent strain \cite{kane_size_1997,manes_symmetry-based_2007}. This gave rise to the idea \cite{guinea_energy_2010} to engineer  constant  pseudo-magnetic fields via strain to observe Quantum-Hall physics that would otherwise require huge magnetic fields, as demonstrated shortly thereafter \cite{levy_strain-induced_2010}.

Like for the other Dirac-based approaches discussed above, pseudo-magnetic fields  can be understood by analysing how the underlying Dirac Hamiltonian  is modified after breaking  a symmetry.
In particular, we consider the same initial setup as in Section \ref{sec:valley-Hall} (wallpaper groups $p6$ or $p31m$ in the presence of the time-reversal symmetry ${\cal T}$), but now we analyze  what happens after breaking the three-fold rotational symmetry ${\cal C}_3$. In this case the cones are not  gapped (because the degeneracy is protected by the two-fold rotational symmetry)  but rather their tips  are displaced away from the high-symmetry points $\mathbf{K}$ and $\mathbf{K}'$. The location of the tip of  the cone, in relation to the high-symmetry $\mathbf{K'}$ ($\mathbf{K}$) point, can be represented as a vector ${\bf A}=(A_x,A_y)$ (see  Fig.~\ref{fig:pseudo_magnetic_field}a). Within a smooth-envelope approximation, the band structure in the vicinity of one of the valleys, say the $\mathbf{K}$ point, is then described by the following 2D Dirac Hamiltonian:
\begin{equation}
\label{Eq:dirac_eqn_pseudomagnetic_field}
\hat{H}_D(\mathbf{k}) = v(\mathbf{k} - \mathbf{K} - \mathbf{A}(\mathbf{x})) \cdot \hat{\boldsymbol{\sigma}}.
\end{equation}
We observe that the quantity ${\bf A}$ is analogous to the magnetic vector potential for the case of a relativistic charged particle in a magnetic field. In order to use this effect to produce a pseudo-magnetic field ${\bf B}=\nabla \times {\bf A}({\bf x})$ for phonons, ${\bf A}$ must vary spatially, i.e. the location of the cone in momentum space needs to shift as one moves along in real space. This can be achieved by having a spatially varying deformation of the ideal phononic crystal geometry (see  Fig.~\ref{fig:pseudo_magnetic_field}b). For a constant magnetic field in the z-direction ${\bf B}=B_0 \mathbf{e_z}$, a typical choice is to use Landau gauge, i.e. to set ${\bf A}=-B_0 y \mathbf{e_x}$, which implies a certain spatial pattern of deformation. The band structure then features Landau levels in the vicinity of the $\mathbf{K}$ and $\mathbf{K'}$ points. This also means that valley-polarized waves in the bulk of the material will experience a Lorentz force. 

Pseudo-magnetic fields have been  originally transferred from the electronic realm into the mechanical domain by \cite{brendel_pseudomagnetic_2017}, considering a nanoscale phononic crystal implementation, see Figure   Fig.~\ref{fig:pseudo_magnetic_field}b). The same concept was also analyzed for a tight-binding (mass-spring) mechanical model of strained graphene \cite{abbaszadeh_sonic_2017}. More recently, pseudomagnetic fields  have been implemented for acoustic waves by \cite{wen_acoustic_2019}. Apart from generating bulk pseudo-magnetic fields, \cite{brendel_pseudomagnetic_2017} have also put forward the idea of using those fields to engineer gapless helical  edge states, analogous to their chiral counterparts in the presence of  (time-reversal-symmetry breaking) magnetic fields.   As for the other design schemes producing helical edge states  based on the Dirac Hamiltonian, successfully achieving robust edge-state transport requires to suppress the coupling between different valleys by means of  smooth domain walls or  boundaries, cf Fig.~\ref{fig:pseudo_magnetic_field}(c,d).

\section{Challenges for Topological Phonon Transport}
\label{sec:Challenges}
\begin{figure}
\includegraphics[width=1\columnwidth]{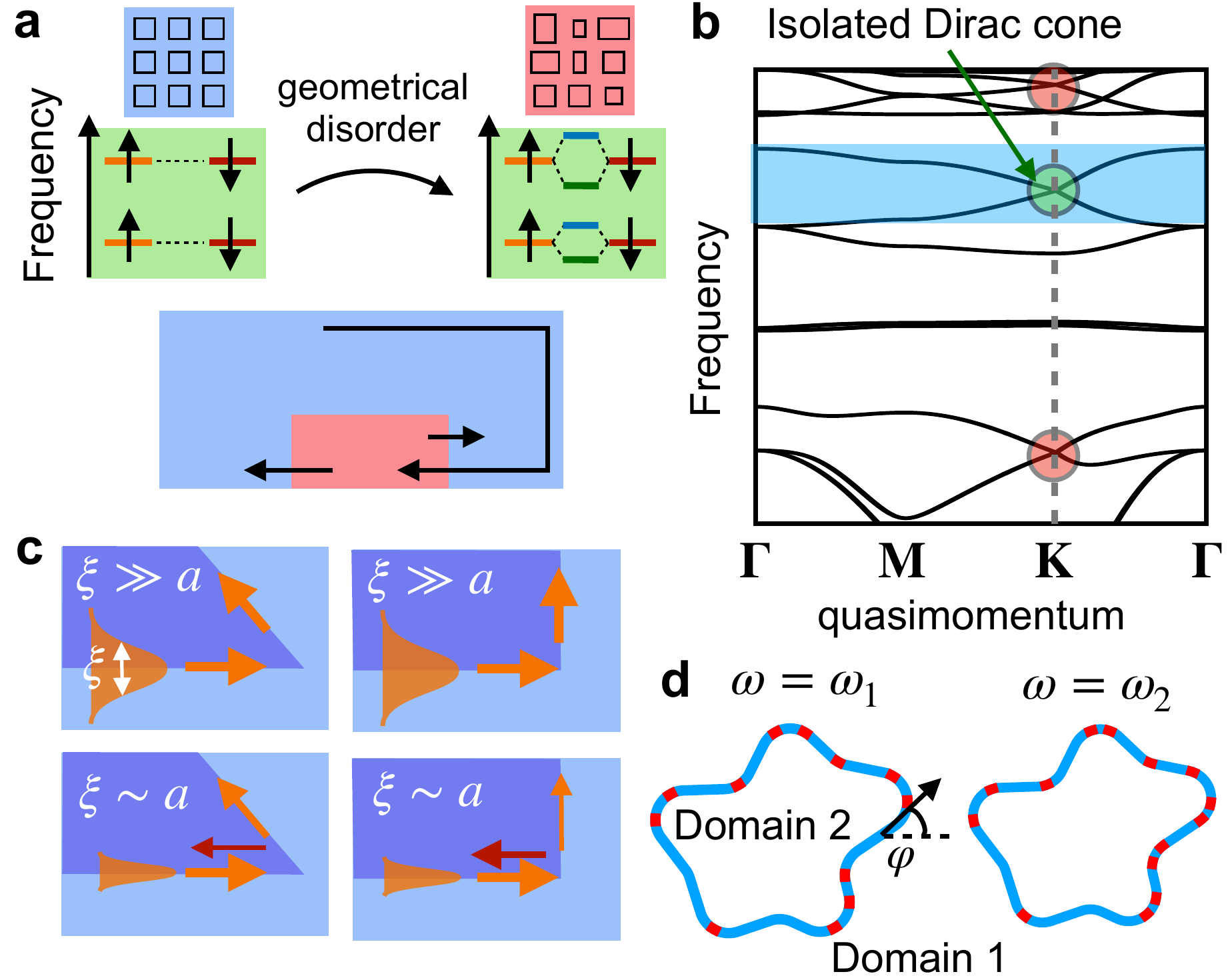}
\caption{Challenges for topological phonon transport. 
\textbf{a}, Effects of geometrical disorder in time-reversal symmetric systems. Disorder coupling opposite pseudo-spin directions also breaks Kramers degeneracy (top) and, thus, could induce backscattering (bottom).
\textbf{b}, Spectral isolation. Phonon band structure of an actual crystal featuring multiple Dirac cones at the $\mathbf{K}$ point in the Brillouin zone. The spectrally isolated Dirac cone is marked in green, others are marked in red. \textbf{c} and \textbf{d},
Backscattering for edge-states based on Dirac Engineering. \textbf{c} Transport with tight vs weak transverse confinement.  The reflection is not guaranteed to be small if the transverse localization length $\xi$ exceeds the lattice spacing $a$. The reflection depends also on the domain wall orientation (thickness of the arrow represents the wave intensity).  \textbf{d} Backscattering for a smooth domain wall and two values of the carrier frequency $\omega$. The reflection is enhanced at certain locations (red)  that depend both on the frequency and the domain wall orientation $\varphi$ (relative to the microscopic lattice).}
\label{fig:challenges}
\end{figure}


Various challenges will have to be addressed to fully realize the potential of topological phonon transport in actual applications.  

For instance, in order to realize Chern insulators for phonons, external modulation schemes are an important possibility, but it is  difficult to engineer those. As most of the future applications of topological phonon waveguides are envisaged in nanoscale devices, these engineering challenges will become even more formidable to solve.  For example, it is  not straightforward to achieve the required control of a platform to implement a Spin-Hall Hamiltonian across the entire BZ. Even when this is possible, Kramers degeneracy is protected by an engineered symmetry and, thus,  simple geometrical disorder could induce backscattering, cf Fig.~\ref{fig:challenges}a. As another example, the very promising schemes of topological transport and phonon control in optomechanical arrays (starting with \cite{peano_topological_2015}) in some cases rely on optical modes being tuned into resonance, which is still a challenge due to unavoidable fabrication disorder, and more progress in this domain is needed, also for many other applications beyond topological transport.


Hence, approaches based on engineered Dirac cones are favoured at the nanoscale \cite{cha_experimental_2018,ren_topological_2022,ma_experimental_2021,ma_nanomechanical_2021,zhang_topological_2021,xi_observation_2021,zhang_gigahertz_2022,wang_extended_2022}. Although for this case one can obtain Dirac cones by enforcing a suitable symmetry in the crystal, one still needs to perform additional engineering to get these cones spectrally isolated from spurious modes, cf Fig.~\ref{fig:challenges}b. Moreover, the smooth-envelope approximation and, thus, the suppression of backscattering relies on the assumption that the edge states are well localized in quasi-momentum. For this reason, tightly confined edge states are  more prone to backscattering, cf Fig.~\ref{fig:challenges}c. For sharp domain walls this translates into a constraint on the bandwidth, see our discussion in section \ref{eq:Dirac_Hamiltonian}.
In addition, the reflection  depends strongly on the geometry and carrier frequency of the wave. For sharp turns connecting straight segments of the domain walls, it is well-known that the reflection is strongly reduced for so-called zig-zag segments, oriented along the primitive lattice vector (of the smaller unit cell in zone-folding schemes), see e.g. \cite{lu_observation_2017}. Indeed, this type of domain walls are used in most experiments. The more detailed analysis in \cite{shah_tunneling_2021} advocated  for smooth domain walls, to eliminate the trade-off between backscattering and bandwidth, cf Fig.~\ref{fig:challenges}d. They further showed  that, in this scenario, the backscattering predominantly takes place at certain locations that coincide with special values of the local angle of the domain wall boundary, cf Fig.~\ref{fig:challenges}d. Future attempts at engineering topologically robust transport along edge channels in time-reversal-symmetric systems should take these observations into account.

Even when  backscattering is negligible, there is one obvious remaining challenge, namely mechanical dissipation: phonons can simply get lost while traveling along the edge channel. This is a common problem for transport in any physical platform without particle number conservation, and it is shared by electromagnetic waves in particular. Fortunately, great effort has been expended over the past two decades to reduce mechanical dissipation in nanomechanical systems \cite{cleland_foundations_2003,bachtold_mesoscopic_2022}, since it represents an important issue for a number of applications, ranging from sensing to quantum information processing. Strategies include the  choice of suitable materials, care in preparing the surfaces, the design of geometries where clamping losses are minimized and strain is distributed in an optimized way, as well as high tensile stresses. Nowadays, mechanical Q factors (the ratio of frequency and dissipation rate, $\Omega/\Gamma$) can reach $10^9$ \cite{tsaturyan_ultracoherent_2017} and more in GHz-frequency phononic crystal platforms \cite{maccabe_nano-acoustic_2020} that may be suitable for being turned into topological waveguides.

\section{Potential Applications and Future Research Directions}
\label{sec:applications}
\begin{figure}
\includegraphics[width=1\columnwidth]{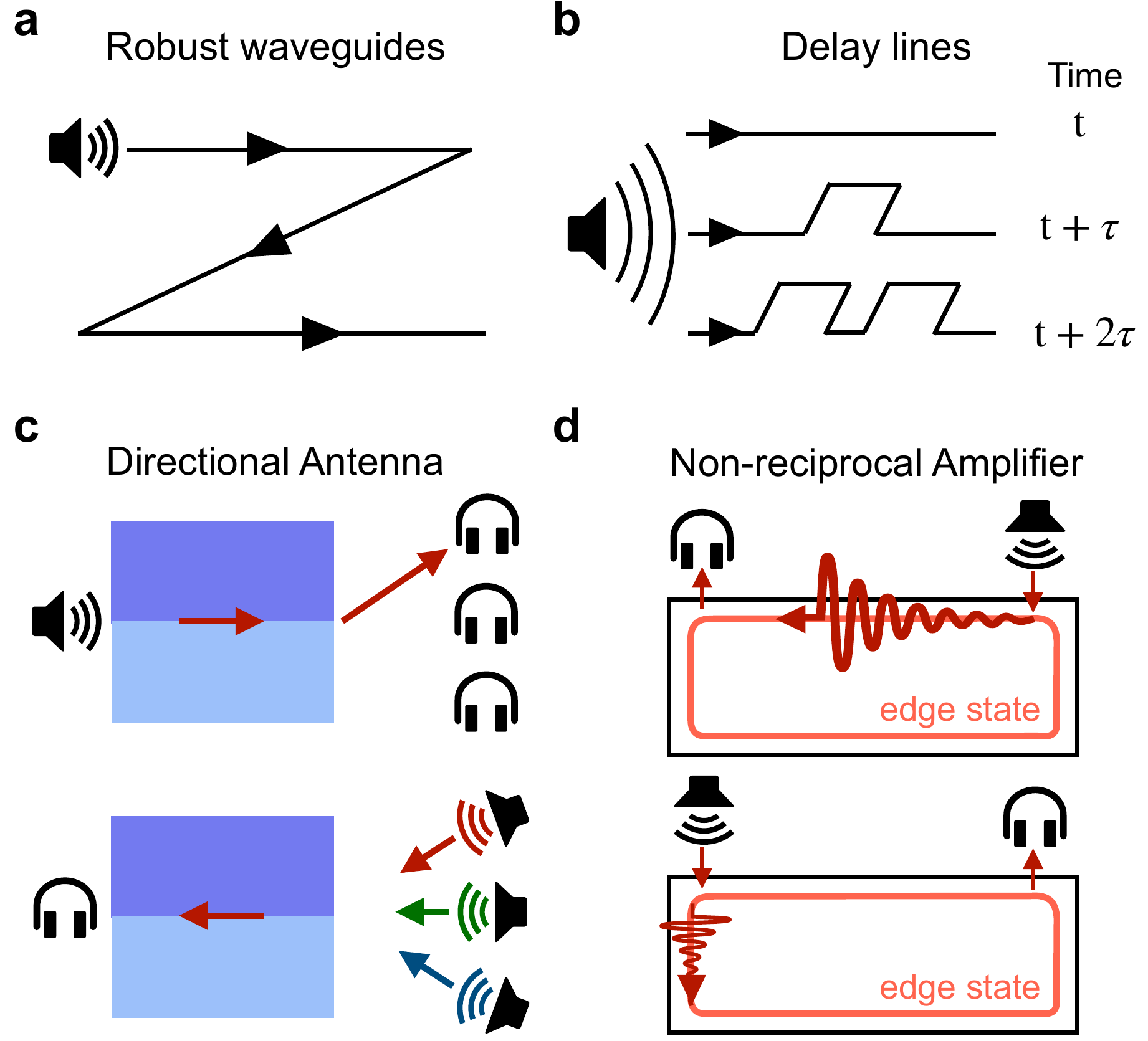}
\caption{Applications of topological waveguides. 
\textbf{a}, The advantage of topological waveguides is that the transport is robust against sharp turns, permitting compact devices and some insensitivity to fabrication imperfections. 
\textbf{b}, Topological delay lines constructed by elongating the path length with multiple loops between source and detector. 
\textbf{c}, Topological directional antenna in a time-reversal preserved topological waveguide. The path between the source and the detector through the waveguide is indicated by a red arrow. 
\textbf{d}, Non-reciprocal amplifier: the signal is amplified when it travels from the source to the detector, but any noise going the other way will be de-amplified, protecting the source.
}
\label{fig:applications}
\end{figure}

Topological transport phenomena have by now been explored in a variety of physical platforms (such as electrons, atomic matter waves, electromagnetic waves, sound waves, magnons). Much of the motivation behind this research is clearly due to the mathematical beauty of the underlying concepts and the fact that topological phenomena are generic and robust, not dependent on small details of the implementation. However, even the first known incarnation of topological transport, for electrons in the Quantum Hall Effect, was quickly understood to be promising for applications. Nowadays the precise quantization of the electrical conductance afforded by the topological protection is used to measure the Planck constant and define the kilogram. This history raises the question which features of phononic topological transport may eventually become useful for real applications. 

Such applications will want to somehow employ the resilience of phononic topological waveguides against backscattering due to disorder and sharp turns, cf Fig.~\ref{fig:applications}a. Additionally, in systems with time-reversal symmetry breaking, one can make use of the non-reciprocal nature of transport along the edge channels as well.

One such potential application is the delay line, which is used to add desired delays for energy to travel from a source to a detector. It can be implemented by simply elongating the length of the waveguide connecting the source and the detector. However, real applications for elastic wave delay lines will typically rely on micro- or nano-scale systems, both to work in the required frequency band set by the application and to restrict the overall size of the device. In order to maximize the ratio of waveguide length to required footprint, one wants to impose several turns of the waveguide, in a meandering structure (cf Fig.~\ref{fig:applications}b). In this setting, topological phononic waveguides can then be employed to have sound travel through relatively sharp turns without any backscattering. This was earlier on already suggested explicitly as an application for the topological transport of electromagnetic waves \cite{hafezi_robust_2011} and then translated to sound waves \cite{zhang_topological_2018-1}.

Another possible application is the so-called superdirectional acoustic topological antenna. The purpose of this device is to transmit sound in a desired very narrowly focused direction, and likewise to receive sound only from a source in the desired direction, cf Fig.~\ref{fig:applications}c. This has been demonstrated based on valley Hall waveguides for sound waves in air \cite{zhang_directional_2018}. Provided the interface at the edge of the device is designed properly, the valley-polarized character of sound transmitted along the waveguide guarantees that there is no undesired backscattering, in contrast to a standard waveguide. The directionality is controlled by the geometry, especially the lateral extent of the edge state wave function.


Another very promising potential application for unidirectional transport along topological edge channels is the non-reciprocal amplifier \cite{peano_topological_2016}. Such an on-chip device amplifies the signal coming from a source and injects the resulting amplified wave into some more conventional second-stage amplifier. The goal is to isolate the source (containing e.g. a fragile quantum device) from any noise that may be injected back from the second-stage amplifier, which can be a serious technological concern. This goal can be realized by employing the unidirectional edge channels of a Chern insulator, with extra nonlinearities in the wave equation ensuring that external driving can provide an energy pump and amplification (cf Fig.~\ref{fig:applications}d). A few years after the proposal in \cite{peano_topological_2016}, a related version of this kind of physics was realized for photonic systems in the form of the topological insulator laser \cite{harari_topological_2018,bandres_topological_2018}, with pumped edge states exhibiting lasing. More generally speaking, tuning dissipation and amplification (e.g. via optomechanical interactions and/or geometrical engineering) leads to the domain of non-Hermitian topology \cite{xu_topological_2016,bergholtz_exceptional_2021,del_pino_non-hermitian_2022}, which likely will become of greater prominence in the future also for topological phononics. 

Being able to control and reconfigure topological edge channels and re-arrange their connectivity can greatly improve their usefulness. In photonic systems, first ideas exist \cite{cheng_robust_2016,zhao_non-hermitian_2019}, e.g. involving macroscopic tuneability of the platform. In vibrational topological systems, several ideas have been explored as well \cite{zhang_topological_2018-1,darabi_reconfigurable_2020,tian_dispersion_2020}. Specifically for nanoscale phonons, an interesting possibility is offered by optomechanics \cite{aspelmeyer_cavity_2014}, e.g. using the optical spring effect to locally tune vibrational modes in and out of resonance, thereby enabling and blocking transport. In this way, light could switch topologically robust phononic edge channels in real time.

Such reconfigurable networks of phononic edge channels could be used to connect on-chip quantum devices like spins, quantum dots, or superconducting qubits (see e.g. \cite{habraken_continuous_2012}). They could also serve to study thermal and quantum transport of phonons in a new and unconventional but very well controlled regime \cite{barzanjeh_manipulating_2018}.

Up to now, we have almost entirely discussed purely linear dynamics, i.e. the propagation of non-interacting phonons (with the exception of the amplifier physics mentioned above, which relies on some nonlinearity for its implementation, but where the equations are nevertheless linear). A very rich set of new phenomena becomes available when nonlinearities do start to play a role. Owing to the typical strength of such nonlinearities in nanomechanical systems, it will be exceedingly hard to observe them on the single-phonon level (unless nonlinearities are drastically increased by coupling to nonlinear quantum devices). However, even the classical dynamics of nonlinear waves in edge channels of topologically nontrivial systems is a fascinating subject of study which has barely begun to be explored. Most existing demonstrations and analysis so far are based on photonic systems \cite{lumer_self-localized_2013,ablowitz_linear_2014,leykam_edge_2016,mukherjee_observation_2020,mittal_topological_2021} (as reviewed in \cite{smirnova_nonlinear_2020}). Nevertheless, first ideas now started to appear for implementing nonlinear topological phononics \cite{pal_amplitude-dependent_2018,chaunsali_self-induced_2019,chaunsali_stability_2021,darabi_tunable_2019}, where the nonlinearity affects the waves propagating along the edge channels or can even be responsible for producing such edge channels.

Enabling all of these applications requires inventive new designs, and the variety of approaches discussed in this review provide examples for that. In recent years, it was realized that deep learning methods can help with analyzing, predicting, and ultimately designing and optimizing topological band structures and topological transport. For example, in \cite{pilozzi_machine_2018}, a neural network was trained to predict topological band structures, taking as input a few parameters for simple tight-binding models. Going a step further, in \cite{peano_rapid_2021}, a neural network learned to predict topological band structures for arbitrary geometries of a crystal's unit cell, provided as an image. The network was trained to output an approximate tight-binding model, taking into account existing symmetries of the problem, and it was used to optimize geometrical designs (made possible by the fact that neural networks are differentiable function approximators). Future studies in this direction will likely be of great help in engineering better platforms.

Regarding experiments, in our view the most important overall trend in topological phononics is the emergence of first nano-phononic realizations of topological transport \cite{cha_experimental_2018,ren_topological_2022,ma_experimental_2021,ma_nanomechanical_2021,xi_observation_2021,zhang_topological_2021,zhang_gigahertz_2022,wang_extended_2022}. These are crucial as they will pave the way towards truly useful devices. Such efforts tie into the overall developments of the fertile areas of nanomechanics \cite{cleland_foundations_2003,bachtold_mesoscopic_2022} and optomechanics \cite{aspelmeyer_cavity_2014}, with ever greater control and ever better mechanical quality factors, encompassing phononic crystals, the use of nanomechanics for sensing, and for coupling to quantum devices.

\section{Conclusion}

In this review we have covered the recent developments of a wide range of ideas for engineering topologically protected transport of vibrations in the solid state and of sound waves in fluids. 

This field of topological phonon transport initially evolved out of the desire to explore and access, in vibrational and acoustic systems, the same conceptually intriguing phenomena that were already established for electrons and which had begun to be analyzed for electromagnetic waves and cold atoms as well. A fruitful combination of theory and experiment has since driven this field forward. Investigations on the experimental side started at the macroscale. At the same time, theoretical works were already exploring possibilities for future nanoscale platforms, where indeed most of the promising applications for topologically protected transport of vibrations reside. 

As devices are miniaturising and becoming increasingly densely packed with time, the robustness of topological waveguides will help to route the vibrations along desired paths without any scattering losses. The non-reciprocity, in situations with engineered time-reversal symmetry breaking, can be exploited for isolation, e.g. in non-reciprocal amplification. Experimental efforts in the field of nanoscale topological phonon transport are currently only in their beginning stages, and the investigation of quantum phenomena is likewise an outstanding challenge.

To add to these promises, such devices can turn into hybrid platforms, by incorporating coupling to microwaves, optical waves, or (e.g.) spin waves. This can enable novel possibilities like re-configurable topological waveguides. All of these functionalities of topological vibrational waveguides make them a promising candidate to be used in future integrated phononic chips. Only time will tell which of the many concepts and types of platforms covered in this review turn out to be the most suitable for future technological applications, and considerable further developments will be required towards this end.




\vspace{2mm}
\noindent\textbf{Acknowledgements}\\ 
T.S. acknowledges support from the European Union Horizon 2020 research and innovation programme under the Marie Sklodowska-Curie grant agreement No. 722923 (OMT). F.M. acknowledges support from the European Union's Horizon 2020 Research and Innovation program under Grant No. 732894, Future and Emerging Technologies (FET)-Proactive Hybrid Optomechanical Technologies (HOT).

\bibliography{bibliography}

\end{document}